# Benchmarking Adiabatic Quantum Optimization for Complex Network Analysis

Ojas Parekh, Jeremy Wendt, Luke Shulenburger, Andrew Landahl, Jonathan Moussa, John Aidun

Sandia National Laboratories

# Benchmarking Adiabatic Quantum Optimization for Complex Network Analysis

Ojas Parekh
Discrete Math & Optimization
Center for Computing Research

Jeremy Wendt
Embedded Systems Engineering

Luke Shulenburger
HEDP Theory

Andrew Landahl, Jonathan Moussa, John Aidun
Advanced Device Technologies
Center for Computing Research

Sandia National Laboratories
P.O. Box 5800
Albuquerque, New Mexico 87185-1326

### Abstract

We lay the foundation for a benchmarking methodology for assessing current and future quantum computers.  We pose and begin addressing fundamental questions about how to fairly compare computational devices at vastly different stages of technological maturity.  We critically evaluate and offer our own contributions to current quantum benchmarking efforts, in particular those involving *adiabatic quantum computation* and the *Adiabatic Quantum Optimizers* produced by D-Wave Systems, Inc.  We find that the performance of D-Wave's Adiabatic Quantum Optimizers scales roughly on par with classical approaches for some hard combinatorial optimization problems; however, architectural limitations of D-Wave devices present a significant hurdle in evaluating real-world applications.  In addition to identifying and isolating such limitations, we develop algorithmic tools for circumventing these limitations on future D-Wave devices, assuming they continue to grow and mature at an exponential rate for the next several years.

# ACKNOWLEDGMENTS

Access and computer time on the D-Wave machine located at NASA Ames Research Center were provided by the NASA Ames Research Center and the Universities Space Research Association.

# Contents

# Figures

# Executive Summary

Our goal is to lay the foundation for a benchmarking methodology that can be applied to assess current and future quantum computers. We pose and begin addressing fundamental questions about how to fairly compare computational devices at vastly different stages of technological maturity. We also critically evaluate and offer our own contributions to current quantum benchmarking efforts, in particular those involving adiabatic quantum computing and the Adiabatic Quantum Optimizers (AQOs) produced by D-Wave Systems, Inc. We find that, over a range of hard complex network analysis problems, D-Wave's AQOs perform on par with classical approaches — this is impressive considering these devices are harnessing an infant technology which is not well understood. However, current architectural limitations of D-Wave devices present a significant impediment to evaluating their performance on real-world applications. In addition to identifying and isolating such limitations, we develop algorithmic tools that may be used to circumvent these limitations if D-Wave devices continue to grow and mature at an exponential rate into the near future. Finally, we offer insight into the challenges of predicting how far into the future a D-Wave AQO might begin to offer a resource advantage over conventional computing technology in solving large-scale problems of practical interest and how such a feat might be accomplished.

We identify and address several obstacles to carrying out informative benchmark comparisons of the performance of an adiabatic quantum computer (AQC) and conventional (*classical*) digital computers. We focus on graph analysis problems inspired by large-scale real-world applications and consider instance families generated from real-world social network data. In addition we offer approaches to generate synthetic graphs that possess real-world structural properties and are amenable to benchmarking on a D-Wave AQO. Obstacles are encountered immediately, starting with two steps that are equivalent to *compiling* and *executing* a program on a classical computer: mapping the problem of interest onto the physical layout of the AQC and devising an algorithm that solves the mapped problem using the operations native to the AQC. In particular, the first of these steps presents a considerable challenge for using the one existing example of an AQC, the AQOs produced by D-Wave Systems, Inc., that is a special-purpose AQC tailored to solve Quadratic Unconstrained Binary Optimization (QUBO) problems.

We conclude that the D-Wave Two AQO (DW2), with 512 quantum bits (qubits), does not yet have enough qubits nor sufficient numerical precision to execute analysis of arbitrary dense and complex graphs beyond 30 or so vertices. Yet looking at the technology trajectory, rather than the current computing device, we estimate that the D-Wave AQO will become more useful for making detailed benchmarking comparisons with classical computers when the system size reaches approximately 2048 qubits, at which point exact classical algorithms will no longer be able to readily obtain optimal solutions for commonly used random QUBO benchmark instances on the *Chimera* graph (the graph representation of the physical layout of the DW2).

We also observe that published benchmarking studies may be requiring an impractical level of performance that happens to underrate the capabilities of the DW2. Those studies all assess the ability of a computer to find an *optimal* solution with some high frequency. An alternate metric that is of practical use is finding a *near-optimal* solution *all* of the time. We observe that the DW2 excels at this problem practically, as well as theoretically, being within 4% of optimal *almost always* on random Chimera instances.

D-Wave Systems, Inc. has achieved an impressive engineering feat in producing a functioning new technology that exhibits essentially the same problem-size-dependent performance (scaling behavior) as the combined, and *very mature*, hardware *and* software technology of classical computing, as indicated by published studies of performance on random Chimera instances. We believe that substantial improvements in AQO performance might be possible for certain kinds of problems, but this will require a broader community of researchers devoting more effort to improving the aforementioned compilation and execution steps for such systems.

We identify some problems for which empirical benchmarking results indicate that a DW2 outperforms the classical algorithms pitted against it. However, we offer evidence that there are better classical algorithms for these problems that are likely to outperform a DW2. We also show that slight changes in problem formulation can lead to widely varying D-Wave performance. In general, the proper selection and configuration of classical algorithms used in quantum benchmark studies is a critical issue. Further, we argue that a fairer and more informative comparison would be between the D-Wave AQO and classical QUBO solvers, rather than against specialized classical algorithms. In addition to providing a more direct performance comparison, this would allow researchers without classical algorithm expertise to benchmark QUBO on a D-Wave AQO. Because it is generally unknown which algorithm will perform best on a given instance of a combinatorial optimization problem, it is reasonable to cast the problem as an instance of QUBO and then, as a practical matter, solve it with a variety of heuristics. In this spirit, the D-Wave AQO should be regarded as one more algorithm for solving QUBO. Planned follow-on work includes creating a suite of classical QUBO solvers for direct comparison with D-Wave's AQO treatment of QUBO.

Our fundamental technical contributions include:
1) **Incorporating more realistic complex network instances into D-Wave benchmarking.** We devised an algorithm to generate, *directly on the D-Wave Chimera architecture*, synthetic graphs with properties of *real-world* complex networks. This algorithm enables benchmarking DW2 on larger real-world-like graphs than is otherwise currently possible. With this method we represent complex-network-like graphs having approximately 280 vertices with the 512 qubits of the DW2. Existing methods generally require up to $N^2$ qubits to represent an arbitrary $N$-vertex problem-domain graph, limiting one to graphs with approximately 30 vertices on the DW2;

2) **Bringing mathematical insights to bear on D-Wave benchmarking.** We show that the most commonly used random Chimera benchmarking instances are indeed (NP-)hard problems, as had been suspected but not proven. As alluded to above, solving a problem-domain QUBO instance with $N$ variables requires up to $N^2$ qubits on D-Wave device using the state of the art in problem compilation. We observe, drawing upon results from theoretical computer science, that this behavior is likely an inherent and inescapable limitation of the D-Wave Chimera architecture rather than the current compilation methods employed. Finally we observe that requiring a near-optimal solution *all of the time* instead of an optimal solution *most of the time* can drastically affect both the theoretical and practical complexity of solving a problem on an AQO;

3) **Developing algorithmic tools for trading off two critical AQO resources: qubits and numerical precision.** D-Wave AQOs are only able to represent numerical values at low precision, and this is often the critical bottleneck in solving real-world problems, which can require QUBO instances with large numerical values. We present algorithmic techniques for allowing a low-precision AQO to accept instances with high-precision data, at the expense of requiring additional qubits;

4) **Developing a Quantum Monte Carlo (QMC) solver to simulate results from an AQC**. We use our QMC simulator to implement hybrid classical and quantum heuristics for solving community detection problems on social networks;

5) **Demonstrating that subtle issues in problem formulation and configuration of classical algorithms can critically impact benchmarking results.** We show that very similar problems with slight differences in formulation can result in widely varying D-Wave performance.

This report begins with a compact introduction to quantum computing and the D-Wave AQO architecture, in Sections 1 and 2, respectively. Section 3 offers our perspective on benchmarking methodology and exposes some subtleties in D-Wave benchmarking. In Section 4 we discuss complex and social networks, and we present algorithms for synthetic generation of real-world-like complex networks; hurdles to representing real-world problems on a D-Wave AQO are also discussed in this context. Section 5 covers our benchmarking study, including our specific benchmarking methodology, problem definitions, and benchmarking results. The report concludes with a high-level and mostly self-contained discussion of D-Wave's technology trajectory in Section 6. Those readers solely interested in our perspective on questions such as: "Can D-Wave systems address real-world applications?", "How do future D-Wave and classical technologies compare?", and "When will D-Wave be able to address more challenging problems?" may skip to Section 6. A significant amount of additional information is provided in the appendices, which cover topics such as a classical benchmarking study of community detection in complex networks (Appendix C and Appendix D) and quantum entanglement in D-Wave devices (Appendix G).

Nomenclature

*AQC* – Adiabatic quantum computer (Section 1.2)
*AQO* – Adiabatic quantum optimization (Section 1.2)
*BGP* – Border gateway protocol (Appendix B.2.2)
*Chimera* – Qubit connectivity graph of D-Wave quantum annealers (Figure 1 and Figure 13)
*DW2* – D-Wave Two quantum annealer (Section 2)
*ETH* – Exponential-time hypothesis (Section 6.4)
*Ising* – Classical Ising spin glass problem, which is equivalent to QUBO (Section 1.3)
*MAIS* – Maximum affinity independent set problem (Section 5.3.3)
*MIS* – Maximum independent set problem (Section 5.3.1)
*NP* – The computational complexity class of decision problems, those with a "Yes" or "No" answer, for which a solution may be verified in polynomial time; a polynomial-sized certificate is provided to aid in the verification process (Technically, optimization problems are not decision problems and do not belong to the class *NP*; however, many types of optimization problems can be converted into polynomial-time-equivalent decision problems.) (Section 1.3)
*NP-hard* – The class of problems that are at least as hard as any problem in the class NP (Section 1.3)
*PTAS* – Polynomial-time approximation scheme (Section 3.3.1)
*QA* – Quantum annealer (Section 2)
*QMC* – Quantum Monte Carlo (Appendix F)
*QSim* – Our Quantum Monte Carlo simulator (Section 5.1.1 and Appendix F)
*QUBO* – Quadratic unconstrained binary optimization (Section 1.3)
*SA* – Simulated annealing (Sections 1.3 and 3.3.2)
*SAT* – Boolean satisfiability problem (Footnote 1)

# 1   Introduction to quantum computing

For many years quantum physics itself was viewed as an obstacle to quantum information-based computing. Over the course of the mid-1980s to the mid-1990s, this pessimistic view softened with a series of discoveries that gave a clearer picture of what quantum physics does and does not allow for computing, with the pessimism essentially vanishing with the discovery by Peter Shor of a fast quantum algorithm for factoring integers into primes [1]. It turns out that quantum physics isn't a barrier to computing — indeed it is quite the opposite — it is an *enabler* that can accelerate some computations far beyond what common sense suggests. By using the new set of rules that quantum physics provides, it appears one can solve some computational problems using fundamentally fewer steps than if one used the classical binary logic. This speedup is qualitatively very different from the more familiar speedups that miniaturization usually affords, such as faster clock speeds. It is more like the speedup one attains when one transitions from using Roman numerals to using Arabic numerals for arithmetic. To be clear, there is no mathematical proof that quantum computers are significantly more powerful than classical computers in solving natural problems. Indeed, it is conjectured that neither quantum computers nor classical computers can efficiently solve *NP-hard* problems (a well-known class of computationally hard problems which will be described in more detail later). However, as researchers continue to develop quantum algorithms that are faster than their best-known "classical algorithm" counterparts (with over 50 known quantum speedups to date [2]), the evidence for the power of quantum computing has mounted to a level that cannot be ignored.

Our main focus is on adiabatic quantum computing; however, before motivating adiabatic quantum computing we briefly describe the seminal circuit model of quantum computing, under which breakthroughs such as the aforementioned Shor's factoring algorithm were forged. This will set the context with key quantum concepts that will be revisited from a different perspective when we discuss adiabatic quantum computing.

## 1.1   The quantum circuit architecture

To be useful, a quantum computer requires a well-defined *architecture* that specifies methods for input, output, and processing. The most widely studied quantum computing architecture is the quantum circuit architecture. The way a quantum circuit works is as follows. First one prepares a set of qubits in some known state. Then one applies a sequence of "elementary" unitary transformations, each of which is called a "quantum gate," to the state, where each gate acts on some small number of qubits, typically one and two qubits. Finally one performs a sequence of measurements, each of which also acts on some small number of qubits. The size and structure of a quantum circuit depends on the specific quantum algorithm of interest. Variations on the quantum circuit architecture revolve around how exactly the input, gates, and output are adaptively chosen as a function of the algorithm instance.

In the "standard" quantum circuit architecture, the state preparation sequence is always the same — each qubit begins prepared in the state $|0\rangle$, where $|0\rangle$ and $|1\rangle$ denote the preferred *computational basis states* of the qubit ("preferred" because we assume the native quantum technology can readily prepare and measure qubits in this basis). The measurement sequence is always the same too, with each qubit measured in the computational basis. The one thing that changes from program to program is the gate sequence.

An important feature of the quantum circuit architecture is that it is *universal*. This means that given any dynamics described by quantum mechanics, a quantum circuit can create an efficient "digital" approximation of the dynamics to any desired precision by using enough qubits, gates, and measurements. In other words, the quantum circuit architecture is as powerful as quantum mechanics will allow.

## 1.2 The adiabatic quantum architecture

Adiabatic quantum computing is an approach to computing that is distinctly different than the quantum circuit architecture. Rather than performing a series of gate operations on a set of qubits to execute a quantum algorithm, an *adiabatic quantum compute*r (AQC) evolves a Hamiltonian governing the physical dynamics of a set of qubits, always keeping the qubits in the Hamiltonian ground state [3]. The *ground state* of a Hamiltonian is the lowest-energy state of the physical system it represents. An initial Hamiltonian, whose ground state is easily attainable, is evolved into a "problem" Hamiltonian whose ground state represents a solution to the problem of interest but can be difficult to obtain directly. A more detailed presentation of an AQC architecture appears in the next section.

While, in theory AQCs could be made to be universal quantum machines, this will require technological breakthroughs that enable a richer set of interactions among the qubits than just Ising two-body (anti-) parallel spin interactions. Currently, a *special-purpose* adiabatic architecture, *adiabatic quantum optimization* (AQO), is under very active development, with D-Wave Systems, Inc. actually selling a device built using superconducting flux qubits and inspired by the AQO architecture. As AQO is the technology most mature and likely the most — or only — available implementation of AQC for the next five to ten years, and we summarize it next. Indeed, as the objective of our investigation is an evaluation of the potential of AQC to benefit analysis of complex network analysis problems in the next five to ten years, our main focus has been on AQO because of its relative maturity. A brief discussion of the challenges of developing a *universal* AQC and its greater potential capabilities appears in Appendix A.3.

## 1.3 The adiabatic quantum optimization architecture

Proposed in 2000 by Farhi, Goldstone, Gutmann, and Sipser [3] as a quantum *algorithm* for "solving" the (NP-hard) Boolean satisfiability problem[1] (SAT), the AQO algorithm lends itself to a novel full-fledged *architecture* for "solving" a large class of combinatorial optimization problems. The word "solving" is in quotes here because, while the AQO algorithm does indeed give a solution to these problems when run for a long enough time, it is an open research question as to just how long the algorithm needs to run as a function of the problem's instance size to give a correct solution with high probability. The perspective of many mathematicians is that, if the required time scales exponentially with the instance size, the algorithm should not be said to be solving the problem. Classical computers can already "solve" the same class of problems in exponential time!

It is worth noting a recent trend of referring to AQO as implemented in the D-Wave devices as "quantum annealing." This is driven in part because of concerns by the research community that the D-Wave devices are not truly utilizing "adiabatic quantum physics" to solve combinatorial optimization problems. Instead, such hardware may be using some kinds of "quantum effects" in a manner reminiscent of the classical algorithm known as "simulated annealing."

---

[1] The Boolean satisfiability problem, known as SAT, is that of determining whether the variables of a propositional logic formula can be assigned values of either $TRUE$ or $FALSE$ so that the entire formula evaluates to $TRUE$. An example of such a formula is, $(x_1 \vee \neg x_2) \wedge (\neg x_1 \vee x_2 \vee x_3) \wedge \neg x_1$, where "$\vee$", "$\wedge$", and "$\neg$" are the logical OR, AND, and NOT operations, respectively. This formula may be satisfied by setting $x_1 = x_2 = FALSE$ and $x_3$ arbitrarily.

Unlike the quantum circuit architecture, the AQO architecture is special purpose and not universal. It is only able to directly solve unconstrained Boolean optimization problems, namely problems of the form

$$\min f(x) \text{ subject to } x \in \{0,1\}^n.$$

We will more precisely describe the class of admissible objective functions, $f(x)$, for AQO shortly. This class of Boolean optimization problems does not represent universal quantum computation, hence one cannot encode an arbitrary quantum algorithm in this form. However, solving this class of problems is highly useful; a great many optimization problems can be cast in this form, including some problems that have very large commercial value. For example, AQO algorithms have been proposed for machine learning [4], [5], satisfiability [6], and graph problems [7], [8]. Quite generally, unconstrained Boolean optimization problems are known to be "NP-hard," which means that they are at least as hard as the hardest problems in the computational complexity class NP. It is widely believed that no NP-hard problem can be solved efficiently (even with quantum computers); there is currently a $1M prize for anyone who can prove one way or the other whether every problem in NP can be solved in *polynomial time*, meaning they can be solved with a time and resource expenditure that grows polynomially with the input size. That said, even if these problems cannot be solved in polynomial time by an AQO device, it might be the case that for some kinds of problems, an AQO device may be able to solve such problems more quickly than any existing or near-term classical machine is able to.

Since AQO devices are able to solve all problems in NP, why would one even want a universal quantum computer? One may certainly encode problems such as integer factoring as a Boolean optimization problem and solve it on an AQO device. However, as noted above, an AQO algorithm may require exponential time to guarantee a solution. Shor's algorithm demonstrates that there is a way to use quantum resources to design an algorithm that solves integer factorization in polynomial time, but we cannot express this quantum algorithm on an AQO. In fact one may view an AQO as implementing just a single quantum algorithm for a specific class of Boolean optimization problems. The way that one "programs" an AQO is by encoding a specific problem of interest as an instance of the class of Boolean optimization problems that the AQO can solve. On one hand this is appealing since one needs to know nothing about quantum computation in order to solve problems; however, this places on the programmer the burden of efficiently expressing problems of interest as Boolean optimization problems. For the case of D-Wave, we will see that the latter can be a significant burden.

We now describe the AQO architecture in more detail. Most of our discussion also applies to the more general AQC architecture. The AQC architecture implements algorithms via a three-step process. For the specific case of AQO, one first prepares $n$ qubits in the state $|+\rangle$ (*i.e.*, the state $|0\rangle + |1\rangle$). This is the lowest-energy configuration, or "ground state," of the Hamiltonian $H_I = \sum_i \sigma_i^X$. This initial Hamiltonian, $H_I$ is turned on and then interpolated to a final "problem" Hamiltonian, $H_F = \sum_x f(x)|x\rangle\langle x|$, where $|x\rangle$ is a shorthand for the $n$-qubit computational basis state whose binary expansion is $x$. Note that the final Hamiltonian depends on the objective function, $f(x)$, of our Boolean optimization problem; the initial Hamiltonian does not depend on the problem, and one may use the same initial Hamiltonian for all problems. If we think of the AQO as operating on a time scale that goes from 0 and 1, then at time $s \in [0,1]$ we have the Hamiltonian,

$$H(s) = (1-s)H_I + sH_F.$$

The ground state of the final Hamiltonian, $H_F$ is a solution to the associated unconstrained Boolean optimization problem for $f$, and the AQO algorithm starts in the ground state of $H_I$. If the interpolation between $H_I$ and $H_F$ is performed slowly enough, then the system will remain in the ground state of $H(s)$ at each time $s$, and the initial state will have transformed "adiabatically" to the ground state of the final Hamiltonian. Finally, the state of each qubit is measured in the computational basis, returning the output of the algorithm. The only difference between the AQO architecture and the more general AQC architecture is the choice of the initial Hamiltonian, $H_I$ and the problem Hamiltonian, $H_F$. While AQO employs the specific $H_I$ and $H_F$ shown above, $H_I$ and $H_F$ are parameters of the AQC architecture, thus the above discussion also serves as a description of AQC. Like the quantum circuit architecture, the preparation and readout steps are problem-independent. Only the dynamics in the middle (described by the interpolating Hamiltonian) depends on the problem.

How slowly does the interpolation have to be run for the output of the AQC algorithm to be correct with high probability? In other words, what is the running time of the AQC algorithm for a given problem? The *adiabatic theorem* (or, more correctly, the *adiabatic approximation*) in quantum mechanics guarantees that as long as the total time, $T$, for the interpolation is chosen to be large enough, the final measurement will yield a value for $x$ that minimizes $f$ with high probability. One may expect that some problems would require more time to solve than others. This is captured by the fact the bounds on the minimum sufficient interpolation time, $T$ tend to depend inversely on the minimum gap between the energy levels of the ground state and first excited state of the Hamiltonian, $H(s)$ over the course of the interpolation. Thus problems that result in Hamiltonians with large energy gaps between ground states and first excited states would require less time for the AQC algorithm to produce an optimal solution with high probability (see Appendix A.1 for technical details).

It is a common misperception to believe that the AQC architecture represents "analog" computing and is riddled with all of the problems that analog classical computing has; most notably a complete lack of robustness. In fact, quite the opposite may be true — the AQC architecture is predicted by some experts to have exceptional implementation robustness, although this is a matter of ongoing debate. Quantum error correction may also aid in effective large-scale implementation of the AQC architecture [9]–[11]; however, our current understanding of error correction for the AQC architecture is more nebulous than for the quantum circuit architecture [12], [13]. The AQC architecture is definitely digital — the input and output are completely digital. Even though time is treated as a continuous variable during the interpolation, the state space of the computer is not described by a continuous variable. In fact, even seemingly discrete-time architectures, such as the quantum circuit architecture, rely on continuous-time evolution when implemented in hardware.

In physical systems, interactions tend to act on just two objects at a time. For example, if the qubits were represented by spin-1/2 particles, then a physically meaningful Hamiltonian would necessarily be restricted to acting on at most two spins at a time. With this restriction in mind, it is natural to consider the AQO architecture that is restricted to the Ising Hamiltonian as a final Hamiltonian:

$$H_{Ising} = \sum h_i \sigma_i^z + \sum J_{ij} \sigma_i^z \sigma_j^z, \tag{1}$$

which arises in many quantum information processing systems including semiconductors, superconductors, and laser-trapped atoms and ions. An advantage of this AQO architecture is that we may characterize the class of problem solved. In particular, the architecture solves *quadratic unconstrained binary optimization* (QUBO) problems, defined as

$$\min f(x) = \sum a_i x_i + \sum b_{ij} x_i x_j, \text{ subject to } x \in \{0,1\}^n.$$

By a linear transformation of variables from 0 and 1 to −1 and 1, an equivalent representation of the QUBO problem is the following, which we refer to as the *Ising problem*:

$$\min f(x) = \sum h_i s_i + \sum J_{ij} s_i s_j, \text{ subject to } s \in \{-1, +1\}^n.$$

In summary, adiabatic quantum optimization as realized by plausible near-term quantum technology will likely be restricted to solving QUBO problems. While seemingly narrow, this class of problems is NP-hard. Billions of dollars have been invested in attempts to solve QUBO problems with classical digital computing. Whether AQO technology will be able to outperform existing or near-term classical methods for solving these problems is an open question; we address certain aspects of this question in this report, focusing on eventual application to big data problems arising in the analysis of complex networks.

## 2   The D-Wave quantum annealer

D-Wave Systems has produced the world's first commercial computational devices that are designed to employ quantum effects to solve optimization problems. D-Wave refers to its devices as *quantum annealers* (QA), which is meant to reflect that, although a D-Wave system does evolve in an adiabatic-like fashion, it may not remain in a ground state throughout the entire evolution. The hope is that the quantum annealing process is able to terminate in a low-energy state of the final problem Hamiltonian, corresponding to a high-quality, but likely suboptimal solution to the optimization problem being solved. D-Wave QAs are not universal quantum computers; they implement an Ising problem Hamiltonian (see Eq. (1) in Section 1.3), hence D-Wave QAs are only able to solve QUBO, or equivalently Ising, problems. As noted, the extent to which D-Wave's QAs are leveraging quantum mechanical effects is currently a contested issue. We discuss the role of entanglement in D-Wave's QAs in Appendix G.

The current generation of D-Wave QA, the D-Wave Two (DW2), has an architectural capacity of 512 qubits; however, delivered systems may have fewer functioning qubits due to fabrication defects (see Figure 1). The physical strengths of the couplers between qubits are tuned according to the $J_{ij}$ values in Eq. (1), which are specified by the programmer. However, the current precision for the problem parameters, $J_{ij}$ and $h_i$, is 16 bits in the range $\{-1, -7/8, \ldots, +7/8, +1\}$ [14]. While the coupler values, $J_{ij}$, are typically used to represent numerical values from the problem domain, common techniques for formulating combinatorial optimization problems as QUBO instances require $J_{ij}$ values with large magnitudes that grow with the instance size. For these reasons, D-Wave's limited coupler precision is one of the most significant barriers to solving real-world problems on D-Wave devices. Moreover, due to control errors and other factors, the actual problem solved by a D-Wave QA may only be an approximate version of the one specified by the user (e.g., [15]).

A D-Wave device performs quantum annealing for a user-specified annealing duration, which is constrained to range between $20\mu s$ and $20ms$ on a DW2. Upon termination of the annealing procedure, a solution is returned to the user consisting of values in $\{-1, +1\}$ for each variable. The solution is not guaranteed to be optimal and a user would typically perform thousands of runs on the same instance to obtain a distribution of solutions, from which one or more are ultimately selected. For the specific parameters used in our D-Wave benchmarking experiments, consult Section 5.2.1.

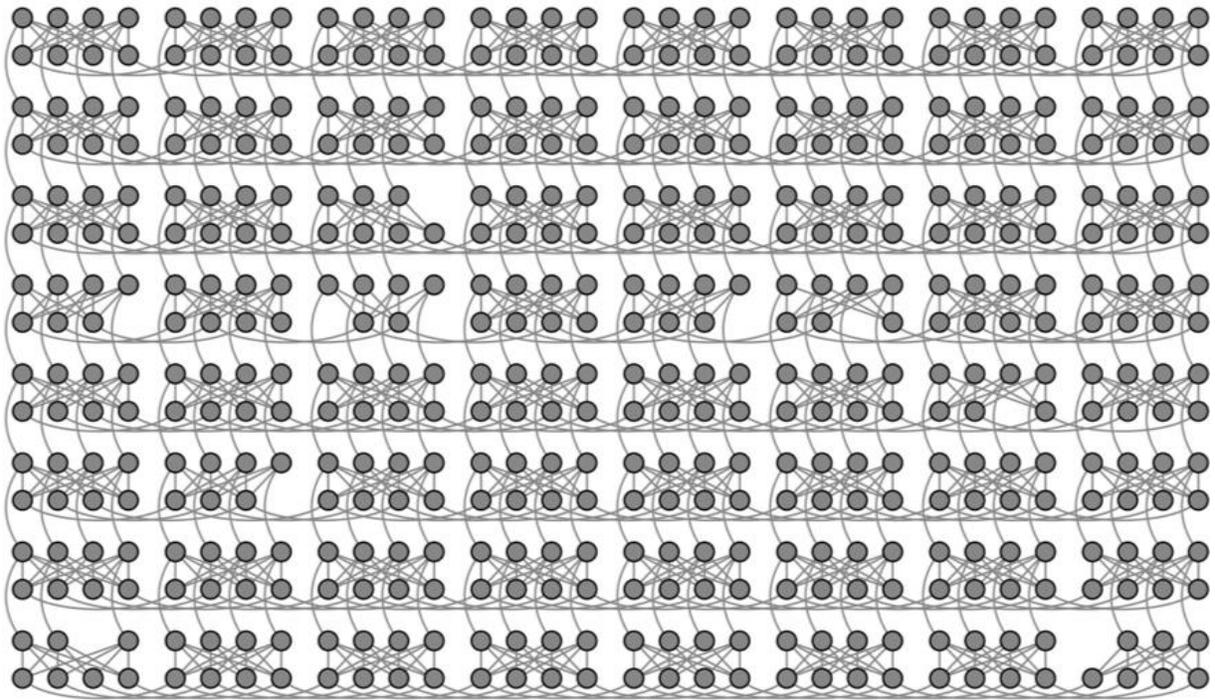

**Figure 1 D-Wave Two architectural connectivity diagram [15], known as the *Chimera* graph.** Nodes represent qubits and links between nodes represent a coupler between the corresponding qubits. Missing nodes and links are an artifact of fabrication defects and vary from physical device to device. Each eight-node cluster is called a *cell* or *plaquette*.

## 2.1 Solving combinatorial optimization problems on a D-Wave device

It is useful to think of QUBO (or Ising problems) as the "programming language" of a D-Wave QA. Although we cannot directly express traditional iterative programs as QUBO instances, QUBO is a rich and expressive NP-hard problem that is able to naturally model many types of combinatorial optimization problems. On classical computers, one solves combinatorial optimization problems by implementing a specific algorithm that takes an instance of a problem as input and outputs a solution that one hopes is close to optimal. In contrast, a D-Wave QA is essentially a hardware implementation of a specific quantum algorithm for solving QUBO problems that directly accepts a QUBO instance as input. Thus the programmer is freed from the burden of understanding and implementing an optimization algorithm; however, this comes at the cost of being tied to a specific implementation (D-Wave's) of a specific algorithm (QA) for a specific optimization problem (QUBO).

Just as classical computers implement programming languages atop an underlying machine language and hardware architecture, D-Wave QAs implement QUBO atop an Ising Hamiltonian that represents the underlying physical layout and connectivity of its qubits. A more expressive machine language may result in shorter programs at the expense of complex hardware that is more difficult to design, fabricate, and verify; a simpler machine language results in extra program overhead consisting of longer sequences of simpler operations. Quantum computers, including D-Wave's, are not immune to such tradeoffs.

D-Wave QAs are special-purpose computers in their physical layout, as well as in the problems they can solve. A qubit in a D-Wave QA cannot interact with every other qubit; it has a specific and restricted connectivity (see Figure 1). D-Wave refers to the class of *graphs*[2] representing the layout of its quantum annealers as *Chimera graph*s, and D-Wave QAs are only able to solve QUBO instances that conform to a Chimera graph. More precisely, in programming a D-Wave QA, a user is only able to specify $J_{ij}$ values for pairs of qubits that are physically connected — all other $J_{ij}$ values are effectively fixed to 0. The Chimera graph represents a balance in the tradeoff described above and is a qubit architecture that D-Wave has been able to successfully realize at successively larger scales. Though one is not able to directly input an arbitrary QUBO instance to a D-Wave system, the connectivity offered by Chimera graphs is universal in the sense that it can "efficiently" simulate solving QUBO on graphs with arbitrary connectivity; i.e., QUBO remains NP-hard when restricted to Chimera graphs. The tradeoff is that the mapping or "compilation" of a QUBO problem of interest (the *problem-domain* QUBO instance) to a Chimera QUBO instance may require, in the worst case, a quadratic increase in the number of variables. As each qubit represents one variable, mapping a problem-domain QUBO instance in $N$ variables onto a D-Wave QA can require up to $N^2$ qubits in the worst case (see Figure 2).

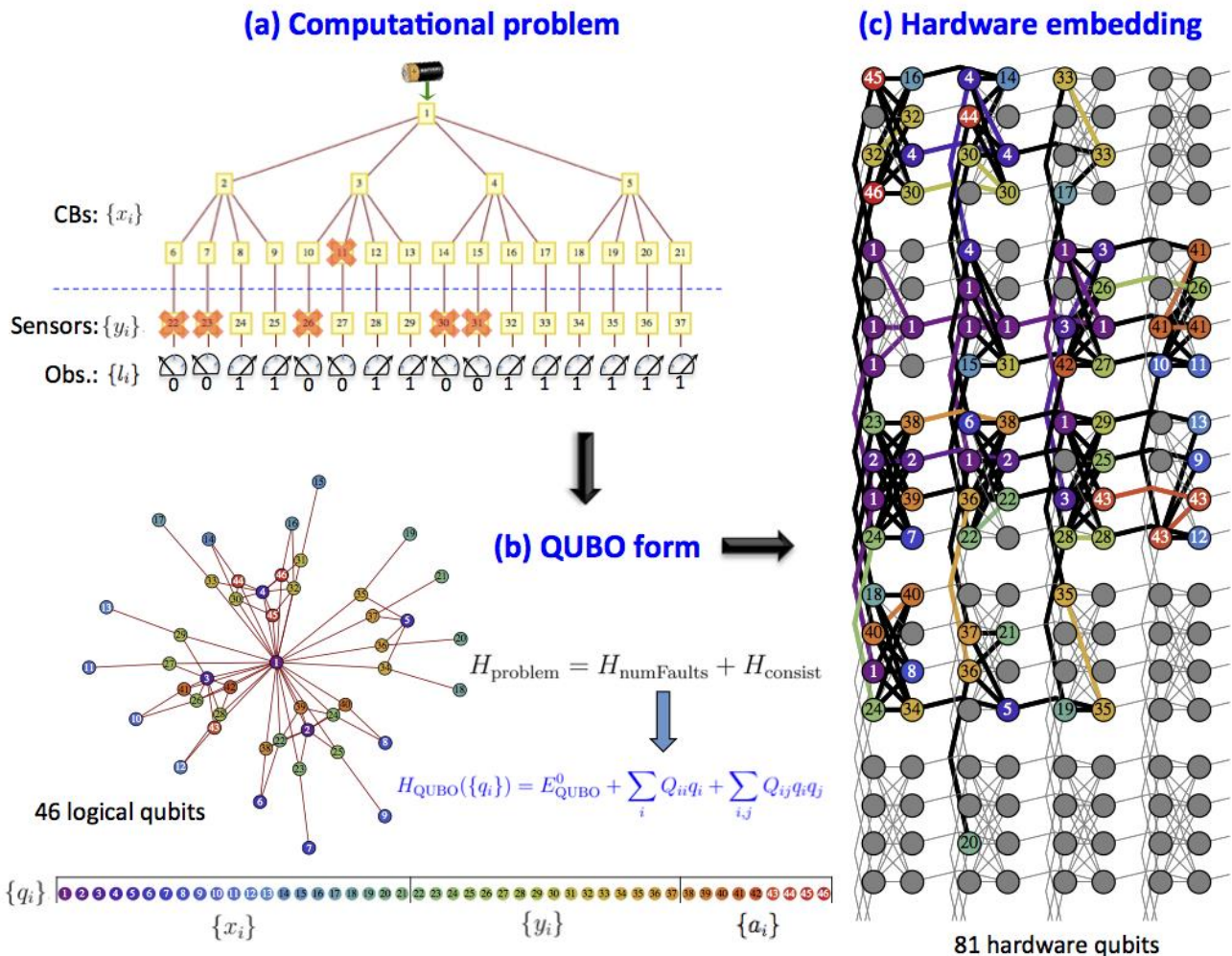

**Figure 2 An example of mapping a problem-domain graph into a Chimera graph [16].** A computational problem, (a) is captured as solving a QUBO problem on a graph, (b); this QUBO instance is then embedded into a Chimera graph, (c). Each vertex of the original graph is mapped to a set of connected vertices in the Chimera graph. Although the original graph has 46 nodes, the embedding uses 81 Chimera graph nodes, which correspond to qubits. A dense, fully connected graph on $N$ nodes requires $\sim N^2$ Chimera nodes when embedded in this fashion.

---

[2] A *graph* or *network*, usually denoted by $G$, in our context consists of a finite set of *nodes* (also called vertices), $V$, and a set of *edges*, $E$, which is a collection of (undirected) pairs of nodes. See Section 4.1 for more details.

This quadratic overhead may be acceptable from a coarse, theoretical outlook, but finer, practical concerns are important to us, especially given the current limited number of qubits available on D-Wave QAs. For a specific problem that we seek to solve, there may be many known representations of it as a QUBO problem. Such representations are likely to result in QUBO instances that are not expressible on a Chimera graph, so we must map the problem-domain QUBO instance onto another representation that casts the problem as a QUBO instance on a Chimera graph. Moreover, selecting different mappings will result in different QUBO instances of different qualities and sizes, and a D-Wave device may be able to solve some of these Chimera instances better than others. It may even be the case that we prefer a representation that requires more qubits if the representation is somehow structurally simpler and solves faster than smaller, but structurally complicated alternatives. As a step toward enabling users to reap the full potential of a D-Wave QA, in the next section we explore the types of subtleties described above from both a theoretical and practical perspective.

# 3   Benchmarking methodology

## 3.1   Benchmarking challenges

There are several layers of challenges in comparing the performance of two algorithms. For the purposes of benchmarking, an algorithm, be it quantum or classical, is simply a black box. When given an instance of a problem and a maximum execution time, the black box either terminates and produces a feasible, but not necessarily optimal, solution or fails to do so in the allotted execution time. Our general goal is to estimate the performance of each algorithm under consideration on problem instances of increasing size. Two questions are immediate: (i) what is an appropriate measure of performance, and (ii) how do we select a family of instances in order to derive a meaningful and valid comparison.

The notion of the performance of an algorithm depends on some resource being measured. The usual notion of performance is with respect to overall running time; however, care must be taken to define "overall running time" properly. For instance, is our concern the CPU time expended or wall-clock time? The former is a fairer comparison metric, but the latter may better reflect practical scenarios. There are cases when running time may not be an appropriate metric at all. Consider comparing a quantum algorithm versus a classical one. The notion of an *algorithm* is distinct from that of a *code* or software that may be readily executed. The former is a more abstract notion of a recipe for solving a problem, whereas the latter is an implementation of such a recipe on a specific platform. In the classical world we have many examples of both algorithms and software implementations, whereas in the quantum world, we essentially have only algorithms. One manifestation of this issue is that overall running time may not be informative in comparing a highly optimized classical code against a seminal implementation of a quantum algorithm on an immature quantum device. In the short run, a well-engineered implementation of a poor algorithm may outperform a poor implementation of an algorithm based on a powerful and clever idea; however, as our data sets grow larger we expect the latter to outperform the former. A benchmark that considers this also allows for the possibility of better-engineered versions of algorithms in the future. *Thus, we focus on relative performance, represented by growth curves indicating scalability over instances of increasing size, rather than absolute performance.*

Instance selection is another critical issue. Because we want benchmarks that are indicative of how algorithms will perform on real-world problems, randomly generated QUBO problems on the Chimera graph are unlikely to provide useful benchmarking results. Instances derived from one-off real-world data sets can also be a poor choice. One reason is that in order to sketch the asymptotic scaling for our algorithms, we need *families* of related instances rather than one-off data sets. Thus we seek families of related test problems of differing size that have real-world character. Indeed, corpuses of data of varying sizes are often available; however, the issue here is that there is possibly no relation between the individual data sets in the corpus. For example, size may not correlate with complexity and we may have a corpus where all the large data sets are "easy," while the small ones are "hard." Our goal is to capture some inherent notion of complexity that is independent of size. Consider a text corpus consisting of Shakespeare's plays and a collection of e-mail messages. This corpus certainly has both large and small instances, but these instances are unrelated. On the other hand consider a corpus consisting of Shakespeare's plays as well as his letters. Now we have a relation among the instances that transcends size – namely, there were all written by a single individual. We may be able to find idealized instances or synthesize families of related instances; however, they may not capture key characteristics of real-world data. On the other hand, real-world data sets tend to be unrelated to one another. This sort of tradeoff between rich but complex versus simper but less accurate models is not uncommon; we saw another example of this type of tradeoff above in CPU versus wall-clock time. The latter is a simpler and more natural model but fails to capture how a computer truly operates.

The complexities described above are precisely why we focus much of our attention on generating and analyzing families of real-world and synthetic social networks (see Section 4). Below we address some of the other issues raised above in the context of recent studies and our present work on benchmarking D-Wave's QAs. We select graph analysis problems on *complex networks*[3] as a relevant and illustrative problem of big data analysis on which to demonstrate our benchmarking methodology. The issues we raise and consider apply broadly to general quantum computing, as well as other types of emerging non-classical technologies.

### 3.2   Our goals and contribution

We view D-Wave's QAs, and more generally AQC, as a technological trajectory. Although we use a current generation D-Wave QA as an experimental tool, the questions we address and techniques we develop apply to future generations as well. Even if we cannot prove that current quantum devices offer a computational advantage on some class of interesting and practically relevant problem, we seek to poise ourselves to answer these types of questions as quantum devices become larger, more robust, and more capable.

Some of the immediate goals of our study, as well as longer-term aspirations, are to:
- Assess the current and projected future performance of D-Wave QAs (and AQCs);
- Compare them against variety of classical algorithms, including real-world heuristics on real-world-like instances;
- Identify barriers and workarounds in formulating and representing problems on D-Wave QAs;
- Place the D-Wave community benchmarking results in context with one another, as well as with our findings;
- Fill the gap between theoretical worst-case analysis of quantum algorithms and quantum software engineering on practical instances with empirical computational results.

In the context of these goals, we consider the following our main contributions:
- We bring real-world algorithms, problems, and instances to bear;
- We introduce more rigorous benchmarking methodology;
- We expose the importance of selecting appropriate metrics, algorithms, and configuration;
- We draw upon the rigor of theoretical computer science techniques and results to offer new insights on the D-Wave QAs;
- We demonstrate the formal NP-hardness of a widely used class of Ising instances;
- We develop new tools and techniques for trading off coupler precision and qubits;
- We explore QUBO-based approaches to complex network analysis problem.

Our contributions collectively offer a methodology to aid one in predicting whether maturing AQCs, such as D-Wave's QAs, are likely to have a near-term impact on real-world applications; we also identify impediments to such an impact.

---

[3] Our notion of a complex network is a graph arising in the modeling of complex social, physical, or other real-world phenomenon. We think of social networks as a kind of complex network, though we often use the two terms interchangeably. See Section 4 for more details.

## 3.3 Subtleties in D-Wave benchmarking

In this section we highlight some of the pitfalls and subtleties that we have both encountered in our own work and observed in recent D-Wave benchmarking efforts [17]. Our goal is simple but important: we seek to alert the reader that while benchmarking itself may very well be a "simple" matter of running computational experiments and processing collected data, drawing valid and meaningful conclusions from benchmarking endeavors is an art that warrants some care.

We bring insights from theoretical computer science and combinatorial optimization to bear on benchmarking quantum algorithms and discuss the following issues.
- What is an appropriate measure of success?
- Which classical algorithms ought to be used for comparison?
- How does one select appropriate benchmarking instances?
- How large are hard instances?

### 3.3.1 What is an appropriate measure of success?

Consider the following success criteria:
(i) for 99% of instances, produce an optimal solution,

or

(ii) for all instances, produce a solution that achieves a value that is 99% of the optimal.

From a practical perspective the two seem virtually indistinguishable, and one tasked with solving a real-world combinatorial optimization problem would likely be happy with an algorithm meeting either criterion. However, surprisingly enough, the theoretical complexities of (i) and (ii) can differ drastically. In fact this is the case with D-Wave's QAs. For solving QUBO problems on the D-Wave Chimera graph, we observe that, from a formal perspective, (i) is likely NP-hard, while (ii) is solvable in polynomial time by a classical algorithm. This is counterintuitive and, as far as we know, has not been previously observed in the D-Wave benchmarking community.

We note, however, that "polynomial time" does not mean practically efficient, as the currently best-known algorithm for criterion (ii) follows from a *polynomial-time approximation scheme (PTAS)* for QUBO on the Chimera graph by Saket [18]. A PTAS is a polynomial-time algorithm that, for any constant $\varepsilon$ in (0,1], is able to always deliver a solution whose value is at least $(1 - \varepsilon)$ of the optimal; however, the running time can depend exponentially on $\varepsilon$ (since $\varepsilon$ is a constant that is independent of the instance size). PTAS's are technically polynomial-time algorithms, but are rarely of practical use since for small values of $\varepsilon$, the constant factors in the running time can be astronomical. However, the existence of a PTAS is encouraging as it leaves open the possibility that a more efficient polynomial-time algorithm exists.

We complement the above observation by demonstrating that criterion (ii) also appears to be easier from a practical perspective. D-Wave performance on random Ising instances on the Chimera graph with no linear term ($h_i = 0$, for all $i$, in Eq. (1) from Section 1.3) tends to scale as $\sim c^{\sqrt{N}}$ where $N$ is the instance size (Section 5.2.3 or [19]). However, we observe that even though finding a precise optimal solution appears to be hard, D-Wave is *almost always* near optimal — *all* runs produced a solution achieving value at least 96% of the optimal for our experiment in Figure 3.

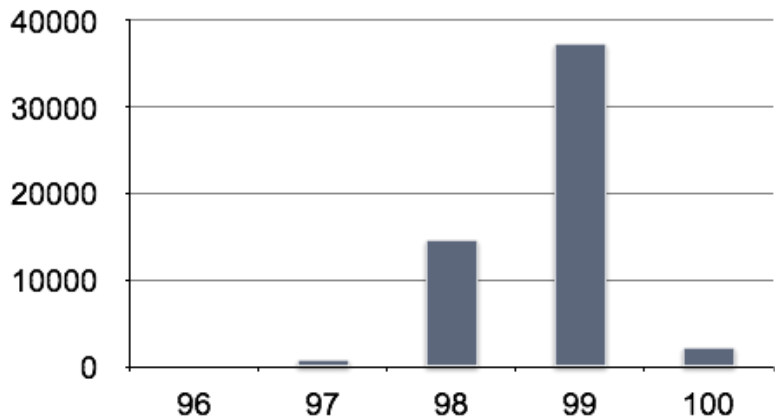

Solution quality (relative % of optimal)

**Figure 3 Histogram of solution quality as relative percentage of optimal (100% corresponds to optimal).** The results are over 55,000 random Ising instances on the Chimera graph with couplers uniformly in $\{-1, +1\}$ and no linear term.

Since both criteria (i) and (ii) seem equally valid from a practical perspective, which should one select? Following the seminal work of Boixo et al. [20], the D-Wave benchmarking community seems to have adopted (i) as its standard, which raises the question: has the community been focusing on an artificially hard success criterion, when one that is potentially easier for classical computers might suffice in its place?

### 3.3.2    Which classical algorithms ought to be used for comparison?

Although simulated annealing seems like a natural classical counterpart and point of comparison for quantum annealing, Hen and Young [21] observed empirically that for a certain variant of the Boolean satisfiability problem, both simulated annealing and simulated quantum annealing scale exponentially, while a reasonably efficient polynomial-time algorithm based on Gaussian elimination exists. We give another such example in Section 5.3.2. The existence of such problems should not be so surprising, since polynomial-time algorithms often employ sophisticated problem-specific insights; in fact it would be more surprising if a general approach like simulated annealing were able to somehow automatically and categorically take advantage of problem-specific insights. The point to be observed is that a suboptimal classical algorithm in a benchmarking study may lend an unfair and artificial advantage to quantum algorithms; yet, at the same time, it is extremely difficult to predict an optimal algorithm for a particular class of instances of a given problem. The polynomial-time algorithm for criterion (ii) from above is another example. Suppose we discovered a practically efficient polynomial-time classical algorithm that was able to always return a solution with value 99% of the optimal for any D-Wave QUBO instance. No doubt D-Wave's QAs would still remain scientific curiosities, but would anyone still be interested in using them?

On the other hand, an expertly engineered classical code may currently outperform a quantum alternative, but may not be able to sustain an asymptotic advantage as problem size grows. How do we compare a technology that we have had decades to hone versus an infant technology that has yet to mature? Even after selecting an algorithm, how do we know that it is configured properly for optimal performance?

We illustrate some of the above issues with an example. McGeoch and Wang [22] reported a 3,600x speedup for a DW2 over the CPLEX integer linear solver. As one of the first published D-Wave benchmarking studies, this work captured the imagination of the popular press:

> D-Wave's machine matched the competition—and in one case dramatically beat it. On two of the math problems, the D-Wave worked at the same pace as the classical solvers, hitting roughly the same accuracy. But on the hardest problem, it was much speedier, finding the answer in less than half a second, while CPLEX took half an hour. The D-Wave was 3,600 times faster. For the first time, D-Wave had seemingly objective evidence that its machine worked quantum magic (*WIRED Magazine*, [23]).
>
> D-Wave has battled to prove that its computer really operates on a quantum level, and that it is better or faster than a conventional computer. Before striking the latest deal, the prospective customers set a series of tests for the quantum computer. D-Wave hired an outside expert in algorithm-racing, who concluded that the speed of the D-Wave Two was above average overall, and that it was 3,600 times faster than a leading conventional computer when working on the specific type of problem that the quantum computer was built to solve (*Nature*, [24]).

This speedup is reduced from a factor of 3,600 to a factor of 2 when CPLEX is configured appropriately using an integer linear reformulation of the problem [25]. CPLEX can solve QUBO problems directly, but its primary design and purpose is in solving integer linear programming problems. Using CPLEX as a QUBO solver without further configuration is arguably fair in a comparison against a DW2 system. Moreover, McGeoch and Wang do employ an overall sound benchmarking methodology and do not claim anything more than their data suggests [22]. However, the overarching goal in quantum benchmarking is to compare quantum systems and algorithms against the best available or possible classical algorithms, and the selection and configuration of the classical algorithms to be pitted against quantum algorithms is challenging, even for algorithms experts. It is unreasonable to expect a few benchmarking studies to provide a comprehensive assessment, and one must remain mindful of the assumptions underlying such studies in order to draw valid conclusions.

Another facet of algorithm selection is considering the mode of operation. For example, there is a fundamental disparity in *heuristic* solvers, such as D-Wave's QAs, and *exact* solvers, such as CPLEX, that are designed to produce an optimal solution. A heuristic's only requirement is producing some feasible solution, with the hope that solution is indeed of high quality. On the other hand, an exact solver must not only produce an optimal solution, but it must in some fashion prove that it has done so. It is not uncommon, in practice, for a solver such as CPLEX to find an optimal solution early in its search process, but require significantly more time in verifying that no better solution exists. McGeoch and Wang addressed this issue by selecting other classical heuristic solvers for their study; however, CPLEX, even as an exact solver, generally performed better than the other heuristic solvers [22]. Subsequent D-Wave benchmarking studies have shown that there are both well-known and simple classical heuristics that perform at least as well as D-Wave on the types of instances used by McGeoch and Wang ([19], [20], [25]; we observe this as well in Section 5.2.3). Hence, the 3,600x speedup vanishes when appropriate classical heuristics are used.

One of our contributions is to address the above kinds of issues, as well as gaps in D-Wave benchmarking. For example, if one is willing to spend about $10,000,000 on a D-Wave QA, one should have no qualms spending another $50,000 for a modest classical computer or small cluster on which to execute available classical algorithms. Then one may simply run *all* available algorithms, quantum or classical, simultaneously and select the best solution returned by any of them. Thus one need not try making a choice in advance between using a quantum (annealing) algorithm or a classical algorithm. Some of the D-Wave studies seem to indicate that problems hard for quantum annealing may be easy for simulated annealing and vice-versa. This type of analysis was used to suggest that quantum annealing is different from and perhaps superior to simulated annealing. However, ironically, if this were true, then by running both algorithms simultaneously and selecting the better solution, we obtain a new algorithm that performs at least as well as either — yielding the best of both worlds. The fact that we do not routinely see such hybrid algorithms, especially in benchmark studies, is perhaps a blind spot fueled by our nature to compete rather than cooperate. Hybrid approaches have proven quite successful in solving Boolean satisfiability problems (SAT). Solving SAT problems has become an international competitive phenomenon [26], attracting experts and amateurs alike. Recently a new algorithm has taken the lead and has proven itself extremely difficult to beat. The approach, pioneered by Xu et al. [27], is simply to use a machine-learning technique that performs a cursory analysis of a given instance and then selects an algorithm to use on the instance from a portfolio of existing algorithms.

### 3.3.3 How does one select appropriate benchmarking instances?

As previously discussed, random instances are easy to generate but are unlikely to be representative of real-world problems. Moreover, random instances can be subject to phase transitions wherein the values of critical parameters can render instances extremely easy or extremely hard. Such parameters are often difficult to ascertain. Hence, by blindly generating random instances, one may be unwittingly solving artificially easier or harder problems than intended. How does one know that one is selecting or generating instances of appropriate difficulty? For example, the random instances used in recent benchmarking seem to have been assumed to be NP-hard based on Barahona's seminal work [28] on the complexity of Ising problems. We observe that the hardness of these instances does not follow directly from Barahona's work, and devise a new proof showing that these random instances are, indeed, NP-hard. The technical details of our proof are beyond the scope of this document and will appear in a technical article that is under preparation [29].

However, this is not the end of the story. The notion of NP-hardness pertains to worst-case complexity, whereas, from a practical perspective, we are more interested in randomized or approximate algorithms, such as simulated annealing or quantum annealing. Also, the instances themselves are drawn randomly (from a uniform distribution). Are such instances hard in some average-case sense? Defining a robust and meaningful notion of average-case complexity is a non-trivial task and computer scientists have struggled in making significant progress on understanding this phenomenon.

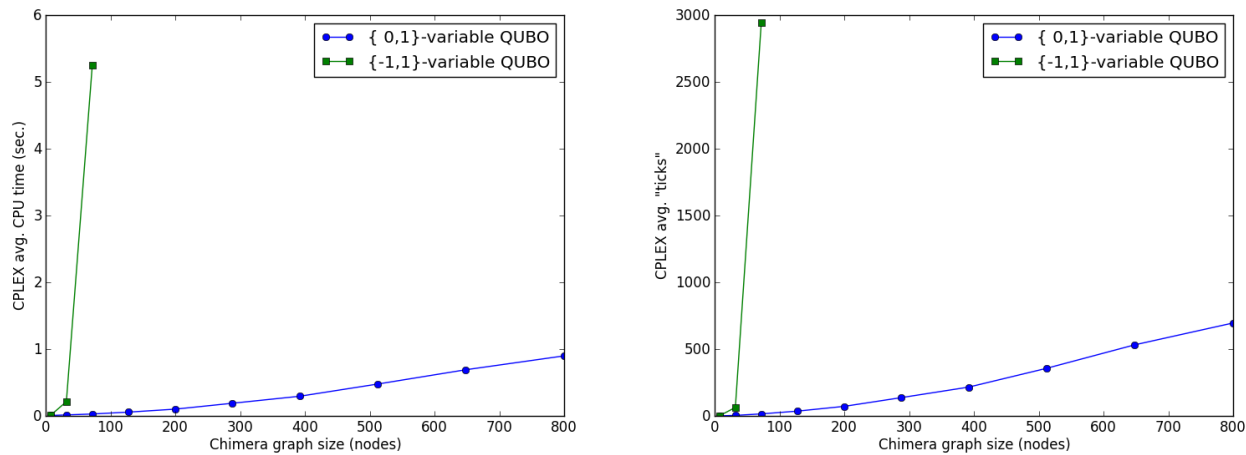

**Figure 4 CPLEX performance on random Chimera QUBO and Ising instances.** We compare the performance of CPLEX v12.4, used as an integer-linear-program solver, on randomly generated QUBO (blue plot) and Ising (green plot) instances. Coupler weights were drawn uniformly from $\{-\mathbf{1}, +\mathbf{1}\}$ for each edge in the Chimera graph, and all linear-term coefficients were set to 0. We generated 100 instances of each size and measured the average time to find an optimal solution on a 16-core machine with 3GHz Intel Xeon cores. Execution time was limited to 15 minutes.

To illustrate points made above, recall that the only difference in the definition of the QUBO and Ising problems from Section 1.3 is that that variables in a QUBO problem take values $x_i \in \{0,1\}$, while the variables in an Ising problem take values $s_i \in \{-1,1\}$. By a linear transformation in variables, $s_i = 1 - 2x_i$, we see that the two problems are equivalent. As random instances are easy to generate and implement, such instances have been used heavily in D-Wave benchmarking studies. Since the QUBO and Ising problems are equivalent, we may expect comparable performance on randomly generated QUBO and Ising instances. Surprisingly, in Figure 4 we observe a sharp performance difference in randomly generated QUBO and Ising instances on the D-Wave Chimera graph. Figure 4 demonstrates a performance gap between these instances for CPLEX, and we note that our experiments indicate a comparable gap on a DW2. An explanation for this discrepancy is that even though the QUBO and Ising problems are equivalent by the aforementioned linear transformation in variables, such a transformation does not preserve the probability distribution over instances when the weight of each coupler is simply chosen uniformly at random. The surprise in this case is not just that these classes of random instances are different, but that there is such a striking difference in performance for both CPLEX and a DW2. Both these classes of instances are quite natural, and there is not a clear choice, at least a priori, for which one lends itself to more interesting or relevant benchmarking. One must be careful of sensitivity of algorithmic performance to seemingly insignificant model and parameter choices, especially for random instances.

More realistic instances are desirable; however, in the case of D-Wave, we must efficiently transform our desired instance into a related instance on the Chimera graph. We summarize our efforts to overcome this barrier. One such approach is embedding the desired instance within a Chimera graph (details presented in Section 4.3). However, doing so is an NP-hard problem, itself, and moreover, such embeddings require ranges of coupler weights beyond the current precision of D-Wave annealers. We have developed an algorithm that is able to generate, directly within the Chimera graph, synthetic graphs with properties of real-world social networks, sidestepping the embedding issue (see Section 4.3.2). We have also designed an algorithm to reduce a weighted instance of QUBO to a polynomially larger, but unweighted, instance of QUBO to circumvent coupler precision limitations. The details of this technique will appear in a technical article that is under preparation [17]; however, Figure 5 gives a high-level schematic view of the different problems and mappings involved. The resulting unweighted instance is larger than the original weighted instance, and one may view such techniques as enabling a tradeoff between coupler precision and qubits. Currently our approach is only practical for relatively sparse instances of weighted QUBO, and the associated qubit overhead renders it impractical for the current generation 512-qubit D-Wave annealers. However, we expect that our type of approach will be a useful practical tool for modeling real-world problems on upcoming generations of D-Wave QAs, for which we expect coupler precision to remain a bottleneck.

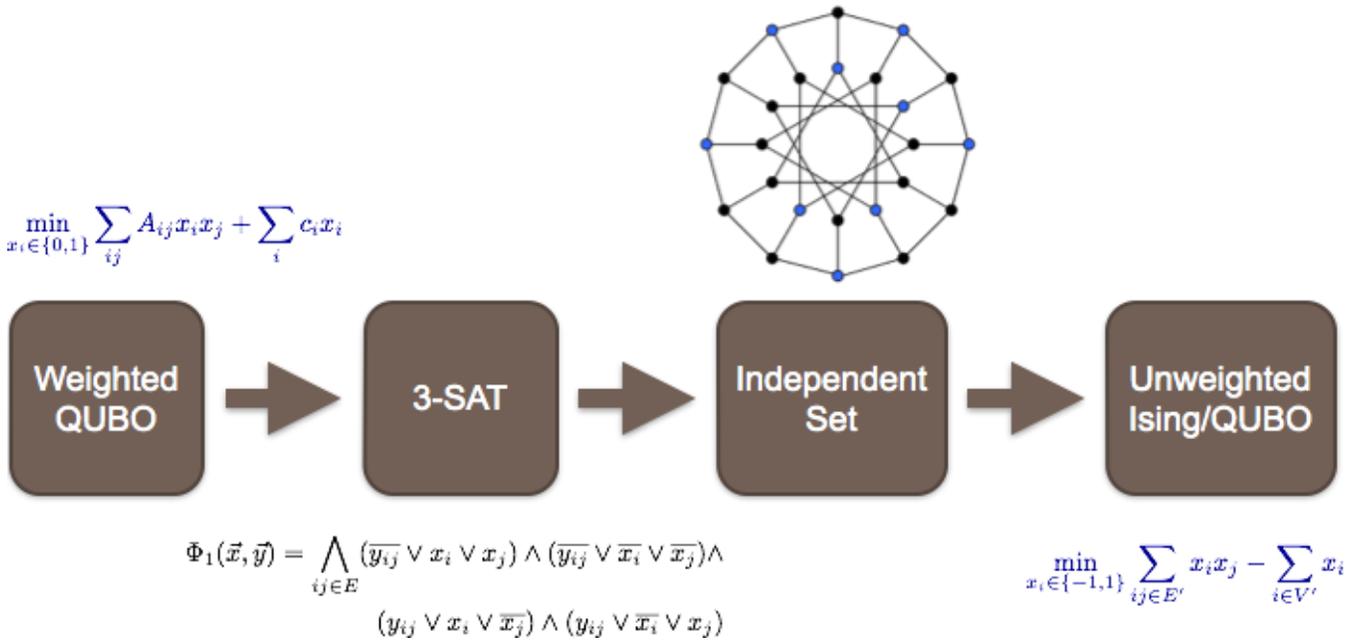

**Figure 5 A high-level schematic view of our reduction from weighted QUBO instances to unweighted Ising instances.**

### 3.3.4  How large are hard instances?

Ideally we would be able to discover a hard family of instances that tax conventional computing approaches but that we can faithfully represent on a current or near-term D-Wave device. This would allow for a somewhat direct comparison between conventional computing and quantum annealing. This would also give us a better sense of what we could accomplish with the nearly 512 qubits of a DW2.

How large would we expect such an instance to be? We demonstrate that this is an extremely difficult question to answer. For example, Figure 6 depicts the evaluation of the performance of a particular classical Ising (equivalently, QUBO; see Section 1.3) heuristic by Bonato et al. [30] on two families of instances. One family consists of randomly generated instances on a grid ranging in size from 900 to 6,400 nodes, while the other consists of a few 250-node instances that were collected by Beasley for inclusion in a library of hard and varied QUBO instances [30], [31].

Some of the random grid instances with thousands of nodes took minutes to solve, while most of the 250-node Beasley problems took hours to solve. Moreover, the Beasley problems represent a family of related instances of varying degrees of difficulty, yet all of them have 250 nodes. This suggests that size is not the only measure of complexity, and that other, more elusive, structural properties can play a key role in the complexity of a problem.

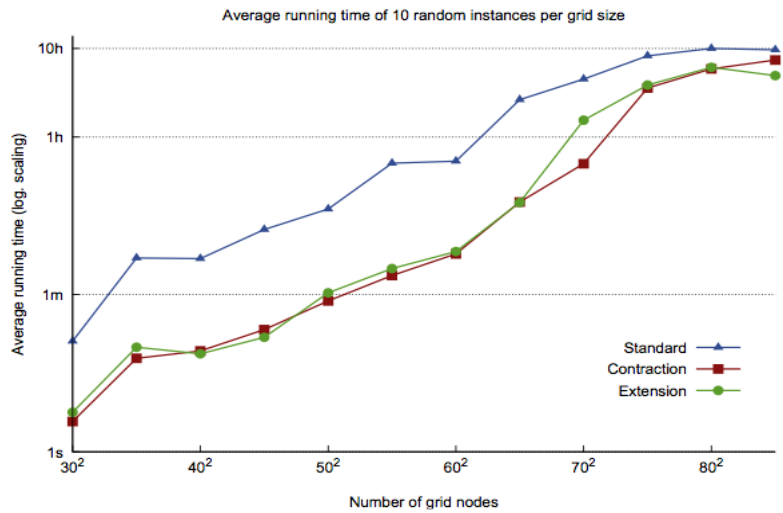

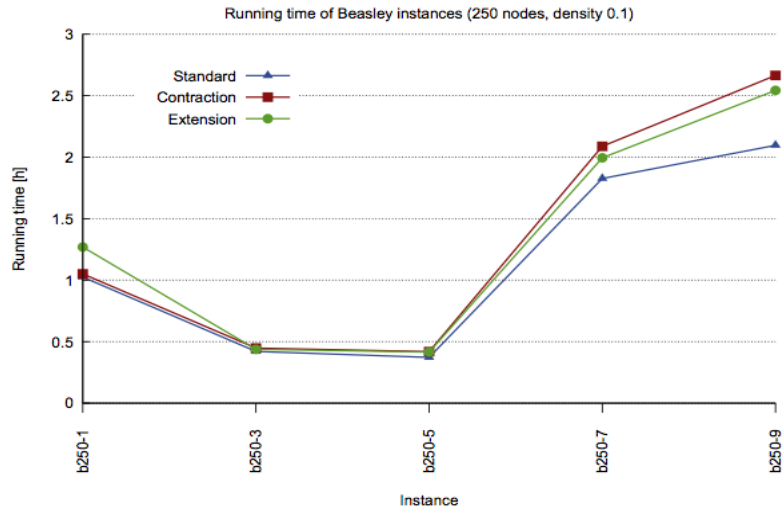

**Figure** 6 **Performance of Bonato et al.'s maximum cut (equivalent to QUBO and the Ising problem) heuristic on random grids and Beasley's QUBO instances [30].**

The above examples are hundreds and thousands of nodes in size. For some problems we will discuss in this report, such as community detection, there are very fast heuristics, such as the Louvain algorithm, that can handle sparse graphs with millions of nodes in a matter of minutes (see Appendices C.2 and D.1). Yet in this case it is not clear what "handle" means, since the Louvain algorithm is a heuristic that comes with no guarantees. Practitioners have widely observed that this algorithm delivers high-quality solutions on large real-world networks; however, it is entirely possible that there exist very small instances, structured in a particular way, on which the algorithm fails dramatically. Thus a quantification of hard instances is as much a function of the problem and solution model[4] as it is the instance itself. This is one of the reasons why the search for a small but hard class of problems that can be represented and executed on D-Wave's QA has been a difficult enterprise. Hardness is difficult to predict and analyze; the few options at our disposal usually entail trial and error.

---

[4] For example, an *exact algorithm* that always delivers an optimal solution is one such model, while a heuristic, with no performance guarantees, is another.

# 4 Complex-network instance families

We motivated the need for a careful consideration and selection of benchmarking instances in the previous section. One of our main contributions is bringing real-world instances into quantum annealing benchmarking. Existing work has focused on random or simple instances, mainly due to the limitations imposed by the D-Wave Chimera architecture (Section 2). We bridge this gap by devising a novel method to generate real-world-like social networks that can be natively embedded within a D-Wave Chimera graph. Our method can also be generalized for other limited-connectivity architectures besides the D-Wave Chimera graph. Although synthetically generating real-world-inspired graphs is an active research area, to the best of our knowledge, our work is the first of its kind, combining synthetic graph generation with the notion of graph embeddings (see Figure 2 and Figure 12). We also develop a method to produce families of social networks derived from actual Twitter and Internet routing data; this addresses a shortcoming in many social network analysis studies, which use one-off data sets, such as Netflix or academic co-authorship data, rather than families of related instances. We use these families of networks for the classical portion of our benchmarking study and as a ground truth against which we validate the structural properties our synthetically generated Chimera social networks.

Though these contributions constitute a significant portion of our effort, the details may be somewhat of a diversion from the overall theme of our investigation. We present highlights below in order to give a flavor of our work and refer the reader to Appendix B for further details.

## 4.1 Introduction

Graphs are commonly used to model real-world data and offer a rich platform for further analysis. Graphs, which are also called *networks*, are composed of a finite set of *nodes* (also called *vertices* and often depicted as points) and edges (links or line segments) that connect pairs of vertices. One may think of the edges as defining a relationship on pairs of vertices, with those pairs that share an edge as being related. In social networks, for example, edges tend to denote similarity, affinity, or perhaps that two nodes have communicated. In case of the latter, one sometimes may want to distinguish the sender of a message from its recipient. In this case we can think of an edge as being *directed* with one of the nodes of the directed edge its *source* and the other its *destination*.

The variable $G$ is usually used to denote a graph, and $V$ and $E$ denote its vertex and edge sets, respectively. For our purposes, undirected graphs will generally suffice, and for nodes $i$ and $j$ that are connected by an edge, we use the notation $(i, j)$ or $ij$ to refer to this edge; in an undirected graph, $ij$ and $ji$ both refer to the same edge.

Graphs serve as a mathematical representation for interesting real-world phenomena: personal interactions become social networks; webpages linking to each other become Internet graphs, and protein interactions in biology become biochemical networks, for example. By way of graphs, real-world phenomena can be encoded and shared succinctly, categorized, and analyzed efficiently.

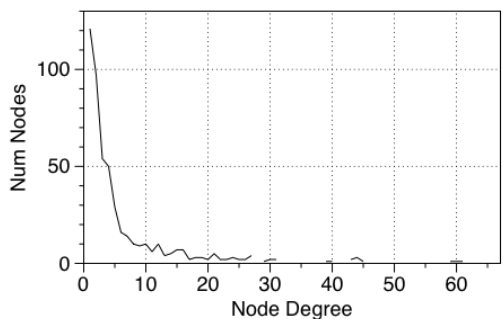

**Figure 7 Degree distribution.** The degree distribution is the histogram of node counts for each possible degree in the graph. This specific degree distribution shows the common "long-tailed" behavior seen in many real-world graphs.

To further categorize and describe graphs, researchers have proposed various graph measures. We present a few of these below with a more detailed list appearing in Appendix B.
- Number of vertices ($n$) – Together with number of edges, describes graph size.
- Number of edges ($m$) – Together with number of vertices, describes graph size.
- Degree ($d$ or $\delta$) – The number of edges incident to a vertex.
- Average degree ($2m/n$) – Describes the average number of edges incident to a vertex.
- Degree distribution (see Figure 7) – The number of vertices for each distinct degree value.
- Clustering coefficient ($c$) – A measure (between 0 and 1) indicating the tendency for nodes to cluster. Specifically, it is the fraction of all node *triples* (three nodes connected by at least two edges) that are connected by three edges, forming a *triangle*.
- Diameter ($d$) – The maximum distance between any pair of nodes, where the distance between two nodes is the number of edges on a shortest path between them.

Measures such as these allow us to better understand graph features despite their inherent complexity. Real-world graphs generated from different sources will often have similar values on some of these measures, but can have very different values on other measures.

Performing a scaling analysis of algorithms requires having graphs of different sizes that possess some inherent or structural similarity; we call such a collection a family of related graphs. The danger of benchmarking on a family of unrelated graphs of different sizes is that performance variation at different scales may not be a function of scale alone and may be related to varying complexities among the instances themselves. We saw examples of this in Section 3.3.4. Moreover, we want our results to be indicative of the scaling of computational effort of analyzing *real-world* graphs; all benchmarking studies to date have treated *random* graphs on the D-Wave Chimera qubit connectivity graph. Therefore, we want a family of graphs with real-world character that grow in size while remaining consistent in the other measures. After examining various possibilities (see Appendix B.3), we determined that growing graphs of different sizes from the same underlying data was our best option for generating a family of such similar graphs. The graph families we generate in this manner serve two purposes: (i) we use them directly in a comparative benchmarking study of community detection algorithms (covered in Section 5.5 and Appendix D), and (ii) we use them to validate a new algorithm we have devised to generate synthetic real-world-like networks that can be easily represented on a D-Wave Chimera graph.

## 4.2 Real-world instances

### 4.2.1 Twitter graphs

Internet-based technologies that enable and record human interactions are often called social media. These technologies include weblogs (blogs), Facebook, Twitter, and YouTube. In social media, people often link to specific other people. The graphs resulting from these interactions can be used to group people in communities, describe human communication patterns, and identify influential people.

We used Twitter data readily available from its data feed. Twitter is a social network where users broadcast short messages (tweets) that can, in practice, be seen by anyone on the Internet. If the sender wants a specific other user to see their tweet they refer to that user's name and precede it with the "@" character; this does not, however, preclude others from seeing the message as well. Twitter does also offer a private direct message facility. These so-called "@-refers" indicate that the sender is "talking" to the receiver. However, as Twitter contains many celebrities to whom many fans @-refer, we did not want to count all @-refers as a "conversation." Therefore, in our Twitter graphs, we include an edge between two nodes, corresponding to users, precisely when both users @-refer each other.

We ran our algorithm, which generates Twitter graphs based on specified time windows (see Appendix B.2.1), for various time periods of the Twitter data feed: from 2 hours to 20 days and with 13 different start dates. This resulted in 92 different graphs. Visual inspection of several of the graphs shows they appear similar with many long chains of vertices and occasional higher degree clumps (Figure 8).

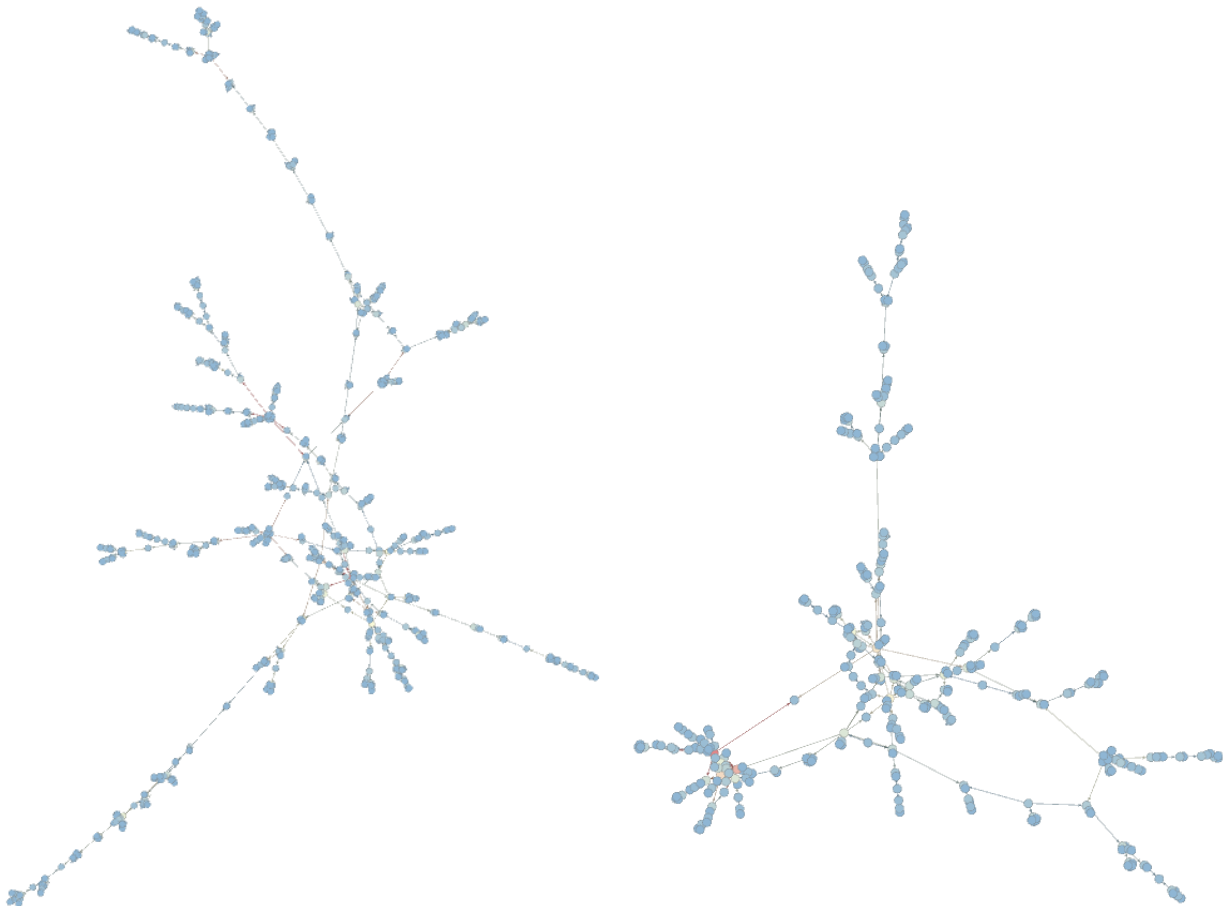

**Figure 8 Two sample Twitter graphs.** The graph on the left has approximately 350 vertices and the graph on the right has approximately 700 vertices. Notice that both have small clusters of connected nodes and long strings of smaller groups of nodes.

These graphs grow drastically in size while maintaining a similar average degree (Figure 9). The average degree remains steady at a value of around two for graphs with tens of nodes through graphs with tens of thousands. As the graphs grow considerably larger, the average degree does begin to smoothly increase until average degree is approximately 4. We note that the graph size is primarily a function of the length of the corresponding time window, and although we hypothesize that degree increase may be related to short-term versus longer-term communication patterns, an analysis validating such a hypothesis is beyond the scope of this work.

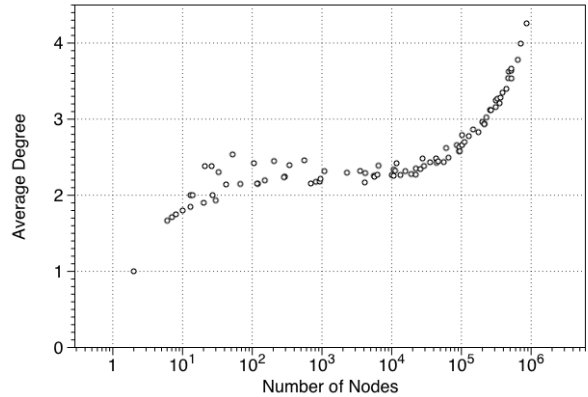

**Figure 9 Average degree for all our Twitter graphs.** Each point displays the number of nodes and the average degree for that graph.

These Twitter-based graphs smoothly change diameter as the size of the graph increases (Figure 10). The diameter increases steadily until around 10,000 nodes, and then the diameter stabilizes and slowly decreases. It is interesting that this is also approximately when the average degree begins to increase, and this could support our hypothesis above in that people may communicate with more distant acquaintances in addition to closer friends over longer periods of time, resulting in long-range edges that might decrease diameter.

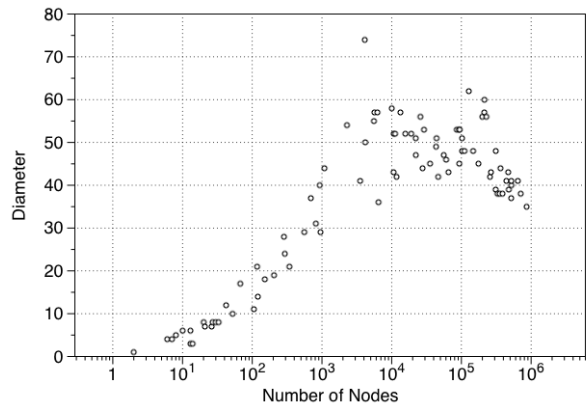

**Figure 10 Diameter for all our Twitter graphs.** Each point displays the number of nodes and diameter for that graph.

These graphs also smoothly transition their clustering coefficient while staying rather low. With the exception of some very small graphs (a couple hundred nodes), the clustering coefficient ranges between 0.05 and 0.10. The smaller graphs with both the highest and the lowest clustering coefficients are more drastically affected by noise where a small number of triangles[5] can have a larger influence on the score.

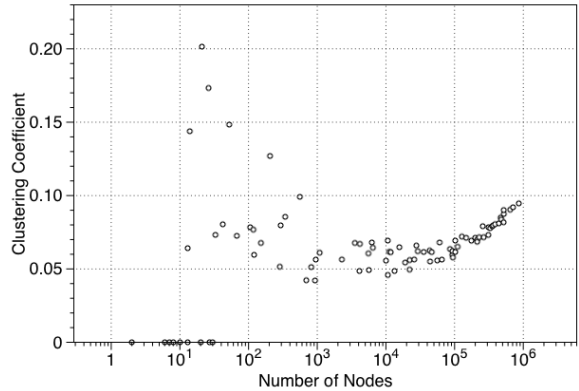

**Figure 11 Clustering coefficient for all our Twitter graphs.** Each point displays the number of nodes and clustering coefficient for that graph.

We also generated a family of graphs from Internet Border Gateway Protocol data; details are given in Appendix B.2.2.

### 4.3  D-Wave-embeddable graphs

For our benchmarking work, graph families that can be run on a D-Wave device are paramount. We briefly describe the issues with representing problem-domain graphs on a D-Wave Chimera graph and present a new algorithm for generating real-world-like graphs that are readily representable on a D-Wave Chimera graph.

#### 4.3.1  D-Wave graph embeddings

A D-Wave QA is equipped to solve QUBO problems defined on a sparse and fixed Chimera graph, representing the connectivity of its qubits (see Figure 1). In order to solve more general QUBO problems on a D-Wave QA, one must somehow reduce the problem at hand to an instance of QUBO on the Chimera graph. One natural approach is called a *minor embedding* in the graph theory literature. The idea is to represent each node of the original graph by a set of connected nodes in the Chimera graph. For each edge in the original graph, we require at least one edge between the sets of corresponding nodes in the Chimera graph (see Figure 12 and Figure 2). By using coupler weights of sufficiently large magnitude on the edges within each of the node sets, we can force the Chimera QUBO instance to set all the variables in any given node set to take the same value. This, in essence, solves QUBO on the original problem-domain graph.

In order to maximize the size of problems we can solve on a D-Wave device, we seek an efficient minor embedding — one that uses as few Chimera nodes as possible. Unfortunately, not only is it intractable to find an efficient minor embedding, it is in general NP-hard to even determine whether a graph is minor-embeddable within another graph. This poses a serious hurdle to using such an approach. We may still appeal to heuristic approaches to attempt to find a minor embedding, and D-Wave provides such a heuristic as part of its software tool set. In general such heuristics, including D-Wave's, are not viable for real-world applications.

---
[5] Recall that a triangle is a set of three nodes where all three possible edges among these nodes are present in the graph.

Another approach is to try and find a universal minor embedding. If one can embed a fully connected or *complete* graph on $N$ nodes within the Chimera graph, then one can embed any graph on $N$ nodes. It is known that a complete graph on $N$ nodes can indeed be minor-embedded into a Chimera graph on $\sim N^2$ nodes (Figure 12) from the work of Choi [32]; conversely, any minor embedding of a complete graph requires $\sim N^2$ nodes[6], so we cannot hope to do better using this strategy. This quadratic overhead may be prohibitively high; for example, we currently only know how to embed a complete graph with approximately 33 nodes in the current generation DW2 QA.

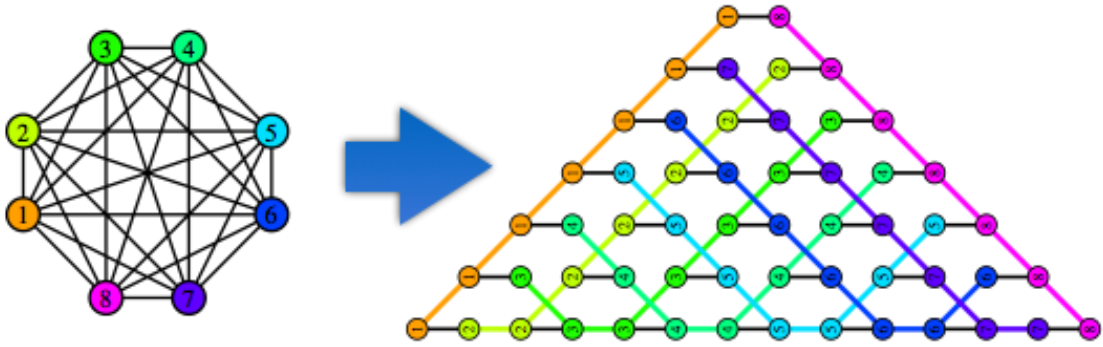

**Figure 12 A minor embedding of a fully connected graph on 8 nodes into a subgraph of the Chimera graph [32].**

Even if one is able to find an efficient minor embedding, a bigger issue is that current D-Wave QAs do not offer couplers with great enough precision (see Section 2) to accurately represent the minor-embedded problem. Thus even if one can embed a problem-domain graph into the Chimera graph, attempting to solve the problem on a D-Wave device will likely result in a solution that cannot be converted into a solution to the original problem. For many problems, the embedding coupler magnitude may need to be as large as the number of nodes in the graph. *This is the most significant hurdle in solving real-world problems on D-Wave annealers.*

Despite these obstacles, minor embedding is currently the only general technique known for reducing arbitrary QUBO instances to Chimera QUBO instances. We leverage the fact that our goal is to benchmark on real-world-*like* graphs, and we turn the embedding problem on its head. For the purposes of benchmarking, we do not have a *specific* instance that we need to embed and only seek to represent families of graphs that resemble real-world social networks. Thus, rather than attempt to embed a given graph into the Chimera graph, we start with a Chimera graph and generate a real-world-like random graph within it. For an $N$-node Chimera graph, our algorithm is able to produce a Chimera-embedded real-world-like social network with approximately $N/2$ nodes. While we can only embed complete graphs with approximately 33 nodes on current D-Wave QAs, our algorithm is able to generate Chimera-embedded social-network-like graphs with more than 250 nodes. For a point of reference, recall from Section 3.3.4 that we know of QUBO instances with 250 nodes that take hours to solve on a conventional computer. Although a few hours may not seem long, if one's task requires solving a large number of such instances, reducing the per-instance run time could make significant difference in overall run time.

---

[6] This follows because the *treewidth* of an $N$-node Chimera graph is $\sim\sqrt{N}$; a discussion of treewidth is beyond the scope of this report.

### 4.3.2 Real-world-like Chimera-minor graphs

We describe our new embedding approach in more detail. Our goal is to compare the performance of both classical and D-Wave systems on real-world graphs; however, as noted earlier, embedding arbitrary real-world graphs within the Chimera graph can impose drastic size limitations. Our approach is to start with a Chimera graph and alter its topology to create a new graph that is automatically minor-embedded within the original Chimera graph and exhibits the targeted characteristics. Of course many such minors[7] are possible, and the challenge is generating one that has properties of a real-world complex network. The Chimera graph has three principal features we seek to change to create a graph minor with more real-world character. In order of our valuation of their importance, they are the following:

- o The Chimera graph has only degree five and six nodes, with the vast majority degree six. This is very different from the long-tailed degree distribution seen in real-world graphs (Figure 7).
- o The Chimera graph contains no triangles and thus has a zero clustering coefficient. Again, this is very different from the clustering coefficients of real-world graphs (Figure 11).
- o The Chimera graph's diameter grows drastically as the number of nodes increases. This is also different from real-world graphs (compare with Figure 10).

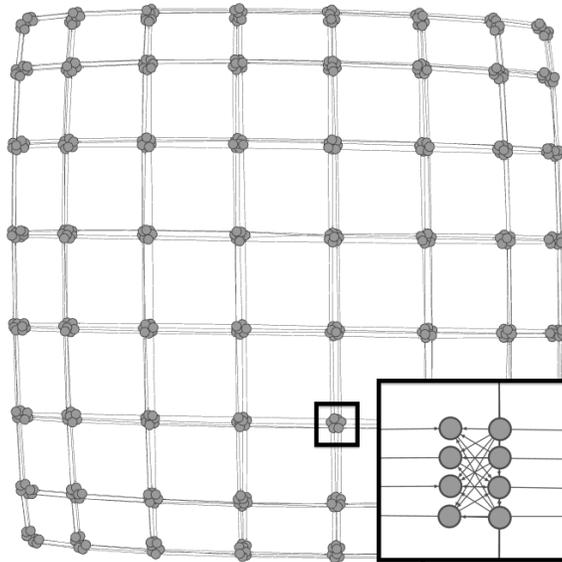

**Figure 13 An alternate drawing of a Chimera graph (compare with Figure 1).** Each eight-node cluster is called a *cell* or *plaquette*.

Our algorithm for generating realistic graph minors from the Chimera graph follows three phases. In the first phase, our goal is to increase the clustering coefficient by introducing triangles. A graph minor is obtained by merging nodes and possibly removing edges or nodes. As removing edges or nodes cannot create triangles, we merge nodes. A good strategy for generating triangles is by merging nodes within the Chimera's eight-node cells (see Figure 13 and Figure 1). In this step, we divide each eight-node cell into four pairs with one node from the left group of four nodes and one node from right group of four nodes (see expanded box in Figure 13). Given a user-specified probability parameter $p_1$, we merge each such pair of nodes with this probability. This results in triangles formed between each merged pair and all unmerged pairs.

---

[7] A graph $H$ is called a *minor* of graph $G$ if $H$ can be minor-embedded within $G$.

The second phase focuses on generating high-degree nodes and decreasing diameter. Again, removing edges cannot create higher degree nodes or decrease diameter, so we employ node merging. We found that merging connected nodes from different cells best fulfilled these goals. As with the previous phase, given a parameter $p_2$, we merge each pair of connected nodes from different cells with probability $p_2$. To extend the reach of some of these merged nodes beyond the immediately neighboring groups, we perform multiple iterations with a lower $p_2$ value rather than opting for a single iteration at a higher probability.

The final phase focuses on creating low-degree nodes. For each edge in the graph, we consider removing the edge with a probability $p_3$. However, we want to ensure that we are not creating too many small components[8] in the graph, as real-world networks tend to have large components. Therefore, after removing an edge from the graph, we ensure that the two vertices in the removed edge can still reach each other through some series of other edges. If not, we reinsert the edge.

We determined suitable values for $p_1, p_2$, and $p_3$ in order to generate graphs that mimic real-world complex network properties, based on analysis of degree distribution, diameter, and clustering coefficient. Our synthetic graphs have much of the long, spindly look of our Twitter graphs (see Figure 8). Unlike the Twitter graphs, these pseudo-social graphs do not often contain loops.

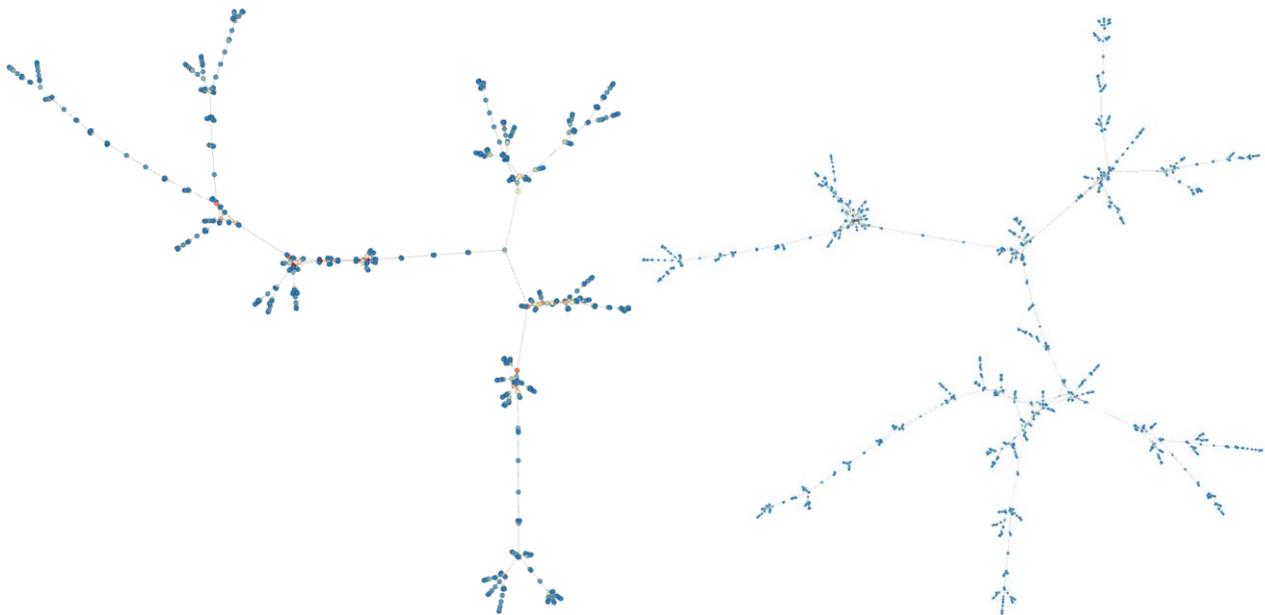

**Figure 14 Two sample randomly generated real-world-like graph minors of the Chimera graph.** The graph on the left was generated from an 800-node Chimera graph and has approximately 450 nodes. The graph on the right, generated from a 1600-node Chimera graph, has approximately 1000 nodes.

We present a quantitative comparison of our synthetic complex-network minors of the Chimera graph with our other families of graphs in the next section. Beyond this, we are also interested in the node utilization of our graph minors, which is the percentage of nodes of the original Chimera graph retained in a minor. As previously mentioned, in a Chimera graph with $N$ nodes, we would expect to be able to embed a graph with only $\sqrt{N}$ nodes, in the worst case. On a 512-node Chimera graph, this represents a node utilization of under 5%. However, our *Chimera-minor*[9] graphs have a nearly constant utilization percentage at just over 55%.

---

[8] A graph is *connected* if there is a path in the graph between each pair of nodes. For a *disconnected* graph, its *connected components* are its maximally connected subgraphs. Every pair of nodes within a component can reach one another, and no two nodes in separate components can reach each other.

We have now presented the three different graph families we will use in our benchmarking analysis: Twitter, BGP, and real-world-like Chimera-minor graphs. In the previous sections (or Appendix B), we showed examples of each graph family along with various metrics for each family. In Appendix B.4 we synthesize these results to more clearly highlight the similarities and differences between each family, with the goal of demonstrating that our Chimera-minor graphs possess real-world graph properties.

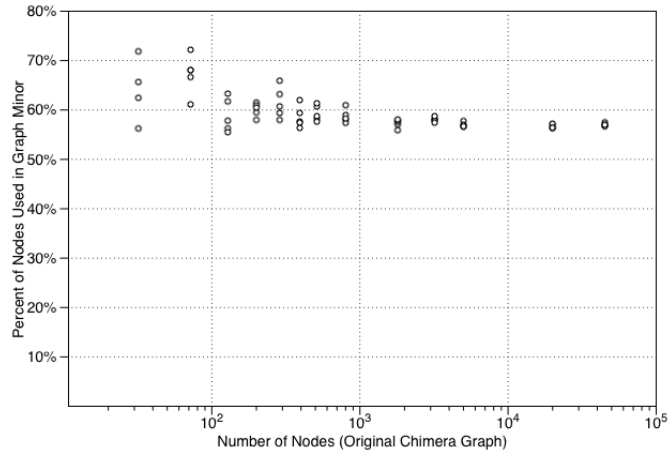

**Figure 15 Percentage of nodes in original Chimera graph retained in real-world-like graph minors.** The utilization remains consistently greater than 55%.

---

[9] By *Chimera-minor* graph we mean a graph that is a minor of the Chimera graph.

# 5 Quantum benchmarking studies

## 5.1 Introduction

This section presents our quantum benchmarking studies, complementing our classical benchmarking studies, presented in Appendix D. We expand our repertoire of the problems considered as well as the algorithms applied. While we focused on real-world classical heuristics in Appendix D, here we present data from experiments conducted on the DW2 QA housed at NASA Ames Research Center at the time of this writing, as well as comparative experiments performed using the classical Chimera QUBO/Ising[10] heuristic by Selby [33], and a Quantum Monte Carlo (QMC) simulator of our design (see Appendix F for details). The Ames DW2 has a 512-qubit "Vesuvius" chip with 509 functional qubits.

We consider complex network analysis problems, as well as random Ising spin glass instances on the Chimera topology. For the former, we employ Ising-based formulations that are described below. The latter class of instances has become somewhat of a standard for D-Wave benchmarking instances following the seminal work of Boixo et al. [20] in the field. Our interest in these instances is to validate our basic methodology by ensuring that we obtain comparable results as previous studies. Moreover, we use these instances to illustrate some of the subtleties in D-Wave benchmarking that we have encountered (Section 3.3). The random spin glass instances also serve as a benchmark for assessing the performance and fidelity of our QMC simulator with respect to a D-Wave QA. We have selected problem sets to evaluate various aspects of solving Ising problems on a D-Wave QA. For example, while our baseline Ising instances have no linear field term, our independent set problems critically rely on a linear term. The planted-solution instances employed are generated using random walks as opposed to selecting coupler weights directly from a random distribution [34], [35].

Although some recent D-Wave benchmarking work has considered instances requiring a minor embedding [16], [36], [37], we have specifically avoided such instances. This is because, even with an embedding in hand, one must determine coupler weights allowing a D-Wave system to accurately represent the embedded instance. This typically requires large coupler magnitudes, resulting in coupler weight ranges that cannot be captured with the precision offered by current D-Wave systems. Recent work by Perdomo-Ortiz et al. [37] presents a heuristic for setting coupler weights, enabling approximate representation of an embedded instance on D-Wave. We could benchmark D-Wave on our real-world-like Chimera-minor instances (Section 4.3.2) in conjunction with such a heuristic; however, doing so would add another layer of complexity and variability that is beyond the scope of our present study. We do benchmark our Chimera-minor instances on our QMC simulator, which does not share D-Wave's precision constraints.

We summarize our problems and algorithms in Table 1; each problem is introduced in detail in its respective section below.

---

[10] As noted in Section 1.3, the two problems are equivalent by a linear change in variables. We use QUBO and Ising problem interchangeably.

|  | Real-world classical heuristic | Selby classical QUBO solver | D-Wave quantum annealing | Simulated quantum annealing |
|---|---|---|---|---|
| **Ising $\{-1, 1\}$ couplers** |  | X | X | X |
| **Independent set** |  | X | X |  |
| **Affinity independent set** |  | X | X |  |
| **Planted solution** |  | X | X |  |
| **Community detection** | X |  |  | X |

**Table 1 Problems (rows) and algorithms (columns) employed in our quantum benchmarking study.** A mark in the table indicates that we ran the corresponding algorithm on the corresponding problem.

Although a variety of classical approaches are known for the Ising spin glass and independent set problems, there is no clear consensus on a widely adopted set of practical classical heuristics, as is the case for problems like community detection (see Appendix C). Thus we have selected Selby's Chimera QUBO solver as our primary classical point of comparison for our problems. The sparse connectivity of the Chimera graph in conjunction with the precision limits of the DW2 are obstacles to a D-Wave implementation of our Ising-based community detection heuristic, which is described in Section 5.5.1; however, as indicated in Table 1, we do implement this community detection heuristic on our QMC simulator.

### 5.1.1 Quantum Monte Carlo simulator

Numerically simulating the operation of a quantum computer presents a host of challenges. The biggest of these challenges may be seen directly by considering the space in which such a computer operates. A quantum computer with $N$ qubits will have $2^N$ possible states. Unlike a classical computer, a truly quantum system may occupy all of these simultaneously, meaning that the cost of keeping up with this information in a direct simulation grows much too rapidly with system size to be practical. For some classes of problem, such as linear chains of qubits, specialized techniques such as matrix product states or density matrix renormalization group are able to directly solve large instances. Unfortunately, none of these methods can be applied to a general topology, such as the Chimera graph. For a reasonably large assembly of qubits, by far the most practical technique for these simulations is to use a stochastic method. Stochastic methods like Monte Carlo sampling are attractive because once converged, the cost of obtaining an answer with a given amount of error does not grow with the system size. For the purposes of this study, more time will likely be spent in converging the simulation than collecting the results, but we still expect the scaling of the cost to be moderate with system size, meaning that simulations with large numbers of qubits should be possible.

We consider two main approaches to the stochastic simulation of qubits. The first approach exploits Feynman's path integral formulation of quantum mechanics to map the quantum system onto an equivalent problem in finite temperature polymer physics. This technique is rather robust, but as it is naturally performed at finite temperature, this method invites significant challenges when studying systems where low-lying excited states may be present, as is expected to be the case in the evolution of an adiabatic quantum computer. The second approach is to stochastically project out the lowest energy solution of the problem. This approach is often less computationally efficient than the path integral formulation, but allows for a much more straightforward interpretation of results.

Our QMC simulator is under active development and will benefit from further tuning; hence we limited its use to our baseline Ising and community detection instances, for which we deemed its contribution critical. We do not report the baseline Ising results here but offer the general observation that our QMC simulation results are well correlated with D-Wave results for up to 288 nodes. For larger instances, our simulator currently rarely finds optimal solutions. The performance of our simulator is not as critical for our community-detection benchmarking, since it is part of a hybrid classical/Ising algorithm, described in Section 5.5.1.

## 5.2 Chimera Ising spin glass instances

In this section we describe the random Chimera Ising spin glass instances that have been commonly used in D-Wave benchmarking studies. We also explain our general D-Wave benchmarking methodology here, which is used for our other problem sets as well.

The specific instances generated draw weights from $\{-1, +1\}$ uniformly on the edges of the Chimera graph. Hence, they can be natively represented on a D-Wave device. This avoids the embedding issue, and the choice of weights sidesteps the precision issue.

We generated 100 random instances, where each instance corresponds to a Chimera graph consisting of a $k \times k$ "grid" of eight-node cells, for $k$ in the range [1,8]. For example, in this taxonomy the entire D-Wave graph depicted in Figure 1 is an $8 \times 8$ grid of eight-node cells (ignoring faulty nodes), hence $k = 8$; setting $k = 7$ would correspond to D-Wave's 128-qubit previous-generation "Rainier" chip.

This results in total node sizes of 8, 32, 72, 128, 200, 288, 392, and 512, allowing us to get a sense of scaling on a D-Wave device. Of course our DW2 QA has a fixed-size Chimera graph (consisting of 509 out of 512 functional qubit nodes). We simulate Chimera graphs of varying size by selecting an appropriate subgraph of the overall 509-Chimera graph. We ignore the three non-functioning nodes.

### 5.2.1 Methodology

We performed 1000 or 5000 anneal cycles for each of the 100 instances of each of the 8 possible sizes. For our purposes, a *gauge* is simply a transformation of the Ising instance that does not alter the landscape of attainable energy values, although the particular solutions corresponding to a specific energy value may change. A gauge transformation results in a mathematically equivalent problem; however, the particular numerical values of the transformed instance may render it easier to solve for a D-Wave QA, due to coupler calibration issues in representing problems. In order to more fairly account for calibration issues with couplers, we adopt two strategies: (i) for each instance we generate 20 random gauges, and (ii) when selecting one of our $k \times k$ grid Chimera subgraphs, we select its position[11] within the overall 512-node Chimera graph randomly. Thus, we performed at least 1000 (D-Wave annealing runs) $\times$ 100 (instances) $\times$ 20 (gauges) = 2,000,000 runs for each of the eight Chimera instance sizes considered. For a given instance, a *trial* is a single D-Wave annealing run with a fixed gauge. In assessing performance, we calculate a mean over all gauges rather than selecting the best one. The latter strategy could be used to obtain an optimistic best-case bound on performance; however, such an assessment is unrealistic as it is difficult to compute or predict the optimal gauge for a particular instance.

---

[11] More precisely we select a cell, $c$, of the Chimera graph randomly as the lower-left endpoint of our $k \times k$ grid Chimera subgraph. We then take the $k \times k$ grid defined by the intersection of: (i) the $k$ columns of cells including and to the right of the cell $c$, and (ii) the $k$ rows of cells including and above the cell $c$. Of course not all choices of $c$ would leave enough room to select a full $k \times k$ grid, and we only pick among the feasible choices.

As with previous studies [16], [19], [20], we opted for the minimum DW2 annealing time of $20\mu s$. The issue with selecting a larger annealing time is that this may expend more time than necessary to solve small or easy instances. For example, in most of our experiments, annealing for $20\mu s$ almost always finds an optimal solution on Chimera instances with eight nodes. Thus $20\mu s$ is likely too large and imprecise an annealing time to obtain meaningful statistics on the performance of the DW2 on such small instances. It may very well be that the annealer performs equally well on these instances when annealing for, say, $10\mu s$, but we would be unable to detect this because of the $20\mu s$ limit on minimum annealing time.

Another technical detail is the implicit assumption that selecting a subgraph of the overall Chimera graph (by effectively setting excluded couplers at a strength of 0) results in an easier instance for a D-Wave QA. To be more precise, a D-Wave QA only accepts an input of a fixed size (e.g., 509 nodes in the case of the DW2 at NASA Ames Research Center); in contrast, we are able to feed inputs of varying size to algorithms running on classical (e.g., von Neumann) architectures. We seek to assess scaling, hence we must find some way to simulate inputs of varying sizes with only a fixed-size D-Wave Chimera graph at our disposal. A natural strategy is to set the strengths on the unused portion of the Chimera graph to be 0; however, this is only a heuristic, as there is no guarantee that a D-Wave device will run faster on such a "smaller" instance. We assume other studies have adopted this approach as well. Fortunately, D-Wave experiments tend to support the above assumption in that the D-Wave annealers do seem to have an easier time finding an optimal solution on smaller instances obtained in this fashion. To be as fair as possible, we follow this same approach, as appropriate, when employing the Selby solver and our QMC simulator.

### 5.2.2 D-Wave run-time estimation

We report comparable results to previous studies on the metrics used therein. However, we also obtain the new result that a different, but perhaps more practically valid, metric leads to a marked improvement in performance across our algorithms (Section 3.3.1). We propose a reexamination or broadening, at the very least, of the metrics considered in future D-Wave benchmarking studies.

As discussed in Section 2, D-Wave QAs run for a pre-specified annealing time and produce a solution. There is no guarantee on the quality of the solution produced, and there is no explicit notion of an execution time necessary to produce an optimal or near-optimal solution. Thus we must estimate the latter. A common technique is to perform many annealing trials on a given instance and count how many times the annealer "succeeds" according to some success criterion. We may take the fraction of successful trials as a success probability and use the latter to derive an estimate for execution time. Two natural questions arise: (i) what is a valid success criterion, and (ii) how do we estimate execution time from a success probability?

The D-Wave benchmarking community has focused on finding an optimal solution as a success criterion, and we also adopt this for most of our experiments. However, as demonstrated in Section 2.3.1, selecting near-optimality as a success criterion can lead to a marked difference in results.

For the question of deriving a run-time estimate from success probabilities, there are two commonly used candidates: (i) the number of annealing trials necessary in order to succeed at least once with 99% probability, and (ii) the mean number of annealing trials necessary to first succeed. Once we have an estimate on the number of annealing trials necessary for each of these criteria, we may multiply by our annealing time per trial of $20\mu s$ to obtain a numerical estimate on running time. More precisely, if $p$ is the observed probability of success for a particular instance, then

$$\left\lceil \frac{\log(1 - 0.99)}{\log(1 - p)} \right\rceil \text{ and } \frac{1}{p}$$

are the functions corresponding to (i) and (ii), respectively. These may be derived using elementary probability theory; we refer to the reader to e.g., [19] for more details. Thus our D-Wave running time estimates are,

$$20\mu s \left\lceil \frac{\log(1 - 0.99)}{\log(1 - p)} \right\rceil \text{ and } \frac{20\mu s}{p}.$$

We note that following the work of Boixo et al. [20], the D-Wave benchmarking community has primarily focused on (i). We cannot overstress that although our plots below depict estimated running times; for the D-Wave system, these values are obtained in the indirect fashion from success counts as described above. Although it is reasonable to expect that the asymptotic D-Wave run-time scaling depicted in our plots approximates the true run-time scaling, drawing conclusions based on direct comparisons between run-time values is a murkier prospect.

Once a run-time estimate is obtained for an instance, we take the mean run time over all instances of the same size. One complication is that, for the hardest instances, one of our heuristics may fail in finding an optimal solution and produce an infinite run-time estimate. We throw out such instances from our data set; hence our results are only over instances for which every heuristic was able to discover an optimal solution at least once. We prefer (ii) as a run-time estimate. One reason is that the function $1/p$ is better intuitively understood and appears more transparent. A more precise and technical reason is that this quantity is easily and directly observable. For a given instance, one could simply perform some number of annealing trials and directly compute the number of trials before success is achieved. Repetitions of this process would produce a mean run time to success. Contrast this with the roundabout way this value is derived from success probabilities above. Although our D-Wave experiments continue to use the latter method in order to maintain consistency with previous work, our run-time estimates for classical algorithms use a variation of the more direct method described above. This is yet another reason one should avoid drawing direct comparisons based on actual run-time values and rather appeal to the asymptotic scaling of the values.

### 5.2.3 Results

Figure 16 summarizes the results of our first, baseline Ising problem set. Our D-Wave run-time estimates are comparable to those previously reported for this class of instances [19], [20]. We plot D-Wave run-time estimates based on both the metrics (i) and (ii) from the previous section. Although they differ by an order of magnitude in value, they scale almost identically. We observe similar behavior between the two in the problem sets to follow, as well. This is further justification that (ii) is a reasonable metric.

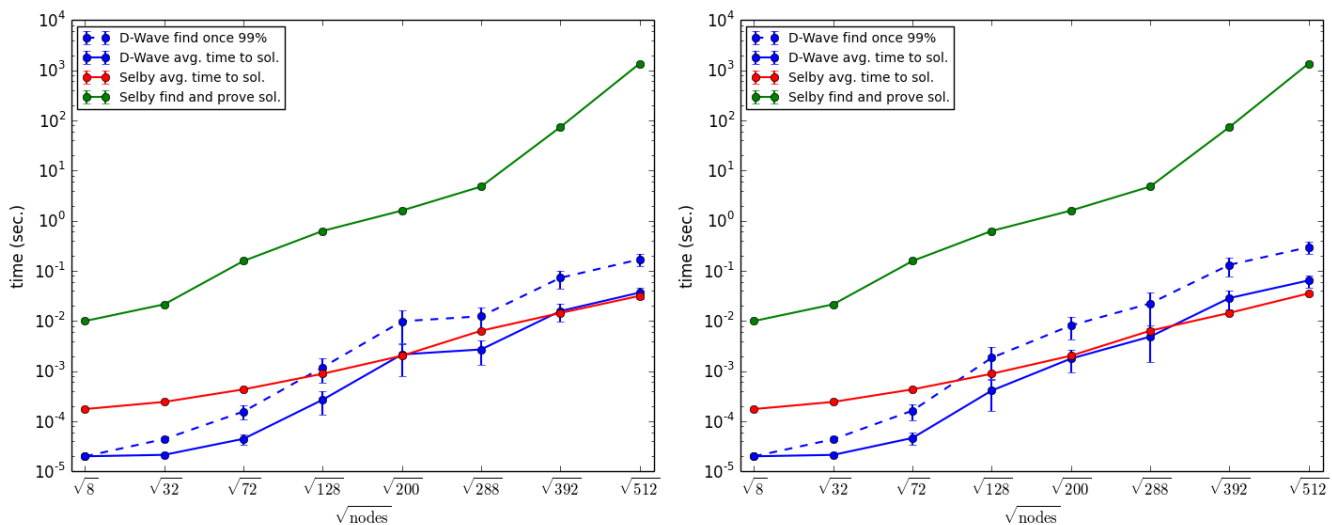

**Figure 16 Comparison of estimated run-time for D-Wave with both Selby's heuristic (red plot) and exact QUBO solvers (green plot).** Instances are randomly generated Chimera Ising problems with -1 and +1 couplers and no linear term. The dashed blue line estimates the run-time for the Ames D-Wave Two to find an optimal solution once with 99% probability, while the solid blue line estimates the mean run-time for this D-Wave Two to first find an optimal solution. The left plot is over 20,000 annealing trials for each of the, at most, 100 instances of each size, while the right is over 100,000 trials. The error bars represent standard error over the, at most, 100 instances of each size.

Selby's classical QUBO solver operates in a number of modes. One of the modes (mode 0) not only attempts to find an optimal solution, but also expends the effort to verify that its solution is indeed optimal. As expected this takes considerably more time than heuristic approaches such as D-Wave's. Another mode of Selby's solver (mode 1) performs a heuristic search of the solution space in hopes of reaching an optimal solution. This is the more appropriate comparison with D-Wave. This mode is also able to report a direct estimate of the mean time to find an optimal solution.

Several features of the plots are worth noting. The first is that one would expect the green plot representing Selby's exact algorithm to asymptotically scale roughly as $\sim c^{\sqrt{N}}$, which would correspond to a linear curve in the log-square-root plot above. This estimate is derived from a worst-case analysis of Selby's algorithm. We do see roughly linear behavior for the corresponding plot above, and this behavior is especially dominant on the three largest instances. Additional data obtained from larger D-Wave annealers would allow us to better evaluate this hypothesis. Such exponential-like scaling is typically indicative of an NP-hard problem; however, conclusions drawn on the theoretical worst-case complexity of a problem based on randomly generated instances are not necessarily sound. As we will see with our independent set problems, below, one can empirically observe exponential-like scaling for problems that are known to be solvable in polynomial time and have efficient practical implementations.

The second feature of note is the dampening effect on left end of the D-Wave plots for the smaller instances. This is very likely an artifact of $20\mu s$ being too large an annealing time for the three smallest instance sizes, as discussed in Section 5.2.1. There may be a similar effect with the classical algorithms as well, and, if we ignore the smallest instances, we do see nearly linear plots for both D-Wave and Selby's heuristic mode. Finally, we note that a crossover occurs between the D-Wave mean time to solution and the Selby heuristic mean time to solution between 288 and 392 nodes. Boixo et al. [20] and Rønnow et al. [19] also report that their classical simulated annealing algorithms outperform D-Wave on these instances. The scaling of the D-Wave seems to be on par with Selby's heuristic for the instance sizes we considered. Whether this trend continues on future D-Wave machines remains to be seen.

## 5.3 Independent set problems

### 5.3.1 Problem motivation and definition

An *independent set* in a graph is a set of nodes such that no two nodes in the set share an edge (see Figure 17). The objective of the *maximum independent set* (MIS) problem is to find an independent set of maximum size. We include this problem in our study for several reasons: (i) it is a fundamental NP-hard combinatorial optimization problem that has been studied for over 40 years [38], (ii) it has applications to complex network analysis, (iii) it has a fairly natural Ising representation requiring a linear term (in contrast to the basic Ising problems above), and (iv) it turns out that the natural Ising representation on a Chimera graph has weight ranges that are representable on the DW2 (for mathematical details see Appendix E.2). Thus MIS helps us cover the gamut in terms of benchmarking problems with linear terms as well as nontrivial, but D-Wave-representable, weights.

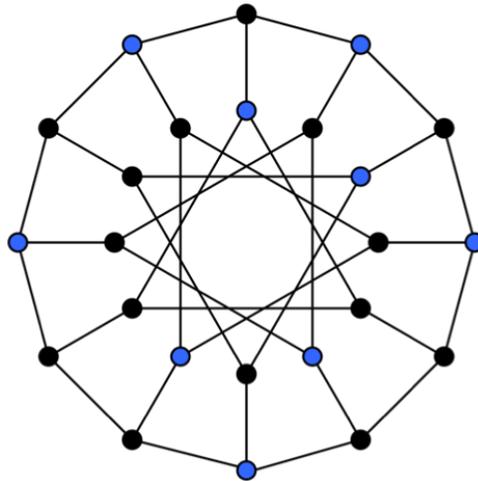

**Figure 17 An independent set in a graph (image from [39]).** The blue nodes form an independent set since there are no edges between any two blue nodes. The black nodes do not form an independent set since some pairs of black nodes are connected by edges.

One application of MIS in a complex network is to find a set of nodes that is likely to intersect many different communities without containing too many nodes from any single community. This is because we expect communities to contain many edges, while an independent set cannot contain any edges. A more precise connection to complex network analysis is via the *maximum clique* problem. A *clique* in a graph is a set of nodes for which there is an edge between every pair of nodes in the set. Another way of thinking of a clique is that it contains every possible edge between its nodes. A clique is in some sense the opposite of an independent set in that a clique contains every possible edge while an independent set contains no edge. In fact, the two problems are equivalent in that one can be solved as an instance of the other by taking the complement[12] of the input graph. In complex networks, cliques represent sets of nodes that are tightly coupled and are of interest in a variety of applications.

### 5.3.2 Independent set results

We generate Ising instances encoding the MIS problem by assigning each edge a coupler value of 0 or 1 with equal probability. The linear term has to be set in a specific manner based on the coupler weights in order for such an Ising instance to correspond to an MIS instance (see Appendix E.2). The range of the required linear term components is $[-2, +4]$ for instances on the Chimera graph. These values may be scaled to lie within the DW2's coupler range and precision (Section 2). Another detail worth noting here is that the edges that are assigned a coupler value of 0 are effectively eliminated from the instance, which is not as obvious as it was with our baseline Ising instances, since a linear term is present in the MIS case. For benchmarking purposes, this is useful as it allows us to represent smaller instances within our full 509-node DW2 Chimera graph.

---

[12] The *complement* of a graph is the graph obtained by deleting all existing edges and adding new edges between nodes that were not previously connected by an edge.

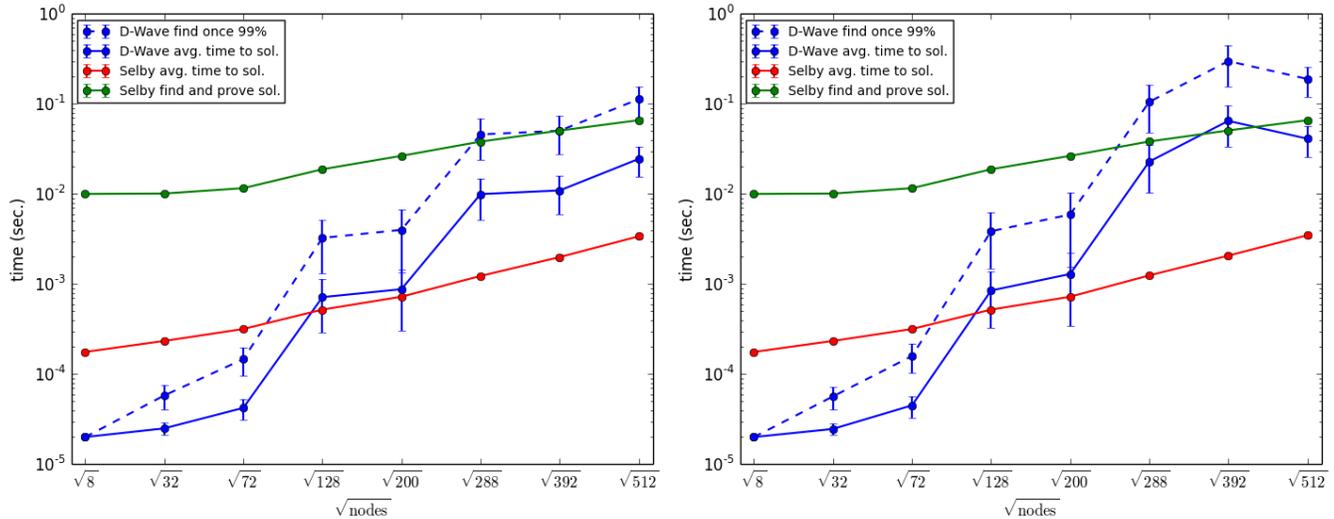

**Figure 18 Comparison of estimated run-time for D-Wave with both Selby's heuristic (red plot) and exact QUBO solvers (green plot).** Instances correspond to randomly generated maximum independent set instances on the Chimera graph, where couplers are assigned a value of 0 or 1 with equal probability and the linear term values are derived from the coupler values (Appendix E.2). The left plot is over 20,000 annealing trials for each of the, at most, 100 instances of each size, while the right is over 100,000 trials. The error bars represent standard error over the, at most, 100 instances of each size.

The performance of Selby's heuristic mode is comparable to its performance on the baseline Ising instances from the previous section; however, the performance of Selby's exact mode is drastically improved and appears to scale better than the heuristic mode, which is surprising. We hypothesize an explanation for this below. D-Wave performs poorly on these instances both in terms of scaling and run times. For the latter, a crossover with Selby's heuristic already occurs below 128 nodes. D-Wave's scaling behavior is both more varied and extreme than on the baseline Ising instances. These are the hardest instances for D-Wave among those we considered.

The D-Wave performance on our independent set instances is particularly striking considering this problem is actually solvable *in polynomial time* on the Chimera graph. This does not contradict the surprisingly poor performance of D-Wave, since polynomial-time solvability only implies that there is some algorithm that solves the problem in polynomial time; the present results indicate that D-Wave is almost certainly not one of them. Here we have an illustration of one of the prime dangers of empirical benchmarking of algorithms — our perspective is limited to only the algorithms and instances that we try. This should be kept in mind when attempting to extrapolate results and draw conclusions.

In fact an optimal independent set can be found in polynomial time for any bipartite[13] graph. Generally, a theoretical result like this may not say anything about the practical performance of algorithms that only theoretically run in polynomial time. For this particular problem, though, algorithms that work reasonably well in practice are also known. This may offer an explanation of why the exact Selby algorithm scales better than the heuristic Selby algorithm. We might expect that the former would have to do more work than the latter, since it has to not only generate a good solution (optimal in this case), but also prove that no better solution exists. The key is that an exact algorithm knows from the outset it is searching for an optimal solution. If it is able to somehow cleverly eliminate large portions of the solution space as being suboptimal without exhaustively searching them, then it may perform quite well. One could even say that this is precisely the goal of an exact algorithm. Heuristics, on the other hand, are generally designed to search only select portions of the solution space, and so they must decide when to stop searching. If a heuristic searches for too long, a smarter exact algorithm may outperform it. Exact algorithms tend to have higher overhead, and we suspect that if we had benchmarked an exact bipartite independent set algorithm, it would have performed similarly to Selby's exact algorithm for these instances.

---

[13] A *bipartite* graph is one whose vertices may be colored red and blue such that all the edges of the graph join a red vertex to a blue vertex. Although it may not be apparent from a cursory visual inspection, Chimera graphs are bipartite.

### 5.3.3 Affinity independent sets

We also consider a variant of the maximum independent set problem that we call the *maximum affinity independent set* (MAIS) problem. In this version of the problem, the input graph contains two types of edges: *plus* edges represent pairs of nodes that we think of as having an affinity towards one another, while *minus* edges represent pairs of nodes that are dissimilar or incompatible with one another. This type of graph is called a *signed graph* in the social network literature. Our goal in the MAIS problem is to find a set of nodes that is large and contains many plus edges but few minus edges. A precise mathematical formulation is given in Appendix E.2. The relationship to MIS is that if our input graph contains only minus edges, then MAIS is equivalent to MIS.

### 5.3.4 Affinity independent set results

Our MAIS Ising instances are generated analogously to the MIS Ising instances above; however, we select each coupler value to be either -1 or +1 with equal probability. This allows us to represent both plus and minus edges. We may think of affinity independent set instances as versions of independent set where we set couplers to be -1 or +1 rather than 0 or 1. The linear terms differ between the two problems, but they are derived in a consistent manner from the coupler values for both problems.

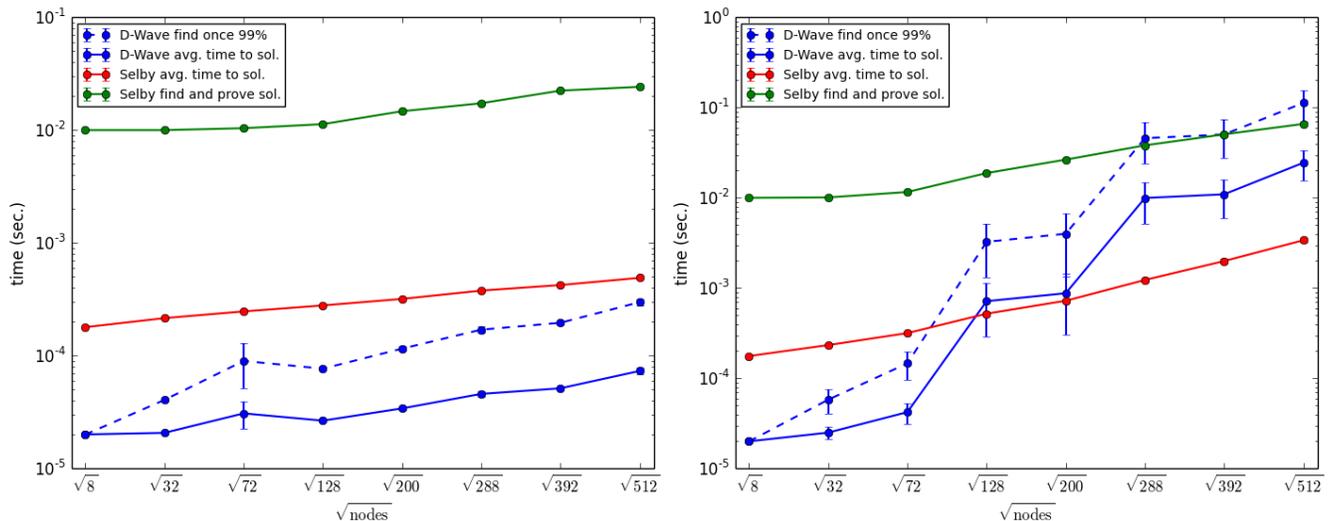

**Figure 19 Comparison of estimated run-time for D-Wave with Selby's heuristic (red plot) and exact QUBO solvers (green plot).** Instances represent randomly generated affinity independent set instances on the Chimera graph, where couplers are chosen to be -1 or +1 with equal probability, and the linear term values are derived from the coupler values. The left plot is over 20,000 annealing trials for each of the 100 instances of each size, while the right plot is the corresponding plot from **Figure 18**, included here for reference. The error bars represent standard error over the, at most, 100 instances of each size.

Our results for the MAIS problem appear in Figure 19, where we have also included the corresponding results for MIS for comparison. The MAIS results complement the MIS results. Although the performance of all three algorithms is better on the affinity independent set instances, the improvement for D-Wave is incredible at up to three orders of magnitude on the larger instances. The performance improvement in the other algorithms is dwarfed in comparison, especially for exact Selby.

These results are quite surprising and relevant for several reasons. The foremost is that because D-Wave is *extremely sensitive* to the difference in the MIS and MAIS problems, while the other algorithms are not, we may have identified a variable that tells us something about the particular strengths and weaknesses of D-Wave's technology. For example, it may turn out that although they are within D-Wave's range and precision, the values used in our MIS instances are an especially bad case for D-Wave for some reason that is not yet understood. Knowing something like this could be tremendously useful. On the other hand, there may be something special about the structure of our MAIS instances that speaks to some yet unknown strength that D-Wave possesses. If we can fathom such a "hidden" strength, then we may be able to apply it to other problems. Clearly, this phenomenon warrants further investigation, which we plan to undertake. We should also note that the outcome of this could be more mundane than either of the options suggested above. It may be that we simply are not using the best classical algorithms or configuring them appropriately (see Section 3.3.2 for an example).

Unlike the bipartite independent set problem, it is not clear what kind of scaling we should expect to see for bipartite affinity independent set. We included the affinity independent set problem in our study as a variation of the independent set problem; we are unaware of its exact theoretical complexity. It is also unclear how to apply the techniques that give a polynomial time algorithm for the bipartite independent set problem to the bipartite affinity independent set problem, nor is it obvious how to show the problem is NP-hard. If we were to casually glance at Figure 19 without any foreknowledge about D-Wave benchmarking results, we would hazard that the MAIS problem is probably not NP-Hard on bipartite graphs. Of course when making such predictions based on limited and perhaps irrelevant data, one must tread carefully, as we have been taking every opportunity to emphasize. Still, it is useful having examples of problems that empirically appear to be easy, but whose theoretical complexity is unknown. If nothing else, we have a clue as to how we might attack the unknown; it may be that understanding why these instances are empirically easy allows us to devise a true polynomial-time algorithm. In our experience such problems are rare gems, and it is our pleasure to make the acquaintance of one more such problem.

## 5.4 Planted-solution instances

### 5.4.1 Problem motivation and definition

As previously discussed, although we have given both empirical and theoretical evidence that finding an optimal solution should not always be one's objective (Section 3.3.1), we still seek to compare performance relative to an optimal solution. Doing so gives us more meaningful and definitive benchmarking results. Fortunately, we were able to use classical algorithms to verify the optimal solution (value) for all of the D-Wave benchmarking instances we considered. In fact, for each of our instances, we were able to find the optimal solution within 30 minutes; the vast majority of the instances took less than one minute to solve. Current incarnations of D-Wave's technology have approximately 512 qubits; however, as D-Wave's QAs grow in size and, presumably, problem-solving capacity, it will be increasingly difficult for one to generate and verify optimal solutions by other means. Although this poses a difficult regime for benchmarking, it appears inevitable and is interesting for assessing the near-term potential of D-Wave's annealing technology. How does one perform benchmarks that require knowing an optimal solution for problems where finding one is extremely difficult?

One elegant means of circumventing the inability to find an optimal solution is to *plant* an optimal solution. The idea is that one first fixes a solution, $x$ and then fabricates an instance where $x$ is in fact an optimal solution for the instance. This endeavor can be as tricky as it may sound: how does one generate an instance around $x$; how does one guarantee that $x$ is indeed optimal for the generated instance, and if one succeeds, how does one ensure that the resulting instance is actually rich and non-trivial? For these reasons, planting strategies are usually highly problem specific. Hen et al. [34] suggested a planting strategy for Ising instances on the Chimera graph[14]. They performed experiments indicating that their generated instances may be tweaked so that they vary in difficulty for a DW2. For easier instances, they give evidence that D-Wave outperforms classical approaches such as simulated annealing and the Selby heuristic that we have used in this study. We find their results compelling and reproduce some of their experiments here. King [35] has obtained comparable results even when considering such instances with range-limited coupler weights. One new discovery on our part is that the optimal solution *value* (but not necessarily the planted solution itself) can be computed in polynomial time [40]. We are in the process of exploring further ramifications of this result and are preparing a technical article describing it.

### 5.4.2 Results

Hen and collaborators' planted-solution instances are defined on a graph that is a collection of cycles[15] within the Chimera graph; refer to [34] for more details. We define the *cycle density, C,* as the ratio of the number of cycles in the collection to the number of nodes in the graph. Hen and collaborators found that $C$ plays a critical role in determining the difficulty of solving their instances, particularly for their DW2. They discovered that $C = 0.2$ appears to be the hardest case. We performed experiments with $C = 0.2$ and $C = 0.5$; the latter is interesting because Hen and collaborators' results indicate that D-

---

[14] Their solution-planting strategy applies to Ising instances on general graphs as well.
[15] A *cycle* in a graph is a path that starts and ends at the same vertex $v$; edges or vertices are not allowed to repeat, except $v$.

Wave outperforms other algorithms at this cycle density. They compare D-Wave against Selby's heuristic solver, but they do not use Selby's exact mode.

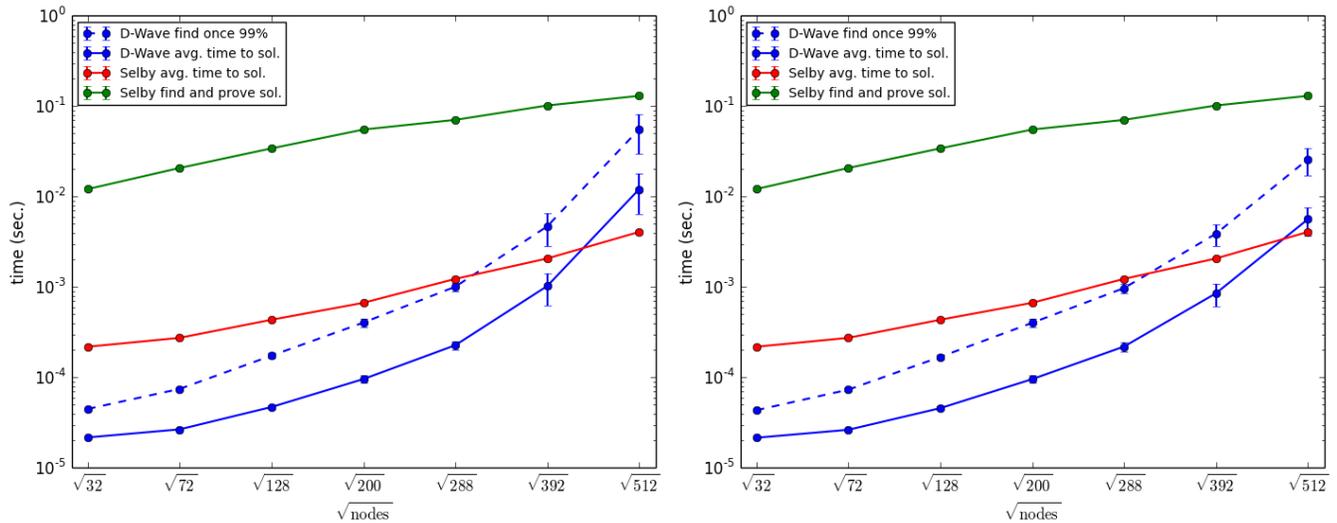

**Figure 20 Comparison of estimated run-time for D-Wave with Selby's heuristic (red plot) and exact QUBO solvers (green plot).** Instances contain a planted solution with a cycle density parameter of 0.2. The left plot is over 20,000 annealing trials for each of the, at most, 100 instances of each size, while the right is over 100,000 trials. The error bars represent standard error over the, at most, 100 instances of each size.

Our results for $C = 0.2$ in Figure 20 are comparable to Hen and collaborators'. One difference is that their results do not show a crossover between D-Wave and heuristic Selby. We obtained slightly larger mean solution times for D-Wave at 392 and 512 qubits. There are many possible explanations for the discrepancy. The most likely is that, although our methodologies are similar, they opted to use the annealing time that resulted in the best performance for each instance. Although this happened to correspond to our annealing time of $20\mu s$ most of the time, this was not always the case. Even though they do not observe a crossover in their data, their data indicates that a crossover is imminent upon extrapolation. The other major difference is that they used the DW2 machine at the Information Science Institute at the University of Southern California. Given that our results are generally consistent, we consider our results a verification of theirs.

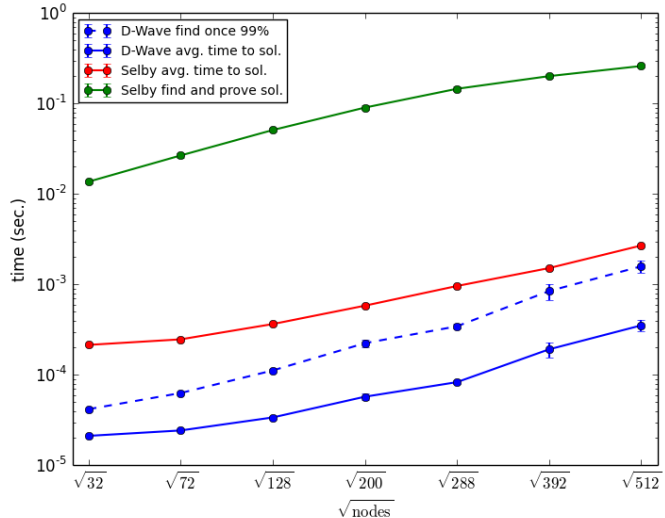

**Figure 21 Comparison of estimated run-time for D-Wave with Selby's heuristic (red plot) and exact QUBO solvers (green plot).** Instances contain a planted-solution with a cycle density parameter of 0.5. The plot is over 20,000 annealing trials for each of the 100 instances of each size. The error bars represent standard error over, the at most, 100 instances of each size.

At a cycle density value of $C = 0.5$, our results are virtually identical to Hen and collaborators'. D-Wave appears to perform an order of magnitude better than the Selby heuristic, and it is not clear whether a crossover might occur. Of course our focus is on scaling, and the plots show nearly identical scaling. The more interesting feature of the results is the scaling of Selby's exact algorithm. It appears to be the only one that may be scaling better than $\sim c^{\sqrt{N}}$ in the long run, though we would need more data to validate this hypothesis. One point possibly supporting such behavior is our aforementioned polynomial-time algorithm to obtain the optimal value of these planted-solution instances.

In the next section we present a different flavor of quantum approach that we are not currently able to effectively implement on a D-Wave system. We frame our results for this approach using our Quantum Monte Carlo simulator.

## 5.5 Community detection

Most types of real-world complex networks contain groups of related or well-connected entities, that one may call a community. For example, a community might represent users who tweet about similar topics in a Twitter network, while in biological networks communities might represent groups of proteins that interact for some specific function. *Community detection* is the problem of identifying interesting or relevant communities in a network. Appendix C offers a more detailed introduction to community detection, and in Appendix D we present a thorough benchmarking study of common classical community-detection heuristics. Our main purpose here is to explain and present our QMC simulation results and compare them with our classical benchmarking results. There are a number of significant challenges to obtaining a fair comparison. The first is that our "quantum" algorithm is actually a hybrid approach that pairs a classical Ising-based heuristic with the QMC simulator as an Ising-solving facility. The second is that the QMC simulator requires a fair amount of tuning and performance is highly sensitive to a number of parameters. Although we have done some preliminary experiments to determine reasonable parameter settings, we suspect further research could have a significant impact on its performance and results.

### 5.5.1 Hybrid community detection algorithm

Although the notion of a community in a complex network is often nebulously defined, researchers have developed several measures that seek to capture the notion of a community. *Modularity* was one of the first such measures considered, and it is the measure used by well-known classical heuristics such as Louvain's algorithm; a precise mathematical formulation of modularity appears in Appendix C.2. Although several issues in using modularity for community detection have been identified, modularity maximization is readily formulated as an Ising problem. For this reason, in addition to those above, we use modularity-based community detection algorithms in our work. Our goal is not to advance the state of the art of community detection; rather, we seek to bring rich and real-world domains to bear on benchmarking quantum annealing.

Our ultimate goal with community detection is to partition a network into many communities so as to maximize the modularity measure. A direct Ising formulation for this problem requires significant overhead in terms of variables, which correspond to qubits for a quantum solver. One way of circumventing this issue is to use a quantum annealer to solve community detection instances where we seek to *bipartition* the graph into two communities. Of course our ultimate goal is to partition into possibly many more communities, and we must resolve this discrepancy. We do so with a hybrid algorithm that has two components: (i) a purely classical component that uses recursive bipartitioning[16] to partition a graph into many communities using, (ii) a black-box Ising solver to actually perform each bipartitioning so as to maximize modularity. We note that any Ising solver, quantum or classical, suffices for the black box in (ii).

We may view our hybrid approach as follows. The classical bipartitioning heuristic, (i) from above, has at its disposal a black box, (ii) from above, that can partition a graph into two communities in a way that approximately maximizes modularity, and it seeks to use this black-box facility to partition the graph into many communities to maximize modularity. The heuristic first starts with the whole graph and evaluates the change in modularity attained by splitting the graph into two parts; if such a split would increase modularity, it is performed. Now the same algorithm is applied on each resulting piece; namely, if a bipartitioning would increase modularity, it is executed. The algorithm continues recursively bipartitioning in this fashion until an increase in modularity is no longer possible. Such a greedy approach is common for solving graph-partitioning problems, but can fail to find an optimal partitioning.

The overall quality of the above approach will depend on two main factors: (i) is recursive bipartitioning the right approach in the first place, and (ii) how well does our Ising black-box solve each bipartitioning sub-problem. Even with an exact Ising solver as our black box, recursive bipartitioning could perform poorly as an overall strategy. On the other hand, it could be that recursive bipartitioning could perform poorly only because we are using an Ising solver that is not effectively finding near-optimal solutions to the sub-problems handed to it. Moreover, these factors will likely vary from instance to instance. Thus we must be careful in doling out blame if we witness poor overall performance from the hybrid algorithm.

### 5.5.2 Execution time results

The specifics of using an Ising solver to find a modularity maximizing bipartition are given in Appendix E.1. A point worth mentioning here is that even if the graph underlying a particular bipartitioning sub-problem is sparse, the resulting Ising instance is dense and essentially requires edges between all pairs of nodes. This precludes an effective D-Wave implementation, as we would have to rely on embedding a fully connected graph on a D-Wave system, affording us approximately 33 nodes. Even if we opted to do this, both the bipartitioning Ising instances and the embedding itself would require coupler weights well beyond the precision the current D-Wave systems offer. Admittedly, this is an artifact of the way in which the problem is formulated as an Ising instance. Other formulations are possible; however, the current formulation is direct and natural. Other formulations are likely to incur overhead as well, but for different reasons. Thus we opted not to pursue a D-Wave implementation and stick to our Quantum Monte Carlo simulator, which we refer to as "QSim" below.

---

[16] *Recursive bipartitioning* is an old and well-known approach that splits a graph into two parts and then recurses on each of the resulting parts until some stopping criterion is met, which is commonly based on the size or number of parts obtained.

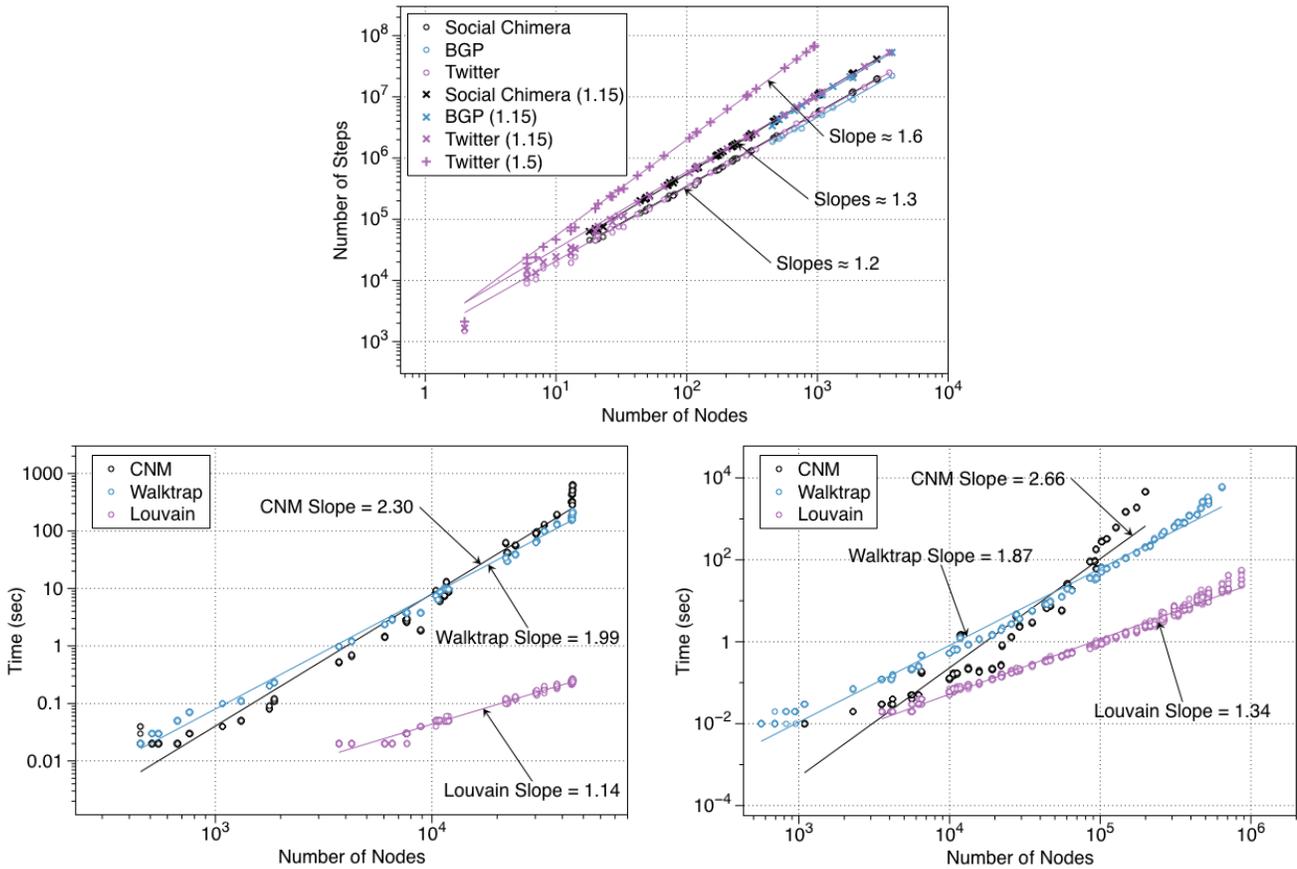

**Figure 22 Comprehensive execution effort results for QSim (top) and for the classical heuristics on BGP instances (left) and Twitter instances (right).** The reported values are the slope of the best-fit line for each algorithm on a log-log plot.

A comparison of the execution times for our simulation and the classical community detection heuristics (from Appendix D, Figure 39) is presented in Figure 22. In order to isolate the quantum system being simulated from the classical simulation overhead, we report the number of Monte Carlo steps as a proxy for run time. This is a reasonable choice since our primary concern is scaling rather than absolute run time; previous studies have adopted this convention as well [19], [20].

The other factor to note is that each time our simulator is called to solve an Ising bipartitioning sub-problem, we control the number of Monte Carlo steps for which it is executed. The number of steps displayed in the plot above is the total number of Monte Carlo steps during the entire execution of the hybrid algorithm. We experimented with the Monte Carlo step parameter in order to obtain total running times comparable to the classical algorithms. The numbers in parentheses next to the instance family name in the legend of the top plot are parameters from which we may infer the number of Monte Carlo steps allowed for each Ising bipartitioning sub-problem executed during the overall hybrid algorithm. The precise correspondence is described by example below.

"Twitter ($c$)," in the top plot, indicates running each Ising bipartitioning sub-problem for $\sim n^c$ iterations, where $n$ is the number of nodes in the subgraph being bipartitioned. Thus $n$ will vary during the execution of the algorithm. For the first subproblem that bipartitions the entire graph, we would have $n = N$, where the latter is the number of nodes in the entire graph. For later subproblems, $n$ could represent just a handful of nodes. As described in Section D.1.2, the slope of the lines in the log-log plot above corresponds to the exponent of the overall running time, as derived from a fit of our data. As an example, from the plot for "Twitter (1.5)," we see that when we afforded each Ising sub-problem $\sim n^{1.5}$ Monte Carlo steps on our Twitter instances, then our overall QSim running time was $\sim N^{1.6}$. We infer from our results that letting QSim run for $\sim n^{1.15}$ steps on each Ising subproblem yields overall run times comparable to Louvain.

### 5.5.3 Modularity results

Having determined how long we should run QSim, we may examine the quality of solutions produced, which we measure using modularity as with our classical community detection study in Appendix D.

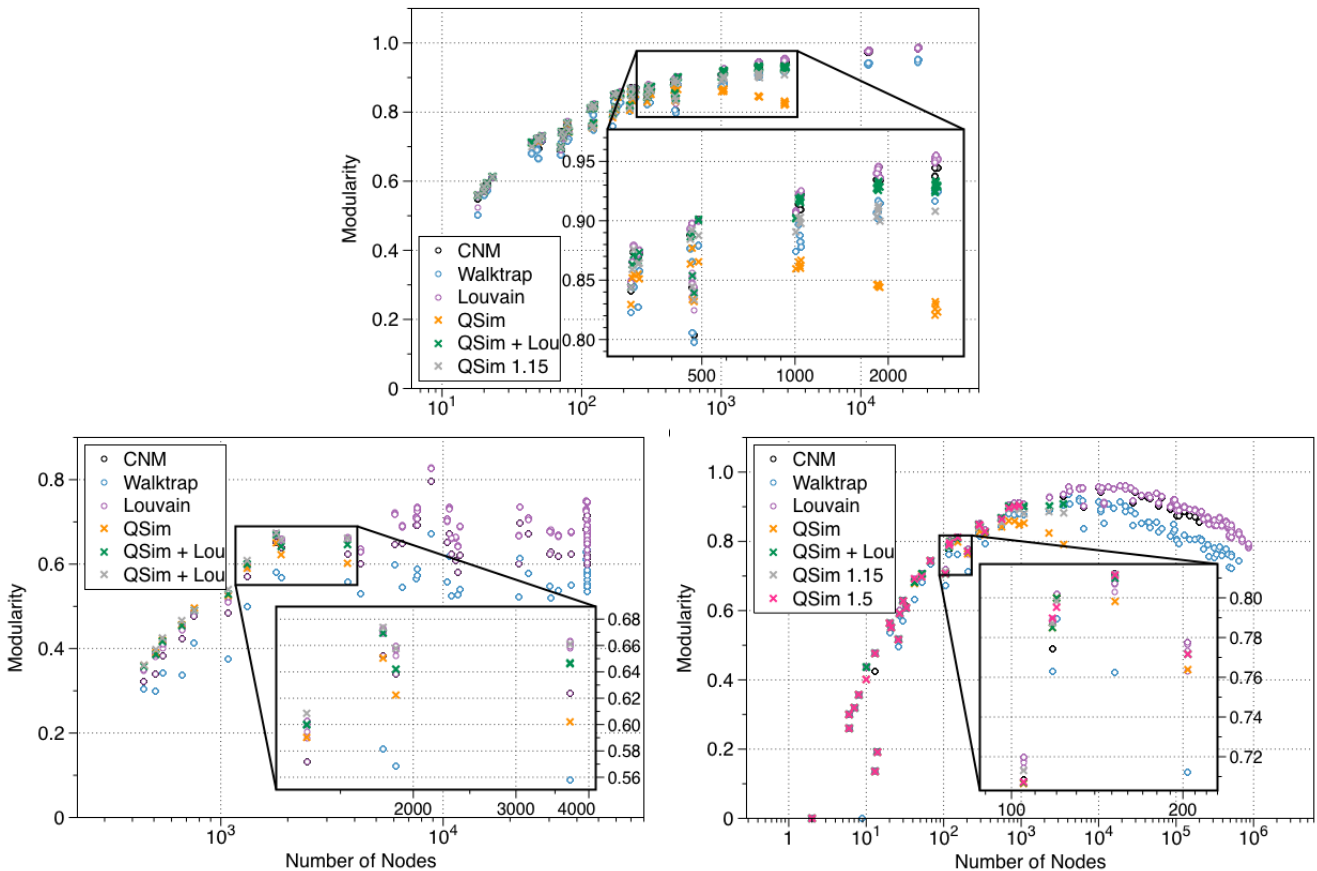

**Figure 23 Overall modularity attained across the classical heuristics, as well as our hybrid quantum simulation approach.** The results for our social Chimera graphs appear at the top, while those for our BGP and Twitter graphs appear at the bottom left and right, respectively.

The modularity of the solutions obtained by our algorithms appears in Figure 23. "QSim" indicates a run time of $\sim n$ for Ising subproblems, while "QSim + Lou" entails first executing QSim, at the aforementioned run time, and then executing the Louvain heuristic seeded with the partition discovered by the former. Louvain is fast, especially relative to the QSim; hence this type of post-processing approach is an inexpensive way to get a potential improvement in solution quality.

For all three social network families we see that the QSim variants are competitive with, if not better than, the others up to approximately 1000 nodes. Yet in all three cases, we see a sharp drop in QSim quality beyond this threshold. This is curious behavior, and the cause is not clear. One possibility is that the graphs are not sufficiently complex until this point. This hypothesis could be tested by further varying the Monte Carlo steps allotted to the local bipartitioning sub-problems, as discussed above. We would expect a QSim run with more allotted steps to perform better than ones with fewer allotted steps. Indeed, for the Twitter instances we were able to collect data allotting $\sim n^{1.15}$ and $\sim n^{1.5}$ steps to the Ising sub-problems; however, we exhausted our data for the $\sim n^{1.5}$ runs soon after the divergence occurs in the above Twitter plot. These runs are time-consuming, with some of them continuing to run for over a week with no apparent end in sight.

Another explanation is simply that our QSim algorithm requires a more favorable setting of parameters or further tuning. We witnessed for our baseline Ising results in Section 5.2 that the current version QSim was not able to produce meaningful results for our 392- and 512-node Chimera Ising instances. Our modularity results could be another manifestation of this phenomenon. If this is the case, then there is hope that future QSim versions or quantum annealers may prove competitive with Louvain on the types of instances we consider.

Our final explanation hinges on the hybrid QSim approach itself. Solving Ising sub-problems to optimality is powerful but expensive. The hope driving the hybrid QSim approach is that it converges to a high-modularity partition of communities quickly without executing too many Ising-based bipartitioning sub-problems along the way. Of course, by setting the number of Monte Carlo steps allotted to these sub-problems, we expect a tradeoff between execution time and the quality of the bipartitioning found by the simulator. Balancing these quantities is key, and a deeper exploration remains to be done.

In contrast, the Louvain heuristic performs basic moves that are simple and inexpensive in that a single node is potentially moved to a community in which one of its neighbors lives (see Appendix C.2 for details). Louvain is more of a "bottom-up" approach that builds larger communities from smaller ones, whereas QSim splits large communities into smaller ones "top-down." Combining these ideas could lead to improvements in QSim, as well as new approaches to community detection that are competitive with Louvain but rely on solving Ising-based sub-problems. For instance, our QSim + Louvain approach uses Louvain as a kind of black-box post-processing step at the end. What if we were to use simple Louvain-like steps during the execution of the algorithm? We intend to explore these types of questions, as well as alternate Ising formulations for modularity-based bipartitioning that may be implemented on larger near-term D-Wave devices.

The aforementioned quality drop notwithstanding, we consider the above results encouraging in that Ising-based approaches amenable to quantum implementation may be able to compete with leading-edge classical heuristics.

# 6   Discussion and concluding thoughts

This section is framed in terms of high-level questions pertaining to D-Wave systems. Although this section may be treated independently of its predecessors, our discussions draw upon and refer to the body of work presented in this document.

## 6.1   Can D-Wave systems address real-world applications?

Although we have developed infrastructure to allow D-Wave systems to address real-world complex-network analysis problems, we feel that the current generation is not up to the task of handling real-world data. Recent investigations attempt to address real-world applications [4], [16], [41], [42], though in general we feel the precision and connectivity limitations are too great for meaningful benchmarks.

## 6.2   How does one assess the growth of D-Wave systems?

Computing is pervasive in modern culture, and jargon pertaining to the anatomy of a conventional computer, such as "CPU" and "RAM," is now mainstream. We, as a computing-savvy culture, also have a basic grasp of the resources underlying traditional computation, such as processing time and memory, at least at a high enough level to guide us when shopping for a new computer. We understand that processing speed is currently measured in gigahertz and that memory is currently measured in gigabytes. We also understand that computers have hierarchies of at least two types of memory[17], with one type being an order or two of magnitude larger (but slower) than the other. Finally, we are aware that modern CPUs are actually composed of several *cores* that are essentially independent processing units that tend to share the same memory. Perhaps to our annoyance as upgrade-averse consumers, we have observed that the capacity for these resources has been growing at a staggering pace.

How do we compare the growth of the above classical resources with the growth of quantum resources? If a conventional computer's CPU is upgraded from 2Ghz to 4Ghz, with all other resources remaining equal, this means that over a fixed interval of time, the computer will perform twice as many basic operations as it could before. This does not directly imply that an arbitrary algorithm will run twice as fast, since other resources may pose a bottleneck; for example, a memory-intensive big-data algorithm may be constrained more by memory access times than processing time. In contrast, what does it mean if a D-Wave QA is upgraded from 128 qubits to 512 qubits? Are there other resources that quantum annealing explicitly or implicitly leverages?

The qubits available to a quantum system allow it to represent a state, which may be a superposition. We may perform physical operations, governed by the laws of quantum mechanics, that allow us to transform one state into another. In this way we arrive at quantum computation. Having access to a greater number of qubits allows us to represent a larger state space and, consequently, larger problem instances. Thus qubits correspond to memory rather than processing speed; qubits are a limited and precious resource, and current quantum systems may be considered memory constrained. If qubits correspond to memory, then what is a quantum analog for processing speed? The speed at which a quantum system solves problems is dictated by its architecture and implementation (see Section 1 and Appendix A).

---

[17] There is generally a hierarchy of memory, with small amounts of *cache* memory offering the fastest access times. One could argue that at the other end of the hierarchy lies the collective memory of the Internet, with relatively glacial access times that require retrieving information over great distances. We will use *memory* to refer to cache and RAM, while *storage* will refer to slower archival systems such as hard drives.

### 6.2.1 Processing speed

For adiabatic quantum computing and quantum annealing (see Section 1.2 and Section 1.3), the "processing" speed is dictated by the particular adiabatic evolution and speed implemented by the system. D-Wave systems do allow some degree of user-level control over the later with the annealing time. However, there is a delicate balance between annealing time and expected solution quality, and the optimal annealing time for a problem is not clear, a priori, and may lie somewhere in between the minimum and maximum allowable annealing times. Empirically, the D-Wave Two's minimum annealing time of $20 \mu s$ appears to perform best for the problems and instances the community has benchmarked [16], [19], [20]. Another complication is that empirical evidence suggests that we must perform upwards of thousands of annealing runs in order to obtain near-optimal solutions with a reasonable success rate. The number of runs necessary is not clear a priori either.

Zeroing in on a concrete and generic analog for classical processing performance is difficult; however, we may circumvent this task by empirically assessing whether D-Wave quantum annealing "processing power" is growing with qubit count. D-Wave qubit capacity has enjoyed exponential growth (Figure 27). Are D-Wave systems becoming faster as they get larger? Although the fundamental quantum annealing architecture remains the same among D-Wave's systems, one might expect engineering improvements to manifest themselves as speedup. To the best of our knowledge, such an analysis has not been previously published. We obtain a rough estimate based on comparing a current-generation D-Wave Two with a previous-generation D-Wave One system, in Figure 24. We compare the D-Wave One results published in the seminal work of Boixo et al. [20] with the corresponding results appearing in the follow-up D-Wave Two work [19] on the largest instance size supported on the former. Since this particular instance size of approximately 100 qubits was not directly used in the D-Wave Two study, we are comparing observed D-Wave One results with interpolated D-Wave Two results. With this said, the comparison indicates that although the D-Wave Two is likely to be about an order of magnitude faster on the hardest 1% of instances, the D-Wave One actually appears faster on the remaining instances. This may be due to the D-Wave One supporting a smaller minimum annealing time of $5 \mu s$ rather than $20 \mu s$. A more conclusive study would incorporate data from previous generations as well, which only D-Wave is likely to possess. For our purposes, this crude comparison is sufficient to suggest that D-Wave systems may not be getting faster as they grow larger. It could be the case that inherent challenges of assembling larger D-Wave systems shadow engineering improvements, or it may be that a more sophisticated study is required in order to bring to light improvements in the current D-Wave systems.

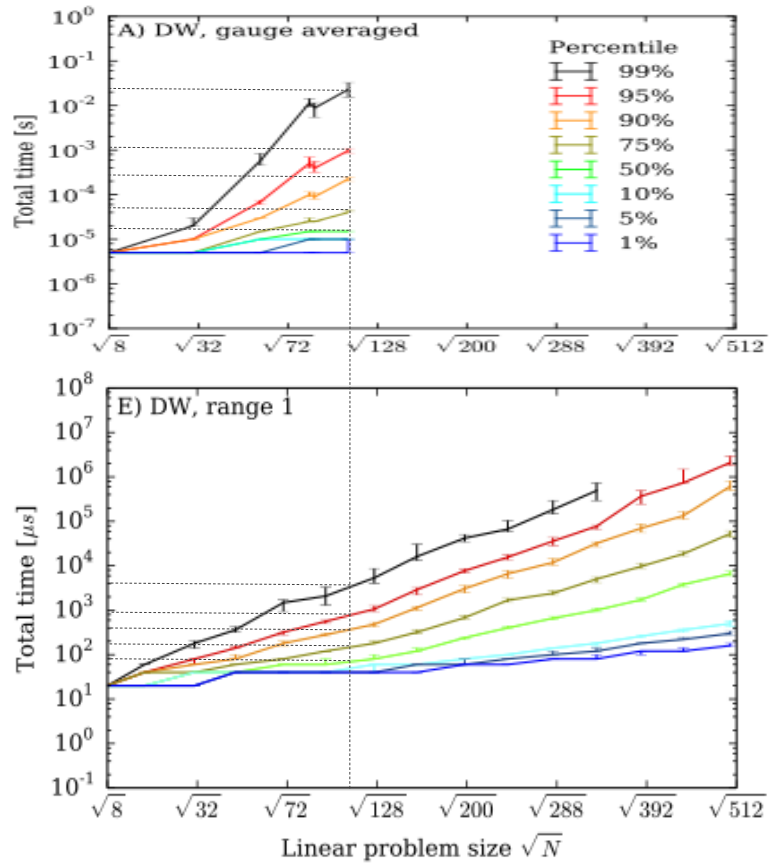

**Figure 24 Comparison of D-Wave One (top [20]) and D-Wave Two (bottom [19]) systems.** Instances are Chimera Ising instances with coupler weights drawn uniformly from $\{-1, +1\}$ and no linear term; performance is plotted according to hardness quantile. The time displayed is the mean time over the different instances to find an optimal solution at least once with probability 0.99. *Note the difference in time units between the plots.* The D-Wave One system employed in the study had just over 100 functional qubits. The minimum annealing times of **$5\mu s$** and **$20\mu s$** were used for the D-Wave One and D-Wave Two, respectively.

### 6.3 How do future D-Wave and classical technologies compare?

If both D-Wave and conventional classical computing technologies continue at their current rate of growth, do we expect D-Wave to offer a computational resource advantage at some point? Although such questions are foremost in the minds of those with an interest in D-Wave's devices, providing a satisfactory, fair, and complete answer is tricky, if not impossible. For one, extrapolating technological growth is ultimately just speculation. The known limits to computation are both broad and deep (Figure 25). Experts have been predicting the end of the golden age of exponential growth in conventional computing, characterized by principles such as Moore's Law and Dennard Scaling[18]. Of course other experts have also managed to creatively and deftly circumvent computing bottlenecks in the form of new paradigms such as massively parallel and cloud computing. Yet new paradigms can demand drastically new algorithms, and a true continued exponential growth in raw problem-solving power draws upon a delicate dance between hardware, architecture, software, and algorithmic insights.

---

[18] Gordon Moore, co-founder of Intel, observed in 1965 that the transistor count in integrated circuits doubled approximately every two years. This became known as *Moore's Law*; it still holds today though many predict a near-term end. *Dennard Scaling* postulates that transistor power consumption is proportional to area. Thus Moore's Law along with Dennard Scaling predicts exponential growth in performance per watt. Unfortunately, Dennard Scaling appears to have ended (see Figure 26).

| Limits | Engineering | Design and Validation | Energy, time | Space, time | Information, Complexity |
|---|---|---|---|---|---|
| Funda-mental | Abbe (diffraction) Amdahl Gustafson | Error-corr. & dense codes Fault-tolerance thresholds | Einstein $E=mc^2$ Heisenberg $\Delta E \Delta t$ Landauer $kT \ln 2$ Bremermann Adiabatic thrms | Speed of light Planck scale Bekenstein Fisher $T(n)^{1/(d+1)}$ | Shannon Holevo NC, NP, #P Turing (decidability) |
| Mate-rial | Dielectric constant Carrier mobility Surface morphology Fabrication-related | Analytical & numerical modeling | Conductivity Permittivity Bandgap Heat flow | Propagation speed Atomic spacing No gravitational collapse | Information transfer between carriers |
| Device | Gate dielectric Channel charge ctrl Leakage, Latency Crosstalk, Aging | Compact modeling Parameter selection | CMOS, quantum Charge-centric Signal to noise Energy conversion | Entropy density Entropy flow Interfaces & contacts Size & delay variation | Universality |
| Circuit | Delay, Inductance Thermal-related Yield, Reliability, IO | Interconnect Test Validation | Dark, darker, dim and gray silicon Cooling efficiency Interconnect Power density/supply 2D or 3D | | Circuit complexity bounds |
| System +SW | Specification, Implementation Validation, Cost | | Synchronization, Physical integration Parallelism, *Ab initio* limits (Lloyd) | | The CAP theorem |

Figure 25 Some of the known limits to computation [43].

With that said, we may still, perhaps naïvely, extrapolate conventional computing and D-Wave technological growth patterns, along with our experimental results, and derive a rough sense of whether a D-Wave system might offer an eventual computational resource advantage on some class of problem or instances. Our goal is to evaluate whether some near-term crossover favoring D-Wave is likely to occur. Although we try to be as precise and accurate as possible, there are many unknowns, and when we must, we generally resolve unknowns in favor of D-Wave. For example, Dennard Scaling appears to have ended (Figure 26); for our purposes, we favor D-Wave and indeed assume that this is the case and that conventional classical computing has plateaued. Thus in looking for a crossover when extrapolating and adjusting our empirical results, we assume no increase in classical algorithmic performance. We observed in the previous section that although D-Wave system qubit capacity has enjoyed an analog of Moore's Law (Figure 27), newer and larger D-Wave systems do not appear solve problems any faster than older systems. Thus we assume no increase in D-Wave's raw algorithmic performance either.

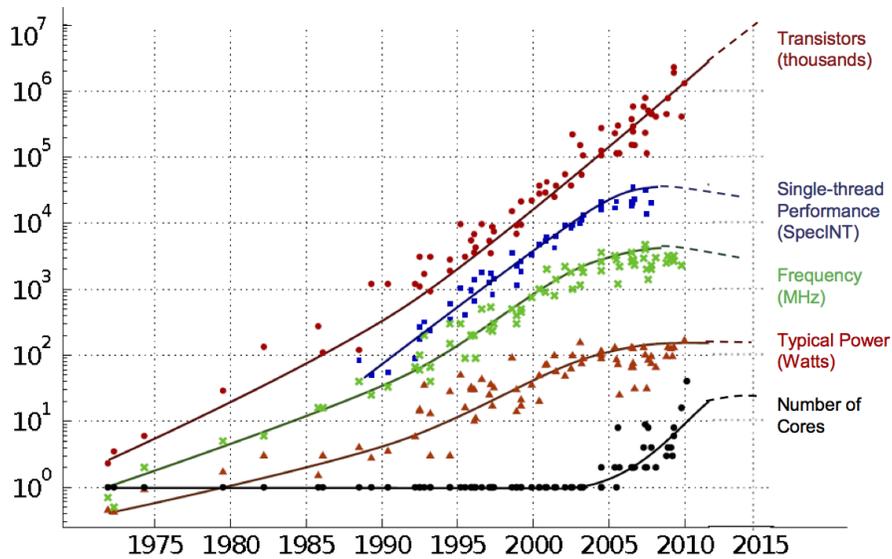

Original data collected and plotted by M. Horowitz, F. Labonte, O. Shacham, K. Olukotun, L. Hammond and C. Batten
Dotted line extrapolations by C. Moore

**Figure 26 The end of Dennard Scaling [44].**

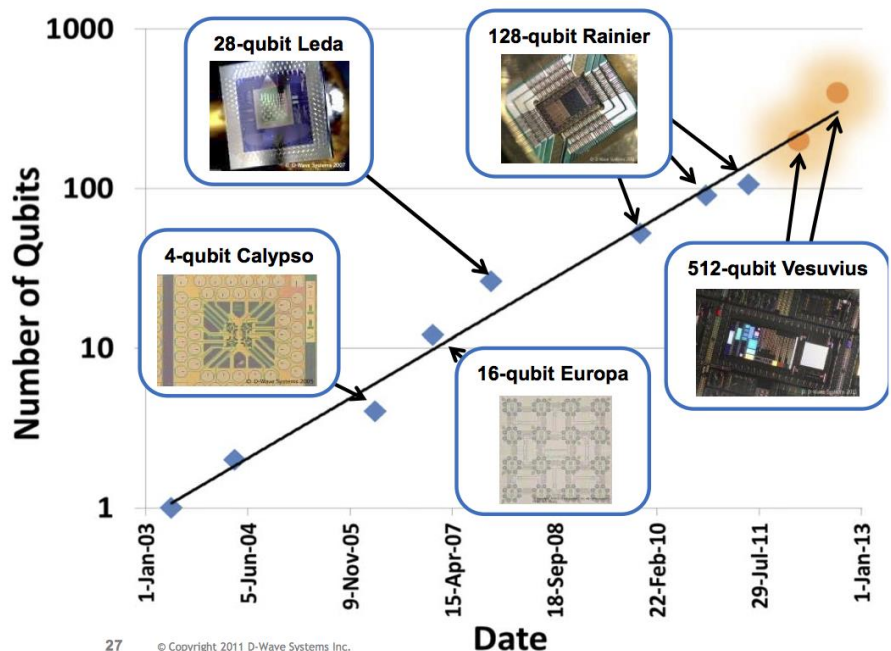

**Figure 27 D-Wave's version of Moore's Law.**

In a regime expecting no performance improvement in future classical and D-Wave devices, we may simply extrapolate our plots from Section 5 without additional adjustment. For the majority of our benchmark instances, we already observe a crossover favoring Selby's classical Ising heuristic. The exceptions are the affinity-independent-set instances (Section 5.3.4) and some of the planted-solution instances (when the cycle density is 0.5; see Section 5.4.2). However, as mentioned earlier, we suspect that these instances are solvable in polynomial time by efficient algorithms and that we are not currently comparing against the best-available classical options [40]. Yet it is not at all clear to us if we may be failing to predict a D-Wave speedup because of underestimating D-Wave system performance.

### 6.3.1 Does parallel computing offer an advantage?

Rønnow et al. [19] suggest that a more accurate assessment of D-Wave system performance ought to take into account that executing instances smaller than system capacity does not fully leverage all the system's resources. In other words, given a D-Wave machine with a total capacity of $M$ qubits (e.g., $M = 512$ for an ideal D-Wave Two), we may actually execute $K = \lfloor M/N \rfloor$ independent instances of size $N$ in parallel on such a D-Wave machine. Rønnow and collaborators thus propose that when assessing a potential D-Wave quantum speedup, D-Wave system performance ought to be scaled accordingly.

We see two flaws in this suggestion. Benchmarking instances of varying sizes on a fixed-size D-Wave system is meant to give us an estimate of how D-Wave performance scales as system size increases. At any given point in time, we assume there is some largest D-Wave system, with machine size $M$, at our disposal. We are most interested in evaluating this D-Wave system's performance on the largest viable instances, of size $M$, rather than smaller instances of size $N \ll M$. An opposing viewpoint is that we have access to some arbitrarily large D-Wave system on which we are interested in executing relatively small instances. As qubit quantity is likely to be a bottleneck well into the future, the latter viewpoint is unrealistic in that instances that are small compared to $M$ are generally not representative of interesting, real-world-scale problems. The other, more critical issue is that to truly take advantage of this factor $K$ in a parallel speedup, one needs to be able to efficiently load all $K$ instances of size $N$ onto a D-Wave system in some sort of parallel fashion, otherwise one might take $\sim M$ steps just to load all the instances. When $N \ll M$, the loading time might dwarf a potential parallel speedup. More to the point, the assumption of solving many independent instances is precisely the type "embarrassingly parallel" task that is extremely well suited to current massively parallel classical systems. Thus, if we assume D-Wave systems are able to take advantage of this scenario, it is fair to assume classical systems are able to as well.

## 6.4 What other resources are relevant to D-Wave computation?

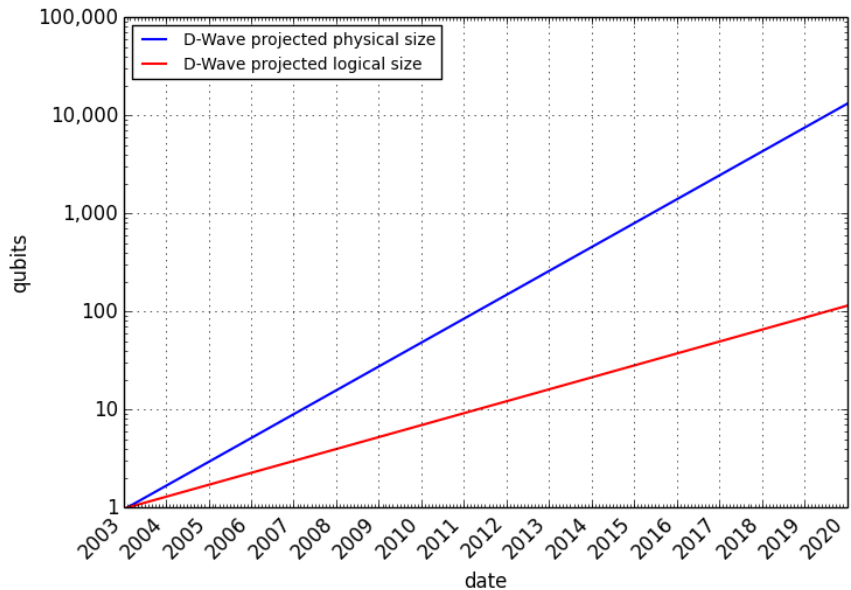

**Figure 28 Projected physical vs. logical qubit capacity for D-Wave systems.** This assumes physical qubit capacity continues to grow at the rate suggested in **Figure 27**.

Although system capacity in terms of total number of qubits is the key D-Wave resource we consider, this does not take into account the significant limitations imposed by D-Wave's restricted qubit connectivity and limited coupler precision (see Section 4.3.1). To factor this in, we consider the difference between *logical qubits*, corresponding to problem-domain variables such as vertices in a graph, and *physical* or *hardware qubits* actually present in a D-Wave system. This distinction is illustrated in Figure 2. If we rely on minor embeddings (defined in Section 4.3.1) to represent logical qubits using physical qubits, then it could take up to $\sim N^2$ physical qubits to represent $N$ logical qubits. Assuming this worst case, Figure 28 depicts projected D-Wave physical and logical qubit capacities. If the current rate of D-Wave system growth were sustained, we would expect approximately 10,000 physical qubits by 2020; however, this corresponds to only 100 logical qubits and ignores the issue of coupler precision, which may require additional qubits to circumvent. This drives home the point we have been reiterating: *the most severe bottleneck in effectively leveraging a D-Wave system is efficiently representing real-world and interesting problems on a Chimera graph*. For benchmarking purposes we have been successful in circumventing these limitations to some extent (e.g., our real-world-like Chimera-minor social networks from Section 4.3.2); however, it is unclear how the situation will unfold for solving real-world problems on future D-Wave devices.

Although minor embeddings are natural, they are perhaps not the best means of expressing problem-domain instances on a Chimera graph. Is there some other approach that would allow one to use fewer physical qubits? Unfortunately, the worst-case requirement of $\sim N^2$ physical qubits for $N$ logical qubits is likely an artifact of the limited connectivity of the Chimera graph itself, rather than minor embeddings. We base this claim on what is known in theoretical computer science as the *Exponential Time Hypothesis (ETH)* [45], [46]. The ETH roughly postulates that there are some fundamental problems for which the best exact classical algorithm, one that always finds an optimal solution, takes no less than $\sim 2^N$ time on an instance of size $N$. One such problem is the maximum cut problem on arbitrary graphs, which is equivalent to the Ising problem on arbitrary graphs (as opposed to Chimera graphs). The Ising problem on an $N$-node Chimera graph can be solved in time $\sim 2^{\sqrt{N}}$, due to the inherent structure of the Chimera graph. This is relevant because, if one could represent an arbitrary $N$-node Ising instance as an $M$-node Chimera instance, where $M \ll N^2$, then one would be able to solve the original instance in time approximately $2^{\sqrt{M}} \ll 2^{\sqrt{N^2}} = 2^N$, violating the ETH. Although no proof of the ETH is currently known, it is widely believed to be true. Thus, in the worst case, we are unlikely to do significantly better than the scenario depicted in Figure 28, but this does not preclude improvements for specific classes of problems to which the ETH does not apply.

## 6.5 When will D-Wave be able to address more challenging problems?

For the majority of D-Wave Two benchmark instances considered, there is a known exact classical algorithm that is able to find and verify the optimal solution in a reasonable amount of time. For our work this is true of Selby's exact algorithm, which typically took under an hour on even the hardest 512-node instances. Being able to find an optimal solution for an instance is critical for benchmarking, although there are alternatives such as planted solutions (Section 5.4.1). On the other hand, if we can find an optimal solution classically, what is the point of using a D-Wave system? From a different angle: if D-Wave systems continue to grow, at what point will we no longer be able to find an optimal solution efficiently with a classical algorithm?

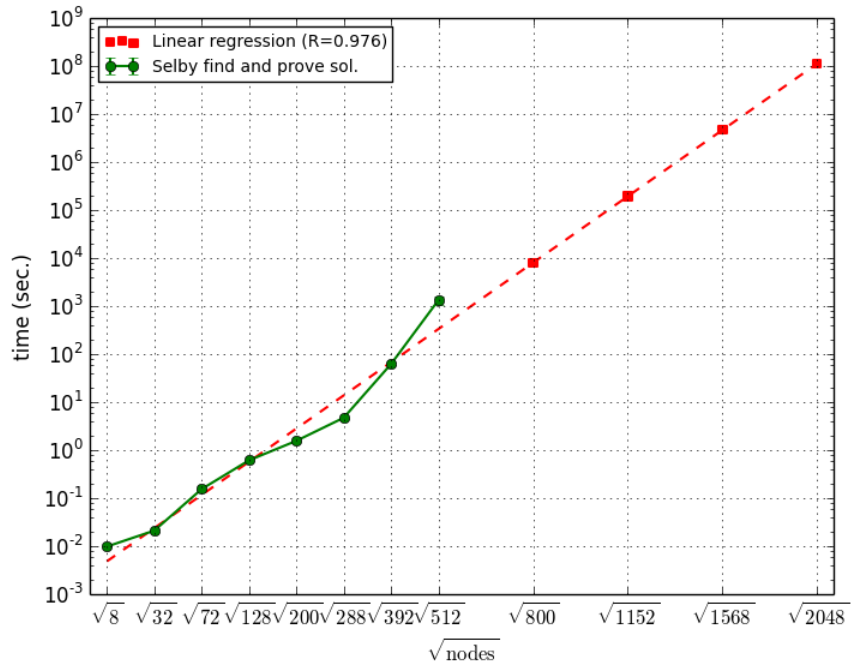

**Figure 29 Projected run time of Selby's exact Ising algorithm on random Ising instances.** Compare with **Figure 16**.

Figure 29 provides a projection of the run time of Selby's exact Ising algorithm on the baseline Chimera Ising instances from Figure 16 in Section 5.2. In our results, the algorithm took a mean time of ~23 minutes for 512-node Chimera Ising instances. At 1152 nodes, which is the approximate size of the next generation D-Wave system, we expect a run time of ~54 hours. We expect ~1321 hours at 1568 qubits, and a whopping ~31,873 hours or ~3.63 years for a 2048-node Chimera instance. From Figure 28, at the current rate of growth, we would expect a 2048-qubit D-Wave system in 2016. In a couple of years we could be in a situation where finding optimal solutions to our staple Chimera Ising benchmark instances is intractable using the current classical approaches. It may be that there are other exact classical approaches that perform significantly better, or that these instances are simply hard for all exact classical algorithms. It will be interesting to see how such a 2048-qubit D-Wave system fares against classical heuristics in addition to exact classical algorithms. We remind the reader that size alone does not dictate the complexity of an instance, and that there may be other problems with much smaller instances that are hard for exact classical algorithms; see Section 3.3.4 for a discussion of this issue.

### 6.6 Should we dismiss D-Wave as a competitive technology?

As far as we know, no one has yet discovered even a single hard problem or class of instances for which D-Wave is the fastest or otherwise best solution method. There are candidates, such as the affinity-independent-set and planted-solution instances mentioned in Section 6.3, but we feel that these have not yet undergone sufficient scrutiny to reach a firm conclusion.

In light of the above, a natural reaction to existing D-Wave benchmarking studies, including this one, would be simply dismissing D-Wave quantum annealers on the grounds that we have not discovered a clear success. We advocate quite the opposite. From a scientific point of view, this uncertainty is one of the more interesting facets of D-Wave benchmarking.

We may seek to know whether a current or future D-Wave system is the best means of solving even a single problem of vital interest, yet a conclusive answer, either way, is unlikely. Even if one were to find an example of such a problem, how would one demonstrate that no classical heuristic could best D-Wave on this problem? On the other hand, what if no one is able to discover such a problem? D-Wave's quantum annealers are a *technological trajectory* rather than a single device. Even if we are unable to find a "killer application" for the D-Wave Two, does this mean that one will not exist for the D-Wave Five? Do the same technological and deeply underlying physical limitations that have precipitated the apparent end of Dennard Scaling ultimately apply to D-Wave's and other quantum devices? This is the long view that has motivated our study. Conclusive answers to such questions are unlikely, and at best, we would expect an "arms race" between algorithm designers aiming to devise classical algorithms outperforming D-Wave and D-Wave engineers and advocates searching for "killer" applications and effective means of efficiently representing problem-domain instances on Chimera graphs. Such a race may lead to advances in classical algorithms, modeling and simulating quantum annealers, and working within the confines of restricted quantum architectures such as a Chimera graph. The lessons we learn may transfer to other quantum devices and architectures. Hence, we see great technical appeal and practical value in continued pursuit of classical-quantum benchmarking comparisons, preferably by an increasingly larger community of researchers.

### 6.7 Conclusion

To us D-Wave is a resource that, if nothing else, is a monumental feat of engineering. We took it upon ourselves to attempt to assess the current computational value of this resource. Thus, our primary contribution is a methodology for benchmarking D-Wave devices. Not all of the techniques and tools we develop apply to the current D-Wave Two system due to its capacity and precision limitations; however, we do not expect this to remain true if D-Wave systems continue along their recent exponential growth trajectory.

The general question we ask is whether systems such as D-Wave's can provide a computational resource advantage, of any kind, in an appropriate context. We do not subscribe to a particular answer to this question and do not believe this question will necessarily converge to a conclusive answer. Our goal is to collect data and provide evidence in order to address this question. We believe that with the current generation of D-Wave system, we cannot adequately do so; however, we have developed an infrastructure consisting of insights, algorithmic tools, and empirical evidence with which the research community may evaluate this question for future D-Wave devices and other adiabatic quantum computers that may be developed.

# Appendix A. Quantum architectures

## A.1 The adiabatic quantum optimization architecture

A natural question from Section 1.3 is, "How slowly does the interpolation have to be run for the output of the AQO algorithm to be correct with high probability?" The *adiabatic theorem* (or, more correctly, the *adiabatic approximation*) in quantum mechanics guarantees that as long as the total time for the interpolation, $T$, is chosen to be large enough, the final measurement will yield a value for $x$ that minimizes $f$ with high probability.

The "rule of thumb" that physicists have used for years for how long "long enough" needs to be [47] is that

$$T \gg \frac{|\langle E_1|\dot{H}|E_0\rangle|_{max}}{g_{min}^2}, \qquad (1)$$

where $|E_0\rangle$ and $|E_1\rangle$ denote the instantaneous ground and first-excited states of the system, whose corresponding energies are $E_0$ and $E_1$ respectively, $g$ denotes the difference in energy $|E_1 - E_0|$, also called the "(energy) gap," $\dot{H}$ denotes the derivative $\frac{d}{ds}H$, where $s := t/T$ is the interpolation parameter, and the maximization and minimization are taken over all $s \in [0, 1]$. In this and subsequent formulas, the units can be corrected by inserting powers of Planck's constant $\hbar$ where necessary; in other words, here and henceforth we will work with units in which $\hbar = 1$.

For most, and perhaps all, AQO problems, it is fairly straightforward to choose an interpolation path without too many wiggles so that the numerator of Eq. (1) does not become too large. The challenge then becomes finding a path such that the minimum gap in the denominator of Eq. (1) does not become too small. For example, if the minimum gap became exponentially small as $n$ increased, then the AQO problem would need to be run for an exponentially long time to ensure that it yields the answer with high probability.

In order to get a better handle on just how long the AQO problem must be run to succeed with high probability, tighter bounds than Eq. (1) have been developed in recent years. Indeed, it is the prospect of adiabatic quantum computing that has motivated mathematical physicists and computer scientists to sharpen this rather loose bound. One of the tightest bounds to date, although somewhat baroque, is the following implicit expression for $T$ [48]:

$$\||\psi(T)\rangle - |E_0(1)\rangle\| \leq \frac{1}{T}\left[\frac{\|\dot{H}\|}{g^2}(0) + \frac{\|\dot{H}\|}{g^2}(1) + \int_0^1 ds \left(\frac{7\|\dot{H}\|^2}{g^3} + \frac{\|\ddot{H}\|}{g^2}\right)\right].$$

## A.2 Alternative quantum circuit architectures

Two alternative quantum circuit architectures are worth mentioning. The first is the architecture variously described as the "cluster-state," "measurement-based," or "one-way" architecture, in which the (entangled) state preparations are problem-independent and there are no gate operations, but the measurement sequence depends (adaptively) on the problem. The second is the "programmable" quantum circuit architecture, in which the input is prepared to hold both the data and the program in a problem-dependent way, but the gate and measurement sequences are problem-independent. These variants are relevant for certain types of adiabatic quantum computing, described below.

## A.3  A universal adiabatic quantum computing architecture

While most of this report is focused on exploring how adiabatic quantum computers might fare when attempting to solve QUBO problems of practical interest, it is widely believed that, at sufficiently large problem sizes, even quantum computers, no matter what architecture they use, will be unable to solve QUBO problems efficiently. What could be a real game changer is a *universal adiabatic quantum computer*—a device that efficiently implements all quantum algorithms that have been or will be discovered.

The basic idea behind universal adiabatic quantum computing architectures proposed to date is to have them directly simulate a quantum circuit by taking a description of that circuit as input. Adapting an idea of Feynman [49], Kitaev [50] and then others [51], [52] proposed constructing a final Hamiltonian of an adiabatic evolution something like the following to simulate a quantum circuit defined by the gate sequence $U_T \ldots U_1$.

$$H = \sum_{t=1}^{T} U_t \sigma_t^+ \sigma_{t-1}^- + U_t^\dagger \sigma_t^- \sigma_{t-1}^+.$$

Numerous improvements to this Hamiltonian have occurred over the years and it is now known that it suffices to have a final Hamiltonian of the following form, with coefficients that can be both positive and negative, to effect universal adiabatic quantum computation [53]:

$$H = \sum_i h_i \sigma_i^Z + \sum_{ij} J_{ij} \sigma_i^Z \sigma_j^Z + \sum_{ij} K_{ij} \sigma_i^X \sigma_j^X.$$

No technology has demonstrated a sufficiently tunable Hamiltonian of this form, but proposals exist for several technologies, including superconductors and semiconductors.

# Appendix B. Complex-network instance families

## B.1 A catalog of commonly used graph parameters

To categorize and describe complex networks, researchers have proposed various measures. The list below extends the one from Section 4.1.

- Number of vertices ($n$) – Together with number of edges, describes graph size.
- Number of edges ($m$) – Together with number of vertices, describes graph size.
- Degree ($d$) – The number of edges incident to a vertex.
- Average degree ($m/n$) – Describes the average number of edges incident to a vertex.
- Degree distribution – The number of vertices for each distinct degree value.
- Clustering coefficient ($c$) – A measure (between 0 and 1) indicating the tendency for nodes to cluster. Specifically, it is the fraction of all node *triples* (three nodes connected by at least two edges) that are connected by three edges, forming a *triangle*.
- Eccentricity ($\varepsilon(v)$) – The distance to the farthest vertex from vertex $v$, where the distance between two vertices is the number of edges on a shortest path between them.
- Radius ($r$) – The minimum eccentricity on the full graph. The vertex with the minimum eccentricity is the *central vertex*.
- Diameter ($d$) – The maximum eccentricity on the full graph. The vertices with the maximum eccentricity are the *peripheral vertices*.
- Number of connected components – A set of vertices where each vertex is able to reach all others in the set by only traversing edges within the set. A *weakly connected component* is one where vertices can reach all other vertices when ignoring edge directions. A *strongly connected component* is one where all vertices can reach all other vertices by only following directed paths.
- Clique – A set of completely interconnected vertices in the graph.
- Assortativity – The Pearson correlation between all pairs of connected vertices' degree. A graph with -1 assortativity is *dissassortative* with low-degree vertices matched with high-degree and vice-versa. A graph with 1 assortativity is *assortative* with high-degree vertices paired with high-degree vertices and low-degree vertices with low-degree vertices.
- Modularity – Given both a graph and a partitioning, the fraction of edges within the partitions minus the expected fraction for a random partitioning. Partitions with high modularity tend to correlate with communities in complex networks.

## B.2 Real-world instances

### B.2.1 Twitter graphs

Our algorithm for growing Twitter graphs is as follows:
```
Function growTwitterGraph(Date startDate, Date endDate, TwitterData data) {
    g ← new graph
    // Create complete graph of all refers
    for each message in data {
        if message sent after startDate and before endDate {
            for each refer in message {
```

```
                        g.addDirectedEdge(message.from, refer)
                }
            }
        }
        // Remove all edges that only go one way
        for each edge in g {
            if g doesn't contain edge with opposite direction {
                g.remove(edge)
            }
        }
        // As there are now many disconnected nodes and separate components,
        // return only the largest weakly connected component
        return g.largestWeaklyConnectedComponent()
}
```

### B.2.2   Autonomous System graphs

The Internet is comprised of a set of separate networks owned by different companies. Internet communications require quickly routing data packets between machines that are possibly very distant from each other. When packets are traveling between machines within one company's network, that company's internal routing policies determine how packets are delivered. However, when the two communicating machines are on different companies' networks, the companies' *border routers* must find a route from the sender's network to the receiver's network. Each border router maintains a table of ways to reach different address families. However, since routers can go down, these tables require constant refreshing.

The Border Gateway Protocol (BGP) was developed as a way for border routers to request and advertise paths between border routers. The advertised paths represent legal peering agreements between the companies that they will route each other's data. BGP logs are available that include these advertise messages. Furthermore, these advertised paths are already bi-directional and can be naturally represented as a graph.

Our algorithm for creating BGP graphs is as follows:
```
Function growBgpGraph(Date startDate, Date endDate, BgpLogs data) {
      g ← new graph
      // Create complete graph of all refers
      for each advertise message in data {
            if message sent after startDate and before endDate {
                  for each edge in advertised path {
                        g.addEdge(edge)
                  }
            }
      }
      // As there could be separate components,
      // return only the largest weakly connected component
      return g.largestWeaklyConnectedComponent()
}
```

We ran this algorithm for various time periods of the BGP logs, from 1 minute to 1 day and with 12 different start dates. This resulted in 43 different graphs. Visual inspection of several of the graphs shows they appear similar having a dense central cluster of vertices with many strands extending off (see Figure 30).

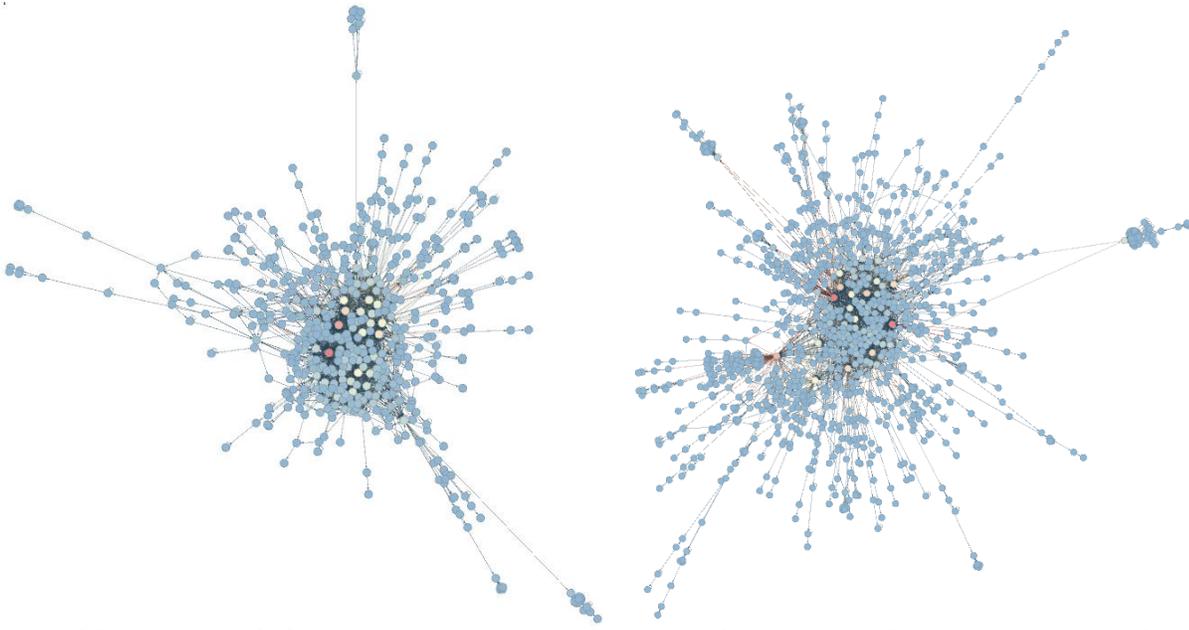

**Figure 30 Two sample BGP graphs.** The graph on the left has approximately 510 vertices, and the one on the right has approximately 1,100 vertices. Notice that these graphs both have a dense central cluster and chains of vertices extending off.

These graphs grow considerably in size (although less than the Twitter graphs from Section 4.2.1) and quickly converge to a fairly consistent average degree staying fairly consistent from 1,000 vertices to approximately 50,000 vertices.

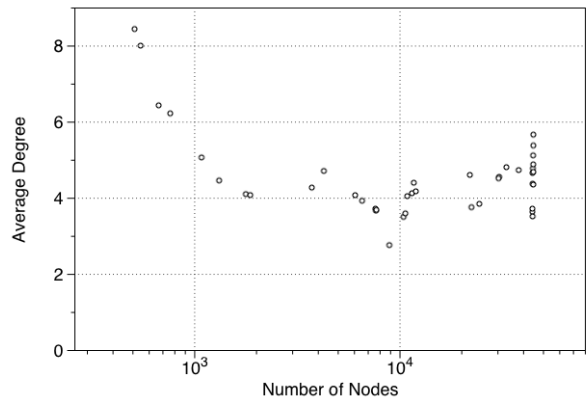

**Figure 31 Average degree for all BGP graphs.** Each point displays the number of nodes and the average degree for that graph.

The average diameter is surprisingly steady varying between 10 and 15 continually throughout the disparate sizes (Figure 31). We believe this consistency is due to the dense central region – few vertices get too far from each other because all quickly connect back to the central cluster.

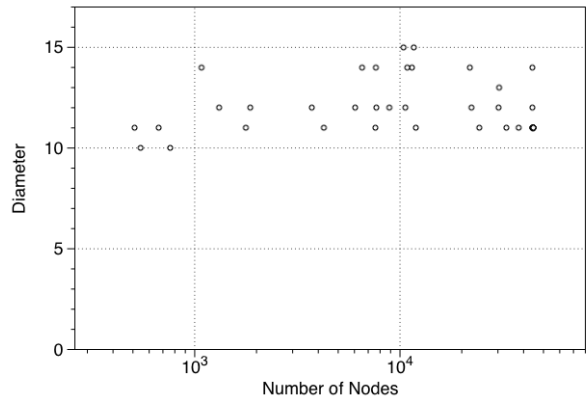
**Figure 32 Diameter for all BGP graphs.** Each point displays the number of nodes and diameter for that graph.

Of all metrics for BGP graphs, the clustering coefficient varies the most (Figure 33). It starts relatively high (~0.2), drops in the middle to a much smaller value (~0.05), and then climbs back up at the end. This variation does surprise us, but it also transitions fairly smoothly.

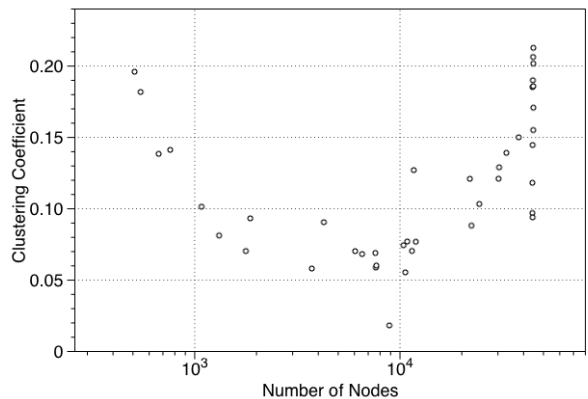
**Figure 33 Clustering coefficient for all BGP graphs.** Each point displays the number of nodes and clustering coefficient for that graph.

With the metrics for both our BGP family and Twitter family (see 4.2.1), we can now compare their results. Although both graph families have some metrics that vary as the graph size increases, the most notable thing is that in both families, all metrics vary relatively smoothly. Recall that each of these graph families was generated by sampling each original data stream from some start time for some period. Thus, the BGP graphs of approximately the same size were generated by sampling different time periods, and so the smooth variation for different time periods for these graphs indicates that the scores at different graph scales appear inherent to the graph family. Moreover, when the two graph families' scores are overlain, it is easy to separate the two families' data points. Qualitatively, these graph families appear internally similar at varying sizes. A quantitative comparison appears in Section B.4.

### B.3 Synthetic graph generation

#### B.3.1 Motivation

We surveyed well-known methods for generating synthetic graphs. Our initial motivation was to build complex network models from our Twitter and BGP data and to then use a synthetic graph generator to produce instances of arbitrary size based on each of these models. We tested some of the techniques below and generally found that real-world properties of interest were not present or did no scale well in the generated synthetic graphs. We ultimately decided to directly generate graphs from the data as described in Section 4.2 and Appendix B. However our foray into synthetic graph generation was useful in developing our algorithm for generating real-world-like Chimera-minor social networks (see Section 4.3.2).

#### B.3.2 Overview of synthetic complex-network generation techniques

***LFR***: Lancichinetti, Fortunato, and Radicchi proposed the LFR graph generation algorithm specifically designed for benchmarking community detection algorithms [54]. The LFR algorithm accepts five parameters: number of nodes, degree distribution exponent, average node degree, community size exponent, and a mixing parameter (between zero and one). Given these inputs, the algorithm assigns each node a degree ensuring the degree distribution and average degree are preserved. It then randomly assigns each node to a community, ensuring that the community is large enough to contain a node of this degree. Edges are randomly assigned within the community – preserving the assigned degree for all nodes. In a final step, random edges are removed from within communities and assigned outside the community until the mixing-parameter-defined fraction of edges occur between communities instead of within communities. While the LFR algorithm generates random graphs with some user-specifiable parameters – which can be set to match those observed in a real-world graph – how to select appropriate values for the non-natural mixing parameter is less apparent.

***MagFit/MagGen***: Kim and Leskovec proposed multiplicative attribute graphs (MAGs) that can represent external features like political affiliation, gender, etc. [55]. Specifically, probability matrices for each feature represent the likelihood that two nodes are connected by an edge. Thus, if one node is assigned a 0 and another a 1 for a feature, the probability they connect based on this feature is stored at entry $A(1,0)$ of the probability matrix, $A$. Nodes are assigned a 0 or 1 for a feature based on a specified probability ($p_i$). They proposed an algorithm, called MagFit, to generate these probability matrices and feature probabilities from existing graphs; a companion algorithm, called MagGen, creates a random graph that follows those probability distributions. Implementations of both are available as part of the SNAP codebase[19]. Given this model, an attribute contributes an average probability ($P_i$) of an edge forming between two "average" nodes as follows:

$$P_i = p_i{}^2 A(0,0) + p_i(1-p_i)A(0,1) + (1-p_i)p_i A(1,0) + (1-p_i)^2 A(1,1)$$

All attributes are combined by multiplication to form the overall probability that an edge will be created between two nodes:

$$\mathcal{P} = \prod_{\forall i} P_i$$

After experimenting with this method, we realized something intrinsic about the MAG model: the above average edge-forming probability ($\mathcal{P}$) in no way depends on the number of nodes. On average, this probability is applied at every opportunity, resulting in $\sim n^2$ expected edges, where $n$ is number of nodes. This means that the expected average degree (number of edges divided by $n$) scales with $n$ – an average node connects to a fixed percentage of the nodes in the graph.

---

[19] http://snap.stanford.edu/snap/download.html

In their study of graphs growing over time, Leskovec, Kleinberg, and Faloutsos found that for real-world citation and affiliation graphs (authors-to-papers and actors-to-movies) average degree does increase with time [56]. The amount of increase varied between $n^{1.08}$ to $n^{1.68}$ (average $n^{1.26}$). Even the most aggressively increasing degree fell well below MAG's $n^2$ rate.

***BTER***: Seshadhri, Kolda, and Pinar [57] proposed a method similar to LFR, but ground their Block Two-Level Erdös-Rényi model (BTER) on several real-world-graph theories. Much like LFR, BTER requires distribution parameters for both the degree distribution and the community size distribution. First, BTER randomly assigns each node to a community. Second, each community is self-connected as an Erdös-Rényi graph. Some number of connections for each node are reserved for cross-community edges, which are added next. These cross-community edges are added randomly. They demonstrate that BTER better matches average clustering coefficient by degree and the eigenvalues of the sample graph than a model proposed by Chung and Lu [58], [59].

Both LFR and BTER should better match the degree distribution as graph size increases than MAG. Unfortunately, LFR and BTER both require community size distribution to generate graphs – a feature that cannot be known for the input graph without having definitive answers for the community detection problem. For this reason, we were unable to use either of these synthetic graph generation methods to demonstrate our benchmarking methodology.

## B.4   Similarity analysis of graph families

In the interest of conducting proper science, we also introduce, as a control, a family of graphs that does not have many of properties of real-world graphs: Erdös-Rényi graphs. Erdös-Rényi graphs are perhaps the simplest model of a random graph and were introduced by Erdös and Rényi over fifty years ago [60], [61]. These graphs are generated by fixing *n* nodes and for each pair of nodes, randomly introducing an edge with a fixed probability *p*. An Erdös-Rényi graph is not necessarily connected; hence we retain the largest component from the graph. We generated Erdös-Rényi graphs by selecting values of *n* and *p* that would produce graphs with large connected components that approximately match the number of nodes and edges for each of our mid-sized Twitter graphs (100 to 5000 nodes).

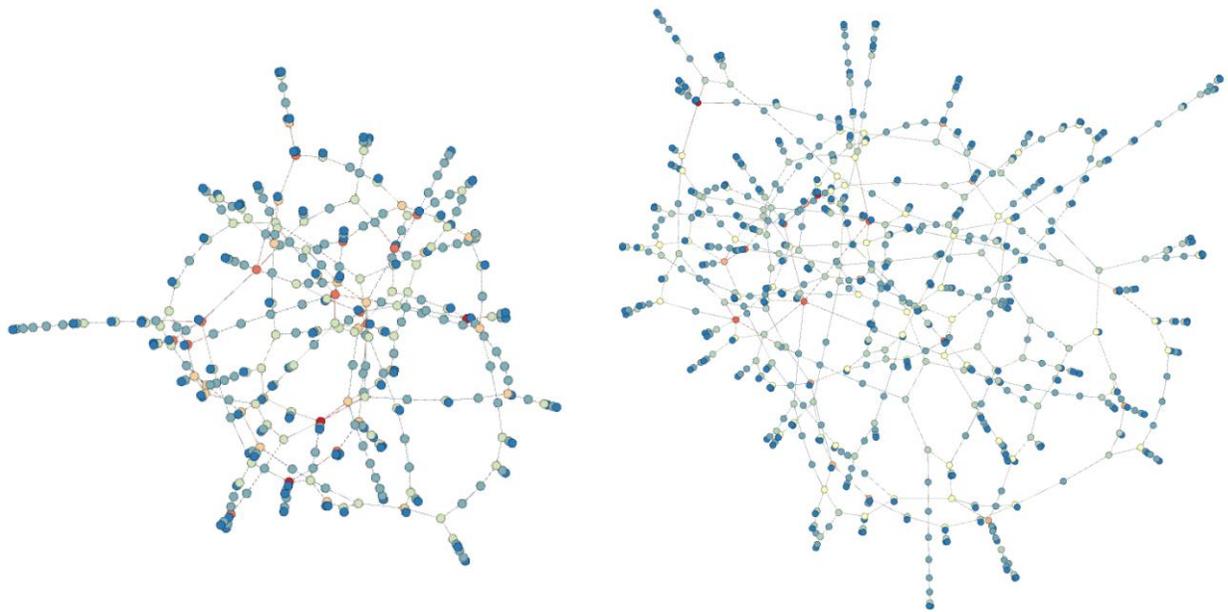

**Figure 34 Two sample randomly generated Erdös-Rényi graphs generated with approximately the same number of nodes and edges in their largest connected component as our Twitter graphs.** The graph on the left has approximately 350 nodes and the graph on the right has approximately 700 nodes.

Comparing Figure 34, Figure 8, and Figure 14, we see that Erdös-Rényi graphs differ qualitatively from our Twitter and Chimera-minor graphs. We present a quantitative comparison of average degree, clustering coefficient, and diameter below. The plots below include standard Chimera graphs as well, for reference.

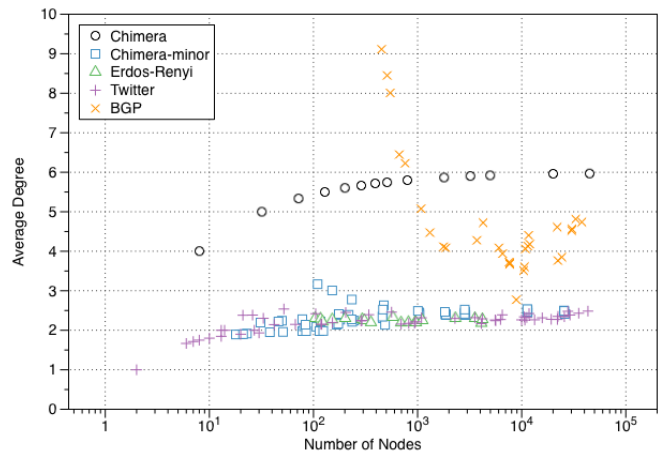

**Figure 35 Average degree for each of our three graph families and Erdös-Rényi graphs.**

Figure 35 shows the average degree for each of our three graph families and the Erdös-Rényi graphs. The Chimera and BGP graphs are internally consistent but are very different from all other graph families. The Erdös-Rényi and Chimera-minor graphs show a similar degree distribution to the Twitter graphs. It is well known that Erdös-Rényi graphs have heavy-tailed degree distributions, as real-world complex networks do, and the former have been used as synthetic models of real-world complex networks. However, Erdös-Rényi is not a completely satisfactory model, as seen below, and other models have been developed (see Appendix B.3).

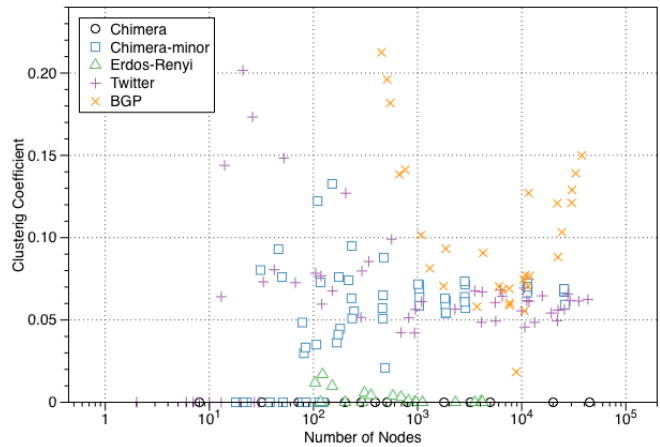

**Figure 36 Clustering coefficient for each of our four graph families and Erdös-Rényi graphs.**

Figure 36 shows the clustering coefficients for each of our four graph families and the Erdös-Rényi graphs. In this case, notice that the two real-world graph families (BGP and Twitter) both show some tendency to form triangles for all but their smallest instances (less than 100 nodes). Also, note that both of these families' scores fluctuate between 0.05 and 0.2 throughout their ranges. Erdös-Rényi graph families have almost no tendency to form triangles, and the standard Chimera graphs have no triangles. However, for all but the smallest graphs, the clustering coefficient of the Chimera-minor graphs correlates very well with the real-world families, especially the Twitter graphs. This is particularly striking as these graphs started from Chimera graphs and formed through a guided but random process.

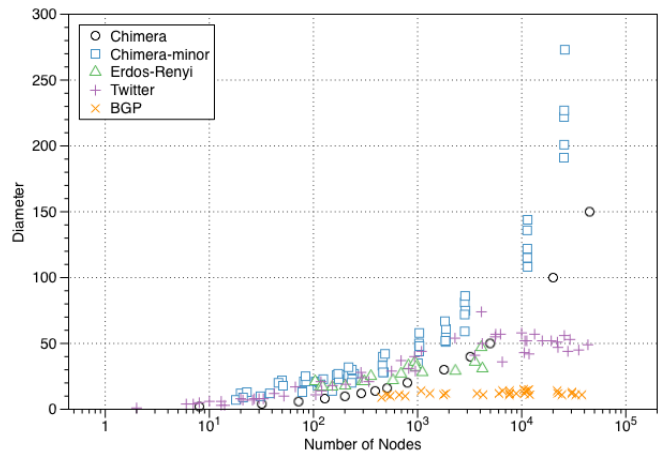

**Figure 37 Diameter for each of our four graph families and Erdös-Rényi graphs.**

Figure 37 shows the diameter for each of our four graph families and the Erdös-Rényi graphs. In this case, BGP is clearly the graph family with smallest diameter; one might expect this, as long-range connectivity is desirable in a communications network. The standard Chimera and Chimera-minor graph families are clearly the largest diameter graph family, with Twitter and Erdös-Rényi graphs lying somewhere in the middle. Reducing the diameter of our Chimera-minor graphs is a subject for future work.

Although our Chimera-minor graphs exhibit larger diameters than the real-world networks, we still feel they have sufficient real-world character to be much more useful for our benchmarking purposes than are random graphs. First, unlike Erdös-Rényi graphs, they are embeddable on the D-Wave's Chimera graph with significant node utilization. Secondly, unlike Erdös-Rényi graphs, they have the real-world tendency to form triangles. Triangle formation is considered a reasonable indicator of community formation and a feature of real-world social networks. We discuss community detection in the next appendix.

# Appendix C. Community detection in complex networks

## C.1  Introduction

Most types of real-world complex networks contain groups of related or well-connected entities, that one may call a community. For example, a community might represent users who tweet about similar topics in a Twitter network, while in biological networks communities might represent groups of proteins that interact for some specific function. The notion of a community in a complex network is often nebulously defined, and pinning down a precise mathematical objective that categorically captures the notion of a community is elusive. In fact Kleinberg showed that for three natural properties that one might expect of a good clustering, it is *impossible* to devise a single clustering function that possesses all of these properties [62]. While defining a community is difficult, we still seek to find naturally occurring communities in social networks, whatever they may be. *Community detection* is the problem of identifying interesting or relevant communities in a network. Given our growing ability to collect and process massive amounts of data, community detection has become quite an interdisciplinary phenomenon over the last decade, spanning sociology, physics, and computer science. We only touch on this rich and vibrant topic here and refer the reader to Easley and Kleinberg's book for an introduction to complex networks [63].

Kleinberg and Lawrence [64] used community detection to identify web pages that shared topics on the Internet. They used this analysis to organize Internet content and proposed that this organization could help identify "hubs" (index pages that identify many pages in a topic) and "authorities" (leader pages in topic areas). They suggest such pages could help improve Internet search engines.

Colbaugh and Glass leveraged community detection in their early warning system for predicting which large-scale movements will gain traction [65], [66]. Their work is based on a principle from social sciences: if one hears of an event at various locations in one's life (work, social functions, sporting events, etc.), then one is more likely to think it important than if one heard of it in only one location. They use blogs and news sites on the Internet as proxy for real-world discussions. They create a graph of the blogs and news sites where edges are based on links between the sites. The communities identified on this graph serve as proxy to the different real-world communities where one might hear of an event. They then search for early distribution of the topic across multiple communities. They found that if the spread occurs before the overall volume on the topic becomes too high, it serves as a good predictor of real-world action.

Ravasz et al. [67] use similar graph metrics to analyze metabolic pathways in cellular organisms. They demonstrate that metabolic pathways create sparse graph structures that are different from the small-world graphs built by social networks. They show that community detection can be used on the metabolic graph to separate largely disparate metabolic processes from each other.

We selected community detection as one of our benchmark problems because, in addition to the above discussion: (i) several classical heuristics exist, there is a general consensus on which heuristics perform best, and optimized implementations are freely available, and (ii) core community detection problems are readily mapped to an Ising problem (see Appendix E.1), suitable for benchmarking on a quantum annealer. With that said, we are currently unable to effectively map community detection problems onto a D-Wave device due to aforementioned Chimera-embedding and limited-precision issues (Section 4.3.1); however, we are able to implement quantum approaches to community detection using a quantum simulator, which we have developed. Moreover, the connection to Ising problems mentioned above makes community detection a prime candidate for benchmarking on future quantum devices.

We present our results benchmarking classical community detection algorithms in Appendix D, and our quantum simulation benchmarking results appear in Section 5.5. Before proceeding to the results, we cover community detection algorithms below.

## C.2 Classical community detection algorithms

***Clauset-Newmann-Moore (CNM)* [68]:** Clauset, Newmann, and Moore proposed an efficient algorithm for a greedy approximation to community detection. They begin by assigning each node its own community. Then, at each iteration they merge the two communities that will most improve modularity. Modularity is high when the number of edges within a community is significantly greater than would be expected in a random reconfiguration of the graph. We describe modularity in more detail below. This algorithm requires no user inputs. By intelligently choosing their data structures, this algorithm runs in $O(n^2)$ time, where $n$ is the number of nodes in the graph.

***Walktrap* [69]:** Pons and Latapy's algorithm begins from the observation that random walks on a graph starting from any node are more likely to stay within that node's community than to leave it. This is due to the fact that most edges should stay within the same community, and fewer leave it. They assign each node $u$ in the input graph a probability distribution over the nodes in the graph, indicating the likelihood of reaching each node in $t$ steps when starting at $u$. The assumption is that members of the same community should have similar probability distributions. They begin by assigning each node its own community. Then, during each iteration they merge the two communities that minimize the change in the probability distribution when merged. This algorithm requires the user to specify the number of steps to take in the random walk. Their algorithm runs in time $O(mn^2)$, where $m$ is the number of edges in the graph.

***Louvain* [70]:** Blondel, Guillame, Lambiotte, and Lefebvre's algorithm performs a variant of CNM's algorithm. They begin by assigning each node its own community. During the first step, they greedily identify nodes that can be moved to another community to increase modularity. This continues as long as modularity increases. When modularity can no longer increase, they create a new graph from the communities created in step one. They repeat these two steps until no communities are merged in the first step. This algorithm requires no user inputs. While they do not provide an algorithmic analysis, they demonstrate that Louvain functions faster than previous techniques and continues to run on massive datasets when previous algorithms fail to complete.

Even though the Walktrap algorithm merges based on a different metric than modularity, all of these algorithms compute their best result using modularity as a quality metric. One may think of modularity as a measure that assigns a score to a particular partitioning of a graph into communities. Modularity was designed so as to favor partitions that represent real-world notions of communities in complex networks [71]. Mathematically, the modularity, $Q$, of a partition into communities is defined as follows:

$$Q = \frac{1}{2m} \sum_v \sum_w \left[ A_{vw} - \frac{k_v k_w}{2m} \right] \delta(c_v, c_w),$$

where *m* is the number of edges, and for any two nodes *v* and *w*, the quantity $A_{vw}$ represents the adjacency between these nodes and is 1 if an edge exists between nodes *v* and *w,* and is 0 otherwise; $k_v$ is the degree of node *v*, and $\delta(c_v, c_w)$ is 1 if *v* and *w* are in the same community, else 0. At a high level, maximizing modularity tends to favor partitions that assign communities in such a way that edges between nodes within communities rather than nodes across communities. A bit more precisely, under an appropriate statistical graph model, one may think of maximizing modularity as maximizing the difference of the actual number of edges and the expected number of edges within each community. Thus modularity maximization seeks community structures that tend to have edges where they are expected (under the aforementioned statistical model). We refer the reader to Newman's seminal work for details [71].

However, maximizing modularity is not necessarily the best way to solve community detection. Fortunato and Barthélemy found that the smallest community that could be found via modularity maximization is bounded by the number of edges in the graph [72]. Specifically, the smallest community that can be found is bounded by

$$m_c \geq \sqrt{\frac{m}{2}},$$

where $m_c$ is the number of edges in a community, and *m* is the total number of edges in the graph. This means that, even when the graph is obviously partitioned into much smaller communities, modularity maximizing methods will merge the smaller "true" communities until the within-community internal edge count reaches this threshold.

*wCNM* **[73]:** Berry, Hendrickson, LaViolette, and Phillips recognize Fortunato and Barthélemy's limitation, but note that it only holds true for unweighted edges. The equation becomes more complicated when edge weights are included, and the resolution limit becomes based on the overall weight on the graph and the weight of the edges within a community. They propose an algorithm that weights edges based on their likelihood of being within-community edges: by adding weights to an edge based on the number of triangles and rectangles it is a member of, they increase the weight of edges with strong interconnections – thus likely within a community. This algorithm requires the user to specify the number of pre-weighting iterations to run. We decided not to include this algorithm in our benchmark study so we could restrict our attention to unweighted graphs.

Further modularity maximization weaknesses are all related to difficulties in finding a single best answer. Even in small graphs, a single vertex that bridges two clusters could be assigned to either neighboring community while still maximizing modularity. This vertex is randomly assigned to one community. When analyzing modularity maximizing community detection, Good et al. [74] found that "it typically admits an exponential number of distinct high-scoring solutions." Moreover, as the size of the graph and number of communities increase, the high-scoring solutions all approach the highest possible modularity score.

Finally, modularity maximization is tied to exclusive community detection – assigning each vertex to a single community. Many researchers point out that exclusive community detection does not faithfully match the real world: For example, a person can belong to a work community, church community, and sports team; exclusive community detection could only assign them to one.

Nonetheless, exclusive community detection is the most mature analysis at present, and modularity maximization implementations are best able to scale.

### C.3  Ising-based community detection

The classical community detection algorithms described above are heuristics that aim to cluster into communities by roughly maximizing modularity; however, as observed above, maximizing modularity is not a completely satisfying objective. In light of this, why consider classical community detection heuristics that seek to maximize modularity? We consider them, because for better or for worse, the above types of heuristics are those currently used in practice due to excellent runtime performance. This presents an interesting benchmarking challenge: how does one compare an arbitrary classical heuristic, which may not necessarily be framed in terms of optimizing a mathematical objective, to a Ising-based algorithm (equivalently, QUBO-based) whose sole purpose is attempting to optimize some given objective function.

For our benchmarking study, we address this question by employing a hybrid community detection algorithm that solves many Ising-based sub-problems in order to solve the overarching community detection problem. One could solve the sub-problems with any Ising solver, including quantum annealing. We defer the details of our approach until we discuss our quantum benchmarking results in Section 5. We cover our classical benchmarking results in the next section.

# Appendix D. Classical benchmarking studies

## D.1 Study 1: Purely classical benchmarking on large-scale families

Our goal is to perform a thorough benchmarking study of three common modularity-maximizing classical-computing community detection algorithms: CNM, Walktrap, and Louvain. To ensure a thorough test, we want to make sure we use datasets with real-world character, that the datasets scale as real-world datasets would, and we collect many different metrics for each test.

Many of the ingredients of our study have been previously discussed. In Section 4.2, we described how we generated two families of real-world graphs (Twitter and BGP). In Section C.2, we described the CNM, Walktrap, and Louvain algorithms. We now describe our testing methodology.

On an otherwise unloaded desktop workstation, we run each algorithm for each graph. We used a Macintosh PowerMac workstation with 2x2.4 GHz Quad-Core Intel Xeon processors[20], and 24 GB 1066 MHz DDR3 RAM. We use the Unix tools "time"[21] and "ps"[22] to record various metrics about the execution of the algorithm (system time, page swaps, memory usage, etc.). To account for both randomness in some of the algorithms, and for random machine usage issues (disk stalls, cache misses, etc.), we run each algorithm and graph instance pair ten times. Finally, as the slower algorithms take too long on the larger graphs, we interrupt the algorithms when their execution time exceeds 2 hours.

### D.1.1 Solution quality and modularity

Before analyzing how quickly the different algorithms complete their execution, we report on the quality of communities found. As described in Section C.2, although modularity has many issues, it is the metric used by each of these algorithms to determine the best community partitioning. Therefore, we report the modularity reported for the final partitioning of each graph/algorithm execution.

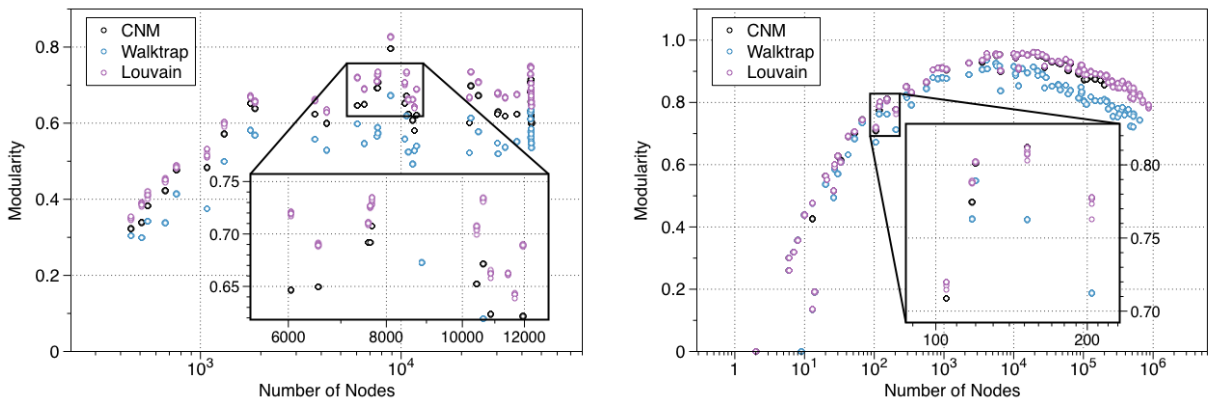

**Figure 38: Modularity results for BGP graphs (left) and Twitter graphs (right).** Insets magnify a portion of the results. Notice that Louvain and CNM perform consistently better than Walktrap. Louvain's modularity is generally equal to or greater than CNM.

---

[20] None of the implementations we used are multi-threaded codes, so we expect that (excepting some multi-threaded system calls) only one processor was used.
[21] Specifically, we used the Mac version of time called "gtime" available through the MacPorts tool.
[22] For an unknown reason, gtime failed to return max resident set size (memory usage) values for one of the algorithms. We wrote our own script that polled "ps –O rss" throughout program execution and stored the maximum value found.

Figure 38 shows the modularity results for all three algorithms on all graphs. First, notice that for any single graph there are different scores for different executions of Louvain. Given the same graph, all executions of Walktrap and CNM produced the same modularity score. This is to be expected as Louvain is a randomized method and Walktrap and CNM are not.

The key modularity result is directly evident: CNM and Louvain return communities with better modularity score than Walktrap. Louvain generally achieves similar or greater modularity than CNM.

For both BGP (left) and Twitter graphs (right), all three algorithms produce lower-modularity partitions for the smaller graphs than for the larger graphs. The scores become more stable (and larger) once the graphs reach several hundred or a thousand nodes. We believe this is a function of true underlying graph features: the smaller graphs have less community structure. We are surprised that the modularity scores fall, for all algorithms, as the Twitter graphs get larger than ten thousand nodes.

Recall that modularity can range from -1/2 to 1. Comparing the maximum BGP and maximum Twitter scores, it appears that Twitter has more clear community structure than BGP, as found by these algorithms and assuming that modularity does correlate with community structure in our graphs.

Finally, although there are sometimes marked differences between the modularity of the resulting communities found, all three algorithms' scores follow similar overall trends – increasing and decreasing together. Thus, they appear to agree on which graph instances demonstrate more clear community structure. The differences in modularity score between CNM and Louvain are so slight as to possibly be acceptable if modularity were the only consideration.

### D.1.2 Execution Time

After quality of the resulting communities, the next most common concern for any algorithm is execution time. The "time" command reports several metrics for execution time. We report the sum of system time and user time.

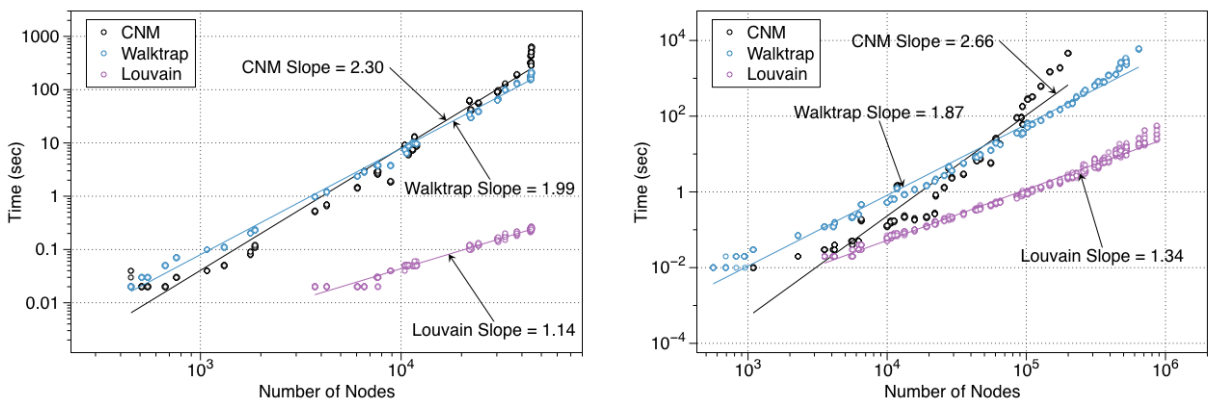

**Figure 39 Execution time results for BGP (left) and Twitter graphs (right).** Reported values are the slope of the best-fit line for each algorithm on a log-log plot.

Figure 39 shows the execution times and best-fit lines for each algorithm on a log-log plot. Since the time command can only measure speed in increments of 0.01 seconds, when any program completed in less than 0.01 seconds, time reports execution in 0.00 seconds. A log-log plot is unable to plot zero-valued points, so all three algorithms' runtimes on graphs below ~500 nodes are not shown (below ~3,000 nodes for Louvain). Moreover, as execution was halted when execution time exceeded 2 hours, there are no scores for CNM on graphs greater than ~20,000 nodes and for Walktrap on graphs greater than ~80,000 nodes.

The key execution time result is that Louvain is markedly faster than both Walktrap and CNM. On smaller graphs, CNM is faster than Walktrap, but CNM scales poorly relative to Walktrap; as graph size increases, CNM becomes slower than Walktrap.

The slope of the line on a log-log plot provides the exponent ($\alpha$) that best fits the following function on a linear plot:

$$\text{Time} = (\text{Number of Nodes})^\alpha$$

Thus, not only does Louvain run faster on all of the instances of both families we tested, the slopes indicate that as the size of the graph increases Louvain will always run faster than CNM and Walktrap, and their relative difference is increasing.

Finally, the quality of the fitted line can help indicate how well the line can extrapolate to future execution speeds. For both Walktrap and Louvain the $R^2$ score of the fit on both graphs (Walktrap $R^2$ = 0.99 (BGP) and 0.98 (Twitter); Louvain $R^2$ = 0.98 (BGP) and 0.99 (Twitter)) indicate a very high quality fit. The corresponding $R^2$ scores for CNM (0.98 (BGP) and 0.92 (Twitter)) are still good, but not as strong. This is visually evident on the Twitter plot where CNM performs better than the line predicts in the middle, but worse on either end.

### D.1.3  Memory Analysis

Memory considerations directly influence computation speeds on classical computers. Classical computers contain a memory hierarchy with small and very fast memory stores (registers) within the central processor, through many different layers, to the truly massive but relatively extremely slow hard drive. Thus, when writing optimized code for classical computers, the programmer must consider many factors related to memory – amount used, how coherent usage patterns are with memory layout, etc. Unfortunately, we were only able to measure maximum memory usage for the algorithms. Fortunately, memory usage smaller than the amount available generally correlates with improved memory coherency.

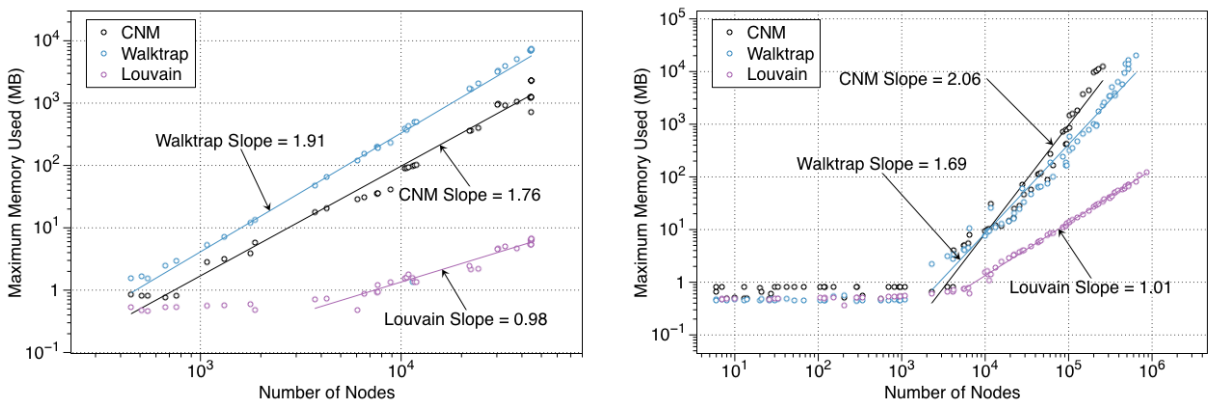

**Figure 40 Maximum memory usage results for BGP (left) and Twitter graphs (right).** Reported values are the slope of the best-fit line for each algorithm.

Figure 40 shows the maximum memory usage for each run of each algorithm on a log-log plot. Loading and running a program requires a minimum amount of memory, and then loading relatively small graphs into memory can seem negligible next to the memory required for the program's instructions. Thus these plots show flat periods of memory usage on the left side of the Twitter plot for all algorithms and for Louvain on the BGP plot. Note that on the Twitter plot, Walktrap and CNM both start increasing from their baselines around 1,000 nodes, which is the approximate size of the smallest graphs in BGP group. Louvain is fairly consistent with the change occurring around 5,000 nodes in both.

The key result from this work is that Louvain uses considerably less memory than both Walktrap and CNM on both graph families. Not only that, but Louvain's slope is again lower than both CNM's and Walktrap's indicating that CNM and Walktrap will use comparatively more memory than Louvain as the graph size increases.

The quality of fit of the lines in these graphs is lower than in the execution graphs. In this analysis, CNM and Louvain have the best $R^2$ fits (CNM $R^2 = 0.98$ (BGP), 0.95 (Twitter); Louvain $R^2 = 0.95$ (BGP) and >0.99 (Twitter)), with Walktrap a little poorer (0.91 (BGP) and 0.97 (Twitter)).

## D.2 Study 2: Classical benchmarking on D-Wave-compatible families

In this study, our goal is to perform a study that permits comparing adiabatic quantum computing capabilities against classical-computing community detection algorithms. Currently the only available adiabatic computing devices are produced by D-Wave and are limited in the size and type of graphs that can be encoded on them. Specifically, the D-Wave Two machine we have access to has 509 functional qubits.

Most of the ingredients used in this analysis have been previously discussed. In Section 2, we described the D-Wave Chimera graph, and in Section 4.3.2 we described our algorithm for generating synthetic real-world-like graphs as minors of the Chimera graph. In Section C.2, we described the classical-computing algorithms we use herein, and in Section 5.5, we describe our approach to solving comparable problems on a quantum annealer. In Section 3, we discussed difficulties with measuring performance on quantum devices in general and D-Wave machines in particular.

### D.2.1 Classical-computing results

For this analysis, we used the same workstation and measuring technique as described in Section D.1. Herein, we will only present and discuss results.

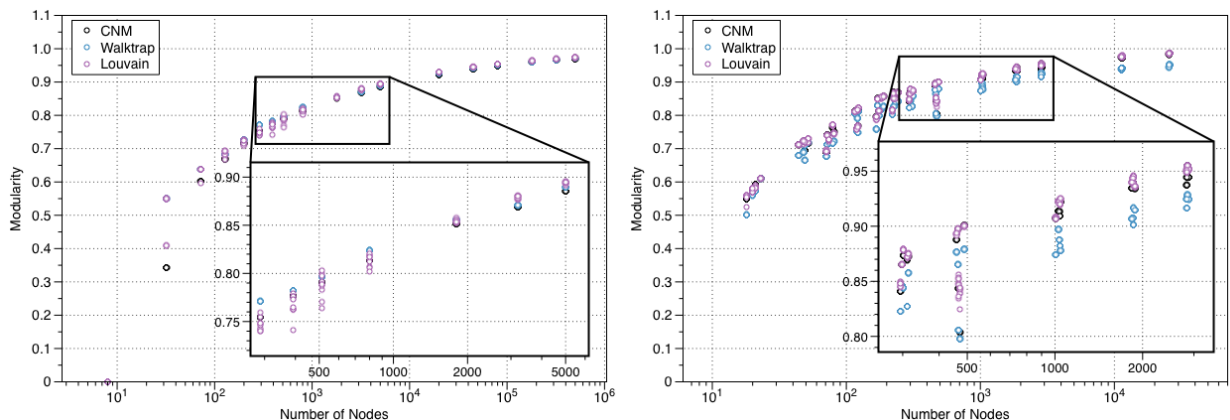

**Figure 41 Modularity scores for standard Chimera (left) and real-world-like Chimera-minor graphs (right).** Insets show a magnified portion of the results.

**Solution quality and modularity**: Figure 41 shows the modularity results for all graphs from both the standard Chimera and real-world-like Chimera-minor graph families. The former graphs are highly structured and do not have social-network properties (Figure 1 and Figure 13) and are included only as a point of reference. Recall that although our Chimera-minor graphs may be embedded within the Chimera graph, they exhibit real-world properties (Section B.4). For the purposes of our classical benchmarks, we need not worry about Chimera embedding issues and simply execute our algorithms on our minor graphs, as depicted in Figure 14.

The principal result on these Chimera-based graphs is less clear than on the real-world graphs. In the previous results, CNM and Louvain consistently found higher modularity results than Walktrap. For the standard Chimera graph, the three algorithms found largely the same modularity results – with Walktrap sometimes finding the best result. In our real-world-like Chimera-minor graphs, it appears that Walktrap finds generally lower-score results, but with considerable overlap.

These graphs show a similar trend as the previous set: all algorithms follow a similar upward trend in finding higher modularity results in larger graphs than in smaller graphs.

Finally, it is interesting that although the standard Chimera graphs have no triangles, they still appear to have clear community structure as measured by modularity.

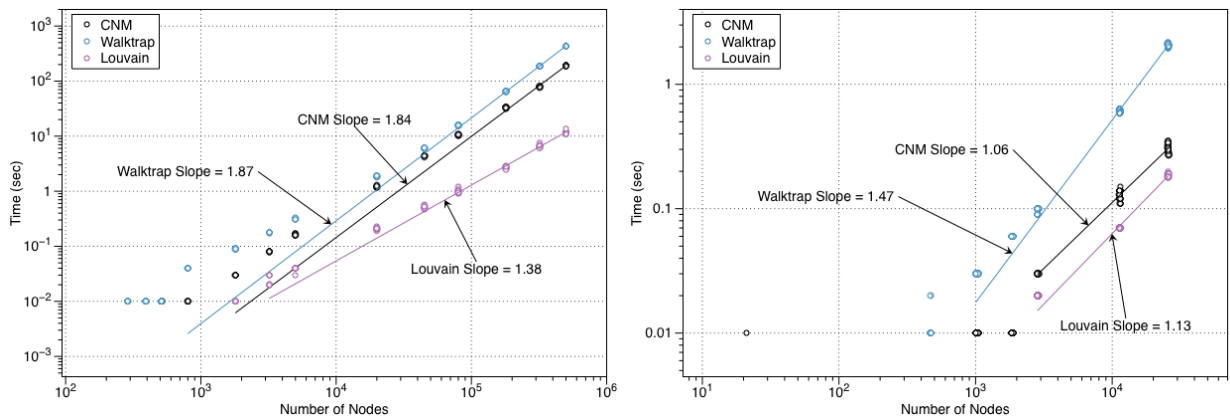

**Figure 42 Execution time results for Chimera (left) and real-world-like Chimera-minor graphs (right).** Note the log-log axes. Best-fit slopes indicate how the algorithm will likely continue to scale on larger graphs.

**Execution time**: Figure 42 shows the timing results for all graphs from both the Chimera and real-world-similar Chimera-minor graph families. As with the previous timing results, if program execution took less than 0.01 seconds, the tests reported a zero running time – meaning that some smaller graphs have no data points on this log-log plot.

The principal result is the same as with the real-world graphs: Louvain runs markedly faster than both Walktrap and CNM. In the case of the real-world-like Chimera-minor graphs, CNM appears markedly better than Walktrap on execution speed for the first time.

Analyzing the slopes, for standard Chimera graphs, we get the same result as in previous tests: not only does Louvain run considerably faster, it scales better and will likely continue to perform much better. However, with the real-world-like graphs, CNM exhibits a slightly better slope than Louvain – opening the possibility that Louvain may run slower than CNM on larger graphs from this family. Unfortunately, our code for sampling these minor graphs does not scale well enough to generate larger graphs in a reasonable time frame. Also, as the slopes for these data are based on so few distinct data points, this trend may not continue.

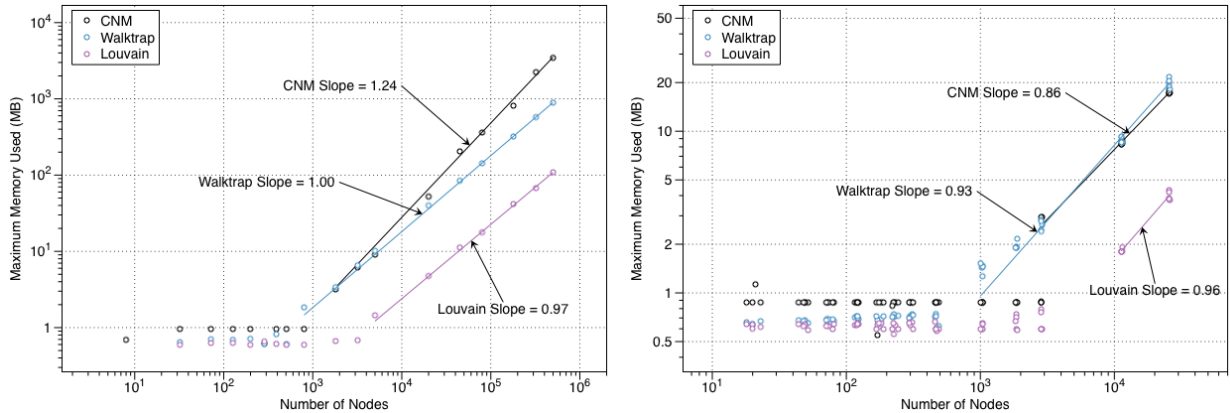

**Figure 43 Maximum memory usage for standard Chimera (left) and real-world-like Chimera minor graphs (right).** Note the log-log axes. Best-fit slopes indicate how the algorithms will likely continue to scale on larger graphs.

**Memory analysis**: Figure 43 shows the maximum memory usage results for all graphs from both the Chimera and real-world-similar Chimera-minor graph families. As with the real-world graphs, Louvain again uses considerably less memory than both Walktrap and CNM. Louvain's slope is also about the same as it was in the previous cases – just under 1.0. However, with these graphs, Walktrap and CNM's memory usage slopes are much lower than on the real-world graphs.

**Conclusion:** Louvain emerges as a clear winner across most of our metrics. If this were the case, then why would someone opt to use anything else? For one, all of these algorithms are heuristics and they come with no guarantee of near-optimality. Although we know that maximum possible modularity is 1, for the instances we used, we do not know the precise maximum modularity. In fact, in Figure 38, we see that modularity drops as the graph size increases across all of the heuristics. This might indicate the larger graphs simply do not have community structure with large modularity, or it may be that our collection of heuristics is unable to find such communities.

Another dimension is that, as noted in Section C.1, a precise definition of a good community is elusive. It could be that a Louvain-user is satisfied with the communities it produces but she would be even happier with some other set of communities of which she is unaware. In many cases such a choice is moot, since Louvain is the only common community detection heuristic that is viable for massive graphs.

# Appendix E. Ising formulations of combinatorial optimization problems

## E.1 Modularity-maximizing bipartition

Here we present a well-known Ising formulation for finding a bipartition of a graph (i.e., a partition into two parts or communities) that maximizes modularity. Refer to Section C.2 for an introduction to modularity; recall that the modularity, $Q$, of a partition of a graph into communities is defined as:

$$Q = \frac{1}{2m} \sum_v \sum_w \left[ A_{vw} - \frac{k_v k_w}{2m} \right] \delta(c_v, c_w),$$

where $m$ is the number of edges, and for any two nodes $v$ and $w$, the quantity $A_{vw}$ represents the adjacency between these nodes and is 1 if an edge exists between nodes $v$ and $w$, and $A_{vw} = 0$ otherwise; $k_v$ is the degree of node $v$, and $\delta(c_v, c_w)$ is 1 if $v$ and $w$ are in the same community, else 0. If we are evaluating the modularity of a bipartition, as opposed to a partition into greater than two parts, we may simplify the expression for modularity. For convenience we refer to the two communities as the -1 and the +1 communities. The definition of modularity now becomes:

$$Q = \frac{1}{2m} \sum_v \sum_w \left[ A_{vw} - \frac{k_v k_w}{2m} \right] \frac{s_v s_w + 1}{2},$$

where, for each node $v$, $s_v = 1$ if $v$ belongs to the +1 community, and $s_v = -1$ if $v$ belongs to the -1 community. Thus, since we only have two communities, we have $\delta(c_v, c_w) = (s_v s_w + 1)/2$. We observe that finding a bipartition that maximizes modularity is equivalent to finding an assignment of either -1 or +1 to each node-variable $s_v$ so as to maximize the value of $Q$. This can be cast as an optimization problem:

$$\max_{s_v \in \{-1,+1\}} \frac{1}{2m} \sum_v \sum_w \left[ A_{vw} - \frac{k_v k_w}{2m} \right] \frac{s_v s_w + 1}{2},$$

which is equivalent an Ising problem, by ignoring the portion of the objective that is constant with respect to the $s_v$ variables, and instead of maximizing the function above, minimizing its negation. We note that one undesirable feature of this formulation is that each term $s_v s_w$ has a nonzero coefficient, $\frac{1}{4m} \left[ A_{vw} - \frac{k_v k_w}{2m} \right]$, even if an edge does not exist between nodes $v$ and $w$ (i.e., $A_{vw} = 0$). In terms of a quantum-annealing implementation of this formulation, one would require couplers with nonzero values between every pair of nodes. This would make a D-Wave Chimera implementation impossible without an embedding or mapping. This is precisely the reason we chose to implement our Ising-based community detection algorithm only on our quantum simulator.

## E.2 Independent set problems

Recall from Section 5.3 that an independent set in a graph is a set of nodes such that no edges exist between any two nodes within the set. The maximum independent set (MIS) problem seeks to find an independent set of maximum size. We are given an Ising formulation of this problem, in the form:

$$\min_{s_i \in \{-1,+1\}} \sum_{ij \in E} J_{ij} s_i s_j + \sum_{i \in V} h_i s_i.$$

The summation is over the edges of the qubit-connectivity graph, which represent the couplers of the underlying quantum annealer. For the case of a D-Wave system, we have the Chimera graph. We observe that by setting $J_{ij} = 1$ for each edge $ij \in E$ in the graph, and setting $h_i = (\sum_{j:ij \in E} J_{ij}) - 2$, we, in effect, model the MIS problem.

When we set some of the $J_{ij}$ variables to be 0, we are in effect solving the MIS problem on a subgraph of the ambient graph. When we let $J_{ij} \in \{-1,1\}$ we obtain the affinity independent set problem. Consult the seminal work of Barahona [28] or the recent work of Choi [75] for more details on the relationship between independent set problems and the above setting of instance parameters.

# Appendix F. Quantum Monte Carlo simulator

The exponential cost of directly simulating an adiabatic quantum computer has necessitated some tradeoffs in the effort to understand how such a computer, if available, would perform on a given problem. This is not to say that simulations cannot play a role in understanding these machines, only that a judicious choice of strategy is necessary. With this in mind we have developed a simulation suite that looks to answer two questions. Firstly, how would a purely classical annealer perform on a given algorithm, and secondly, what is the structure of the quantum problem at any given instant during the annealing? With these two pieces of information, we can hope to answer whether there is an advantage to using an adiabatic quantum computer on a given problem; this would also allow us to compare different implementations of the same problem.

The simulation suite developed for this work is optimized for flexibility with the goal of allowing discovery-based simulations. Specifically the interactions between sites are allowed to take on arbitrary strengths and couple arbitrary pairs of spins. This flexibility allows for a much more general exploration of hardware in designing and evaluating algorithms than would be available if the simulation suite would have been confined to the Chimera graph. The tradeoff is that this precludes many of the optimizations that would be possible with a less flexible framework. As such, we emphasize that this simulation suite is *not* suitable for directly answering the question commonly posed in the literature about whether a highly tuned classical algorithm would be faster than a given candidate adiabatic quantum computer; it can only address scaling behavior of a given problem to help determine whether it may be advantageous to solve the problem on an adiabatic quantum computer.

More concretely, the tools we have developed allow simulation of a Hamiltonian (set of interactions) of the following form:

$$H = \sum_{ij} T_{ij} \sigma_i^z \sigma_j^z + F(t) \sum_i \sigma_i^x + \sum_i L_i \sigma_i^z$$

The indices $i$ and $j$ run over the sites, the $T_{ij}$ are the coupling strengths between the spins, $F(t)$ is the (potentially time dependent) transverse external field, and $L_i$ are local magnetic fields that may be tuned individually for each spin.

An adiabatic quantum annealing calculation proceeds as follows. The device is first cooled to a low temperature and then an external field ($F(t)$, above) is slowly tuned from an initially large value to a small one. If the external field is initially much larger than all of the other interactions above, then all spins will align with the field. Furthermore, if the field is reduced slowly enough, the system will stay in the lowest energy state during the entire evolution, thereby finding the minimum energy configuration for the problem Hamiltonian, *i.e.*, for the set of interactions of interest to the programmer.

There are several pitfalls that may occur during such an evolution that we would like to study with this simulator. These may be more easily understood thinking of an analogous problem of a ball placed on a rough landscape that may be tilted to one side. The location of the ball on the landscape is a proxy for the orientations of the spins in the quantum computer. The heights of the bumps and wiggles in the landscape encode the interactions and local fields on the spins and the tilt is the external magnetic field. The goal is to get the ball to the lowest point on the landscape. Analogous to the operation of the quantum computer, the system is initialized by tipping the landscape so far to one side that the ball is forced to that side of the landscape. Then the tilt is slowly reduced. If the system is purely classical, than the ball will follow a path set by the nearby dips and valleys as the tilt is reduced and in the end will end up at the lowest point on its path. The goal is that this lowest point is the lowest point in the entire landscape.

There are two fairly simple ways in which this method can fail to get the ball to the lowest point on the map. The first of these is that the landscape is not amenable to the problem. As a concrete example, imagine that the landscape is a topographic map of the United States and tipping the map all the way towards the east initializes the experiment. As the map is slowly brought back to level, the ball will never be able to make it over the continental divide to death valley on the western side of the map. This scenario where the basins of attraction for different local minima in the landscape are separated by a barrier is the most problematic for the operation of the quantum computer. The second way in which the method can fail is if the tilt in the landscape is reduced too quickly. Then, the momentum of the ball can carry it out of the local dips and valleys in the map. In addition if there is not enough friction to slow the ball down, it will never settle into the lowest point on the map.

In order to test for these conditions, we developed a purely classical capability to simulate the dynamics of the spins. In this case, after programming in the interactions, the spins are initialized as directions in a three dimensional space. At each point in time, the force on each spin is calculated due to these interactions and then all of the spin directions are updated, synchronously.

This classical technique can be used in two different ways. First, the dynamics of the spins can be calculated ignoring the momentum of the spins as they rotate. By always moving in the direction of the steepest change in the energy landscape, we can find out if an arbitrarily slow classical evolution of the computer could reach the optimal solution. If this works, the implication is that no quantum entanglement is necessary for a D-Wave type device to find the optimal solution. The second technique is to use the momentum on the spins to construct a real time dynamics. In this case, it is possible to get information about how slowly the external transverse field needs to be decreased in order to avoid leaving the local environment of the lowest energy state of the system, again assuming a purely classical evolution of the spins.

These methods provide insight about the evolution of the spins in the absence of noise (or, equivalently, at zero temperature). In order to assess the role of the thermal effects, we also implemented a Metropolis Monte Carlo sampling scheme. This method works by proposing random perturbations of the spins and then accepting these perturbations with the probability $\max[e^{-\frac{E'-E}{T}}, 1]$ where $E'$ is the energy of the perturbed system, $E$ is the energy before the perturbation and $T$ is the temperature. In doing this, the simulation ceases to be an approximation of the evolution of the system with time, but as the perturbations may be quite general, it is much more probable that small energy barriers will be crossed than in the classical methods. This method has also been shown to correlate with the results of simulations on the D-Wave machine by Smolin and Smith [76].

The more interesting case is to perform quantum calculations to see whether the presence of entanglement eases the task of finding the global minimum of the energy landscape. We analyzed several quantum Monte Carlo methods to perform these calculations. We concentrated on zero-temperature projector methods that evolve a trial wave function in imaginary time in order to select out the ground state of the system. These all rely on mapping the time-dependent Schrödinger equation into imaginary time:

$$i\hbar \frac{\partial}{\partial t}\Psi(t, s_1 \ldots s_N) = H\Psi(t, s_1 \ldots s_N) \rightarrow \hbar \frac{\partial}{\partial \tau}\Psi(\tau, s_1 \ldots s_N) = H\Psi(\tau, s_1 \ldots s_N).$$

Now the solution to this partial differential equation can be written as

$$e^{-\tau H}\Psi(\tau, s_1 \ldots s_N)$$

It is apparent in this form that, as imaginary time increases, eigenstates of the Hamiltonian will acquire an exponential damping according to their energy. If the wave function is kept normalized during the evolution this will effectively select out the lowest energy (ground) state.

The difficulty comes in applying the propagator $e^{-\tau H}$, which is a $2^N$ x $2^N$ matrix. For this reason it is applied stochastically. There are several options for how this stochastic projection is implemented. One can either sample terms in the Hamiltonian, as in stochastic series expansion, or sample directly the wave function. For the case of the D-Wave style Hamiltonian, the lack of symmetry imposed both by the configurable interactions between the qubits and the local fields make cluster updates much more complicated to apply. Therefore, we have opted to sample the wave function directly. The wave function is represented in the $\sigma^z$ basis rendering all of the operations implementable as binary arithmetic and random number generation.

Specifically, we make a Taylor series expansion of the propagator above:

$$e^{-\tau H} \rightarrow 1 - \tau H$$

Clearly this is valid only in the limit where $\tau$ becomes small. The advantage of this formulation is that the application of each term in the Hamiltonian operator to a state in the $\sigma^z$ basis is to either keep the state the same, due to the local fields, or to flip a single spin, due to the transverse field. Exploiting this, the simulation now works in two parts. First, a random term of the Hamiltonian is selected to apply to each state in the population according to its strength. If the term is not due to the transverse field, the state is unchanged. If it is, then the appropriate spin is flipped. The second part of this operation is to add a weight to each state due to the non-spin flip terms. This has the effect of making configurations that have a lower energy more important and those with a higher energy less important.

In order to use this machinery to find the lowest energy solution for a given set of interactions with the transverse field set to zero (the final state of the quantum annealer), we initially start with a set of configurations with spins randomly oriented and with some relatively large transverse magnetic field. The transverse field is necessary because without it, the algorithm would never flip any spins and would only select the lowest energy of the random states that we had started with. As the calculation progresses, the transverse field is gradually decreased until its effect is negligible, in analogy with the D-Wave machine.

An important consideration in developing this, or any other stochastic algorithm, is the development of test cases with known solutions to which the machinery can be applied. Our algorithm was tested in two independent ways. The first of these was by comparison to an exact evolution in the full state space that can be accessed by such a machine. Specifically, we store the entire $2^N$ dimensional state vector describing every possible orientation of the spins, including the phases between them. Obviously this limits the size of the system that can be studied (30 qubits requires storing $2^{30} \approx 1,000,000,000$ numbers), but for arbitrary interactions between the qubits the system can be solved exactly. We have tested our implementation for 500 different configurations of interactions between 5 and 25 qubits and have obtained the optimal solution in each case.

Testing this Monte Carlo method on larger systems posed a challenge because of the difficulty of obtaining difficult problems and exact solutions against which the simulator could be tested. However, a recent talk by Itay Hen at the Third Workshop in Adiabatic Quantum Computing, 2014 [34] provided an interesting possibility. His collaborators and he suggested that one could plant a solution of up and down orientations of the qubits and they build a set of interactions around them that would be both difficult to solve and would be unlikely to be degenerate. His method was to take the planned solution and then perform a random walk between the spins until a loop was formed. Once this loop was formed, interactions would be chosen to decrease the energy of the system due to this spin configuration except that one of the interactions would randomly have the opposite sign to introduce a frustration that would make the problem harder. This procedure is repeated many times, building an increasingly complex Hamiltonian until the planted solution is likely to be unique. We discuss our D-Wave benchmarking results on these instances in Section 5.4.

We also performed some quantum simulator tests using these planted solutions on the Chimera graph for problems with square arrangements of plaquettes between 2x2 and 10x10 (32 to 800 qubits; see Figure 1). For graphs that are 4x4 or smaller, we were almost always able to find the optimal solution on the first attempt. As the problem became larger, however, errors began to occur. The nature of these errors was interesting in that they were almost always due to having entire plaquettes of spins flipped relative to the exact solution. We theorized that this was due to the connectivity of the Chimera graph being much greater within plaquettes than between them. In order to test this, we controlled the size of the loops out of which the interactions were generated. Some instances were generated with all loops connecting only four spins. In this case, the interactions were much more likely to remain entirely on a single plaquette and coupling between the plaquettes was therefore rather weak. The instances of errors in our solutions was much higher for this class of problems than the general case and the errors were almost entirely due to flips of entire plaquettes. Conversely, we generated problems where all of the loops connected more than 8 qubits. By definition all of these spanned multiple plaquettes and as expected, our algorithm performed much more successfully than in the case of small loops. Additionally, the errors were fewer and generally due to single spins that were flipped rather than entire plaquettes.

After establishing the accuracy of the quantum simulator for small systems and understanding its failure modes for larger ones, we turned to the problem of benchmarking our algorithm against the D-Wave machine on random instances generated on the Chimera graph. The results of these are detailed in Section 5. Finally, we have used the generality of our simulator to study interactions on topologies more complicated and dense than would be allowed by the D-Wave machine. This has allowed work on community detection algorithms that is also detailed in Section 5.

# Appendix G. Entanglement in the D-Wave quantum annealer

Local quantum entanglement has recently been demonstrated in a D-Wave quantum annealer [77]. Concurrently, there is an ongoing debate [76], [78]–[81] as to whether or not a purely classical model is sufficient to describe the output distribution of the annealer. Certainly, there is some degree of accuracy beyond which any microscopic time-resolved model of the D-Wave device must include quantum effects. However, identifiable quantum effects do not automatically guarantee quantum computational advantages. The semiconductor transistors in modern classical computing devices are small enough to necessitate a quantum description in accurately simulating their behavior and yet they remain purely classical computing devices. There is a clear intention in the design of the D-Wave quantum annealer to leverage quantum effects to enable large but restrictive quantum computations, but these restrictions cast significant doubts on the theoretical and practical advantages of such a device. Even in an ideal, universal quantum computer, the relationship between entanglement and quantum speedup is unclear and under active debate [82].

## G.1 Entanglement in the D-Wave device

Quantum entanglement is a very broad and general concept [83]. It refers to partitioning a physical system into two or more disjoint pieces and identifying correlations between the pieces that are forbidden in classical physics. Usually this is reduced down to an entanglement measure that assigns a numerical value to the "amount of entanglement" in a quantum state that is zero when quantum effects are not necessary to describe the correlations. The most popular entanglement measure in modern quantum information theory is the entanglement entropy of a pure state $|\Psi\rangle$ relative to a subsystem $A$. When examining the state only on this subsystem, the pure state $|\Psi\rangle$ appears to be a mixed state $\rho_A$ and its entanglement entropy is defined as the von Neumann entropy of $\rho_A$,

$$S(|\Psi\rangle, A) = -\text{tr}(\rho_A \log \rho_A). \tag{1}$$

Zero entropy means that the state still looks pure when confined to $A$, which means that $|\Psi\rangle$ must be in a simple product state, $|\Psi_A\rangle \otimes |\Psi_{\bar{A}}\rangle$, that contains no entanglement between $A$ and the rest of the system (represented by $\bar{A}$). This entanglement measure only applies to pure states, but there are many measures that are also suitable for mixed states.

The entanglement in the D-Wave device has a simple origin. It evolves a quantum state by continuously altering the weights ($a(t)$ and $b(t)$) of the two components ($H_x$ and $H_z$) of the device Hamiltonian,

$$H(t) = a(t) H_x + b(t) H_z. \tag{2}$$

The device is initialized at $t_0$ in the ground state of $H_x$ with $a(t_0) > 0$ and $b(t_0) = 0$. $a(t)$ is decreased and $b(t)$ is increased monotonically in time until the end of the device evolution at $t_1$ with $a(t_1) = 0$ and $b(t_1) > 0$. Because of the significant coupling between the device and its thermal environment, the quantum state of the D-Wave device is not a pure state described by a wave function, but rather a mixed state described by a density matrix at time $t$, $\rho(t)$. $\rho(t)$ is believed [84] to be a diagonal matrix in the basis of energy eigenstates of $H(t)$, which means that the system can only be as entangled as its instantaneous eigenstates. For example, there is no entanglement at times $t = t_0$ and $t = t_1$ when all the eigenstates have zero entanglement. The largest amount of entanglement in the D-Wave device is observed [77] at intermediate times when the energy gap in the device goes through a sequence of avoided crossings.

### G.2 Computational power of entanglement

While nonzero entanglement is necessary for any sort of quantum computational advantage, it is not possible to directly infer the latter from the former. There exist highly entangled quantum states such as stabilizer states that can be efficiently simulated classically, thus entanglement is not sufficient for a quantum advantage. Conversely, it is possible to systematically reduce the amount of entanglement within a quantum computer while preserving quantum advantages and only reducing the success rate of the computation [82]. The present view is that some measure of entanglement integrated over time and space is necessary to assess the maximum computational power of a quantum device, but such a performance metric has not yet been established.

It is important to mention the role of classical simulation of quantum states as tools for establishing boundaries between classical and quantum computation. If a quantum state or computation can be efficiently simulated by a classical computer, then it does not offer any computational advantage. In the case of Hamiltonians with local couplings on a linear graph, near-adiabatic time evolution starting from a ground state can be efficiently classically simulated with matrix product states (MPS) [85]. A linear graph can be embedded in the Chimera graph of the D-Wave device, and in this mode of operation it is expected that MPS simulations can efficiently model the behavior of the device. Other graphs, such as a Chimera graph, require a generalization of MPS called tensor network states [86]. While there have not yet been any reported attempts to simulate the D-Wave device with tensor network states, certain basic caveats are expected. The cost of these simulations grows rapidly with the amount of entanglement in a system, particularly if there is long-range entanglement on the graph. However, the coupling between the D-Wave device and its thermal environment might suppress long-range entanglement, which would then reduce the computational cost required to simulate it accurately. Thus it is important to address the open question of whether or not the D-Wave device is able to produce *long-range* entanglement, the absence of which might enable efficient classical simulations of the device.

So far, the most comprehensive efforts to assess the role of entanglement in the D-Wave device have combined theory, experiment, and simulation. The behavior of the D-Wave device has been experimentally characterized by a Markovian quantum master equation [77]. Actual devices have inhomogeneity in the coupling between qubits and with the thermal bath, therefore they do not exactly implement this master equation. Efficient classical simulations of the device must make approximations in order to simulate the master equation on many qubits. Thus, both the physical device and classical simulations are approximate implementations of an underlying reference quantum master equation. From this point of view, the D-Wave device has intrinsic errors that can only be reduced by better manufacturing to more accurately implement the underlying master equation or further refining the master equation to better reflect the actual performance of the physical device. It may be possible for classical simulations to be more accurate at representing the underlying master equation than the physical device if simulation methods improve more rapidly than device technology.

For a small number of qubits, the D-Wave device and approximate classical simulations have been compared with accurate solutions of the quantum master equation on an equal footing [67]. In order to distinguish between the full master equation and simpler models, this study needed to select a special set of problems that enhance quantum effects and study more observables than QUBO success probabilities (e.g. excited state populations and entanglement measures). Under these conditions it was possible to distinguish between the D-Wave device and a simple model of its qubits as classical rotors. Otherwise, the rotor model is able to reproduce QUBO success probabilities for previously studied problem sets [68], and a simple extension is able to reproduce qualitative trends in the success probabilities even for the special set of problems [78]. With a simple rotor model devoid of quantum entanglement able to reproduce QUBO success probabilities of the D-Wave device, it is possible that quantum entanglement serves no useful computational purpose in this type of quantum device. As a result of these studies, research activities have begun to focus more on identifying new problems that might somehow benefit from whatever quantum entanglement exists within the D-Wave device. Once such problems are identified, it is hoped that some subset of these problems will have solutions that are of practical value.

### G.3 Quantum advantages for the QUBO problem

It is also important to mention the most optimistic plausible outcome of a quantum speedup on the D-Wave quantum annealer. Strictly speaking, the annealer outputs samples from a distribution of low-energy eigenstates of $H(t_1) = H_z$ that have energies of the form

$$E_i = \sum_{j=1}^{N} f_j \langle \sigma_j^z \rangle_i + \sum_{j=1}^{N} \sum_{k=1}^{N} g_{j,k} \langle \sigma_j^z \rangle_i \langle \sigma_k^z \rangle_i, \qquad (3)$$

where $\langle \sigma_j^z \rangle_i \in \{-1, +1\}$ is the measured spin in the $z$ direction of eigenstate $i$ on qubit $j$. Ideally, we want to determine the minimum value of $E_i$ and the configuration of spins that generates it. The annealer will output this minimizer with a probability equal to the ground state occupation in $\rho(t_1)$, but nothing guarantees that this probability will be high for a specific problem. In fact, the lower bound on this probability must either decrease rapidly as $N$ increases or else the performance of this device will have major unforeseen implications in quantum complexity theory. This is because the minimization of $E_i$ in Eq. (3) is an NP-hard problem and there are no known or expected quantum algorithms to solve NP-hard problems efficiently [87]. In the circuit model of quantum computation, only a polynomial speedup, comparable to that offered by Grover search, is expected over classical algorithms for this problem. A brute-force classical algorithm would require $\Omega(2^N)$ evaluations of Eq. (3) in order to simply test the energy of each possible state to determine which attain the minimum energy. On the other hand there is a Grover-inspired quantum algorithm that requires $O(\sqrt{2^N})$ evaluations [88]. In an adiabatic model, which is more applicable to the D-Wave device, this Grover-like quantum speedup takes the form of a minimum energy gap in $H(t)$ that would scale as $O(1/\sqrt{2^N})$ [89]. This is an asymptotic quadratic improvement over classical algorithms.

Having no direct connection between entanglement and computational power, most recent performance assessments of the D-Wave quantum annealer have been relative to solvers run on serial classical computers [19]. The most recent and careful of such assessments, including ours (Sections 5 and 6), have found no evidence of a speedup, but it is difficult to compare such dissimilar things as a quantum annealer and algorithms run on classical computers, which are at a later and more advanced stage of their development. Perhaps the most direct way to assess a quantum annealer is relative to a comparable classical annealer. For example, if the operation of the D-Wave device was modified such that $a(t) = 0$ and $\rho(t_0)$ is the maximum entropy state, then it would perform as a purely classical thermal annealer. While the device is not setup such that users can do this, it is in principle a matter of simply turning couplings off. In both classical and quantum modes of operation, the device should, in principle, prepare $\rho(t_1)$ that is the Gibbs state of $H(t_1)$ if run sufficiently slowly. The quantum advantage here is expected to occur when the device is run at a speed that balances thermal and non-adiabatic depopulation of the ground state and maximizes the ground state occupation. The classical mode of operation would contain no quantum entanglement -- this is a very focused and well-posed experiment within which one might establish a correlation between the presence of quantum entanglement and a computational advantage (i.e. an increase in the final occupation of the ground state).

### G.4  Tunneling in the D-Wave device

Besides entanglement, the other quantum effect invoked in discussions of the D-Wave machine is tunneling. Tunneling is the generic term for a transition between two states by a quantum process that would be impossible for a purely classical process. Whether this is beneficial or detrimental for quantum annealing depends on the states under consideration. We consider two points of view with opposite conclusions. First, one might consider the state of the D-Wave device in the computational basis. In this basis, the device starts as a quantum mechanical superposition of all possible configurations and the $H_x$ term in Eq. (2) mediates "tunneling" between configurations that differ by a single spin. This process enables the uniformly distributed weight in the density matrix to focus on low-energy configurations of $H_z$ as the system progresses through the annealing schedule. Tunneling in this basis is then the primary quantum mechanism by which the quantum annealer functions. Alternatively, one might consider the state of the D-Wave device in the instantaneous $H(t)$ eigenstate basis. In this basis, the device starts in the ground state of $H(t_0)$ and the "tunneling" to excited states is an undesirable non-adiabatic effect that is minimized by reducing the rate of change of the Hamiltonian. Ideally, we desire no tunneling in this basis. However, while slower evolutions minimize tunneling to excited states, they allow more time for thermal processes that cause transitions to excited states. Thus, tunneling is a subjective concept and its effect on quantum annealing can change depending on the point of view that one takes.

Because of its subjective nature, quantum tunneling is not usually invoked in the technical discussion and analysis of quantum computing once the hardware has been abstracted away. It is more typically used as a convenient descriptor in the classical computing device physics literature. However, it persists as a compelling concept for describing how the state of a quantum annealer spreads out over the computational basis despite energy barriers and has been demonstrated to provide computational advantages over thermal annealing in special cases [79].

# Distribution

1    Dr. Steve Adachi
     Lockheed Martin Space Technology
     Advanced Research & Development Laboratories
     3215 Porter Drive
     Palo Alto, CA 94304-1215

1    Dr. Mark C. Anderson
     Weapons Physics Directorate
     Los Alamos National Lab
     P.O. Box 1663, M/S T086
     Los Alamos, NM 87544

1    Dr. Rupak Biswas
     Deputy Director, Exploration Technology Directorate
     Manager, High End Computing Capability (HECC) Project
     NASA Ames Research Center
     Moffett Field, CA 94035-1000

1    Chief Technology Officer
     D-Wave Systems Inc.
     3033 Beta Avenue
     Burnaby, British Columbia
     Canada V5G 4M9

1    Prof. Endre Boros
     Department of Management Science & Information Systems and RUTCOR
     100 Rockafeller Road, Room 5180
     Piscataway, NJ 08854

1    Dr. William T. Polk
     Assistant Director, Information Technology Research & Development
     Office of Science and Technology Policy
     Executive Office of the President
     Eisenhower Executive Building
     1650 Pennsylvania Avenue
     Washington, DC 20504

1    Dr. Randall Bryant
     Assistant Director, Information Technology Research & Development
     Office of Science and Technology Policy
     Executive Office of the President
     Eisenhower Executive Building

1650 Pennsylvania Avenue
Washington, DC 20504

1   Dr. William J. Camp
   33 Ranch Road
   Cedar Crest, NM 87008

1   Dr. Frank Gaitan
   Quantum Computing Group
   Laboratory for Physical Sciences
   8050 Greenmead Drive
   College Park, MD 20740

1   Dr. T.R. Govindan
   US Army Research Office
   PO Box 12211
   Research Triangle Park, NC 27709

1   Dr. William Harrod
   Advanced Scientific Computing Research
   U.S. Department of Energy
   SC-21/Germantown Building
   1000 Independence Ave., SW
   Washington, DC 20585

1   Thuc Hoang
   U.S. Department of Energy, NNSA
   Office of Advanced Simulation and Computing, NA-114
   1000 Independence Ave., SW
   Washington, DC 20585-0104

1   Dr. Robinson Pino
   Advanced Scientific Computing Research
   U.S. Department of Energy
   SC-21/Germantown Building
   1000 Independence Ave., SW
   Washington, DC 20585

1   Dr. Vadim Smelyanski
   NASA Ames Research Center
   Moffett Field, CA 94035-1000

1   Dr. Ceren Susut-Bennett
   Advanced Scientific Computing Research
   U.S. Department of Energy
   SC-21/Germantown Building
   1000 Independence Ave., SW

Washington, DC 20585

1 Dr. Greg Tallant
Lockheed Martin Flight Control and VMS Integration
P.O. Box 748
Fort Worth, TX 76101

1 Dr. Carl Weir
Lockheed Martin Information Systems & Global Solutions
230 Mall Boulevard
King of Prussia, PA 19406

| | | | |
|---|---|---|---|
| 1 | MS0110 | Wahid Hermina | 1200 |
| 1 | MS0110 | David Womble | 1220 |
| 1 | MS0351 | Robert Leland | 1000 |
| 1 | MS0621 | Dallas Wiener | 5632 |
| 1 | MS0621 | Kim Denton-Hill | 5638 |
| 1 | MS0801 | John Zepper | 9300 |
| 1 | MS1027 | Jeremy Wendt | 5632 |
| 1 | MS1071 | C. Andy Boye | 1720 |
| 1 | MS1079 | Gilbert Herrera | 1700 |
| 1 | MS1082 | Michael Descour | 1725 |
| 1 | MS1189 | Luke Shulenburger | 1641 |
| 1 | MS1231 | James Peery | 5000 |
| 1 | MS1248 | Steven Rinaldi | 5643 |
| 1 | MS1316 | Andrew Landahl | 1425 |
| 1 | MS1322 | Ken Alvin | 1420 |
| 18 | MS1322 | John Aidun | 1425 |
| 1 | MS1322 | Jonathan Moussa | 1425 |
| 1 | MS1324 | Bruce Hendrickson | 1400 |
| 5 | MS1326 | Ojas Parekh | 1464 |
| 1 | MS1326 | Cynthia Phillips | 1400 |
| 1 | MS1327 | Jonathan Berry | 1464 (electronic copy) |
| 1 | MS1327 | William Hart | 1464 |
| 1 | MS9151 | Leonard Napolitano | 8900 |
| 1 | MS9159 | James Costa | 8950 (electronic copy) |
| | | | |
| 1 | MS0899 | Technical Library | 9536 (electronic copy) |

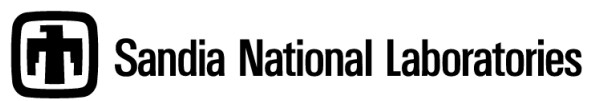